%% file: Master_thesis.tex
\documentclass[a4paper,12 pt,twoside]{book}

\usepackage{verbatim}
\usepackage[italian]{babel}
\usepackage{amsmath,amsthm}
\usepackage{graphicx}
\usepackage{subfigure}
\usepackage{wrapfig}
\usepackage{fancyhdr}
\usepackage{appendix}
\usepackage{pdflscape}
\usepackage{longtable}
\usepackage{color}
\usepackage[top=4 cm, bottom=5 cm, left=4.2 cm, right=3 cm]{geometry}

\begin{document}
%%%%%%%%%%%%%%%%%%%%%%%%%%%
%Titolo
%%%%%%%%%%%%%%%%%%%%%%%%%%%
\input{titolo.tex}

\frontmatter
\input{definizioni.tex}
%%%%%%%%%%%%%%%%%%%%%%%%%%%
%Introduzione
%%%%%%%%%%%%%%%%%%%%%%%%%%%
\tableofcontents
\chapter{Presentazione del lavoro di questa tesi}
\input{introduzione.tex}

%%%%%%%%%%%%%%%%%%%%%%%%%%%
%Parte prima
%%%%%%%%%%%%%%%%%%%%%%%%%%%
\mainmatter
\part{Theoretical background}
\input{convenzioni.tex}
\chapter{La fisica della materia}
\input{many_body.tex}
\chapter{La teoria del funzionale densit\`{a}}
\input{dft.tex}
\chapter{La teoria della funzione di Green}
\input{mbpt.tex}
\chapter{Approssimazioni}
\input{lda.tex}
\chapter{Spettro di eccitazione. L'accoppiamento particella buca.}
\input{excitations.tex}

%%%%%%%%%%%%%%%%%%%%%%%%%%%
%Parte seconda
%%%%%%%%%%%%%%%%%%%%%%%%%%%
\part{Studio di sistemi isolati }
\chapter{La formulazione di Casida della TDDFT}
\input{casida.tex}

\chapter{Risultati}
\input{sistemi.tex}

\chapter*{Conclusioni}
\addcontentsline{toc}{chapter}{Conclusioni}
\input{conclusioni.tex}

%%%%%%%%%%%%%%%%%%%%%%%%%%%
%Parte finale
%%%%%%%%%%%%%%%%%%%%%%%%%%%
%\backmatter
\appendixpage
\begin{appendices}
\input{appendix.tex}
\end{appendices}

%\printindex

%%%%%%%%%%%%%%%%%%%%%%%%%%%%%%
% Conclusioni
%%%%%%%%%%%%%%%%%%%%%%%%%%%%%%
\pagestyle{fancy}
\fancyhf{}
\fancyhead[LE,RO]{\bfseries\thepage}
%\fancyhead[LO]{\bfseries\rightmark}
%\fancyhead[RE]{\bfseries\leftmark}
\renewcommand{\headrulewidth}{0.5pt}
\renewcommand{\footrulewidth}{0pt}
\addtolength{\headheight}{0.5pt}
\fancypagestyle{plain}{%
              \fancyhead{}
              \renewcommand{\headrulewidth}{0pt}}

\input{ringraziamenti.tex}

\end{document}

%% file: titolo.tex
\begin{titlepage}

\begin{center}
%Immagine e commento sopra il titolo
\includegraphics[width=3 cm]{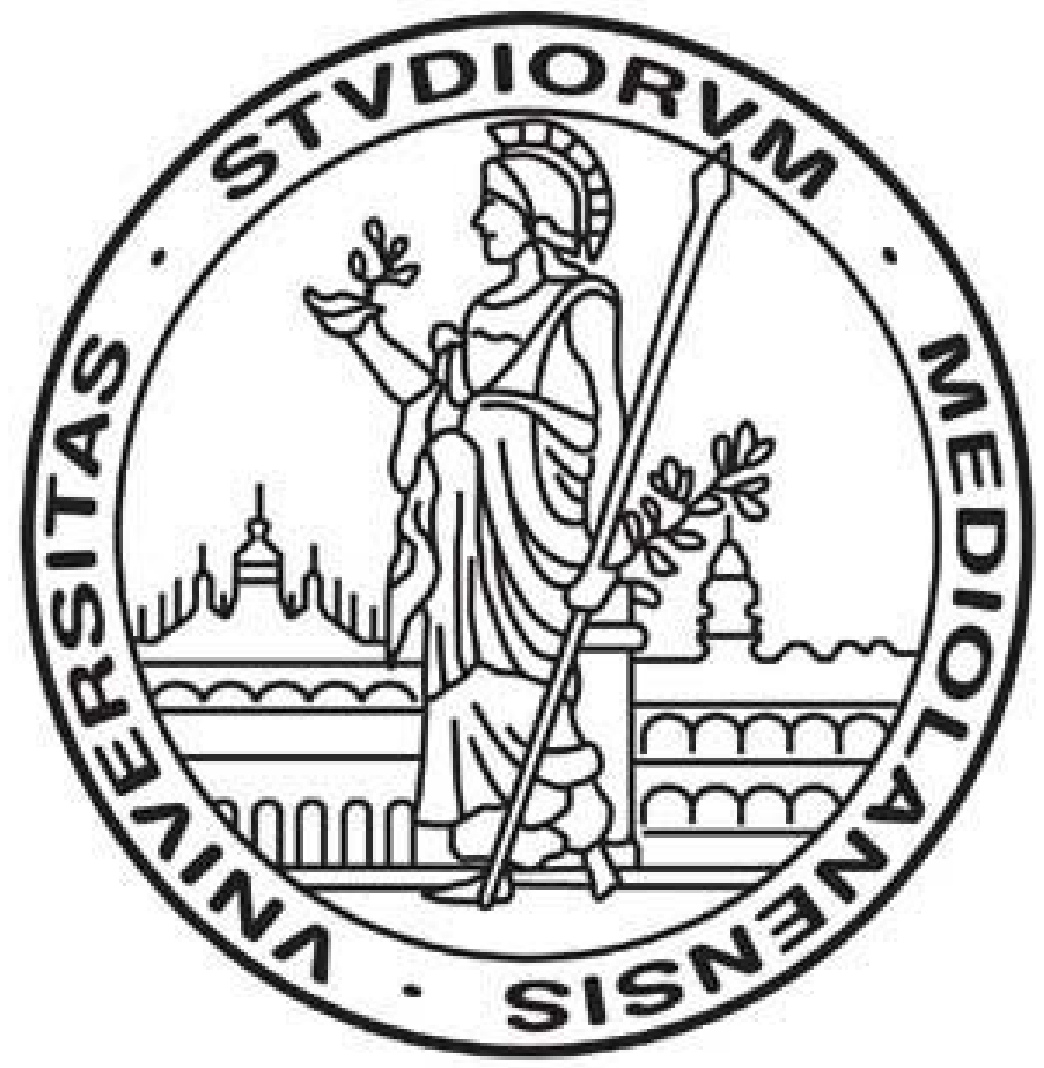}
\\
\textbf{Universit\`{a} degli studi di Milano} \\
\textbf{Corso di Laurea Magistrale in fisica} \\
\vspace{1.5 cm}
\rule{\linewidth}{0.5mm}
\\[0.5 cm]
%Titolo
\Huge
Metodi innovativi per il calcolo della risposta lineare nell'ambito della teoria
del funzionale densit\`{a} in sistemi spin polarizzati
\rule{\linewidth}{0.5mm}
\\[1.5 cm]
\end{center}

% Relatori
\begin{flushleft} \large
\emph{Codice PACS:} 71.15-m \\
\emph{Relatore:}  Giovanni ONIDA     \\
\emph{Correlatore:}  Nicola MANINI  \\
\end{flushleft}

\vfill
\begin{center}
\large
\textbf{\emph{Davide Sangalli}}
\end{center}

\end{titlepage}
\thispagestyle{empty}

%% file: introduzione.tex
La spettroscopia \`{e} lo studio della materia e delle sue propriet\`{a} realizzato
analizzando la luce, il suono, o le particelle che vengono emesse, assorbite o riflesse
dalla materia che si vuole descrivere. La spettroscopia pu\`{o} anche
essere definita come lo studio dell'interazione tra luce e materia \cite{Wikipedia}.
Storicamente la spettroscopia era riferita a un tipo di scienza
in cui la luce visibile veniva usata per studi teorici sulla struttura della materia
e per analisi qualitative e quantitative. Recentemente, comunque, la definizione ha assunto
un carattere pi\`{u} generale con lo sviluppo di tecniche sperimentali che utilizzano non
solo la luce visibile, ma anche molte altre forme di radiazione elettromagnetica e non:
microonde, onde radio, raggi X, elettroni, fononi (o onde sonore) e altre.
Tali tecniche vengono utilizzate nel campo della fisica e della chimica per l'identificazione
di sostanze attraverso lo studio dello spettro emesso o assorbito da esse; in astronomia la
spettroscopia viene fortemente utilizzata e molti telescopi sono dotati di spettrografi utilizzati
per misurare la composizione chimica e le propiet\`{a} fisiche di oggetti astronomici.

\begin{figure}[t]
\begin{center}
\includegraphics[width=12 cm]{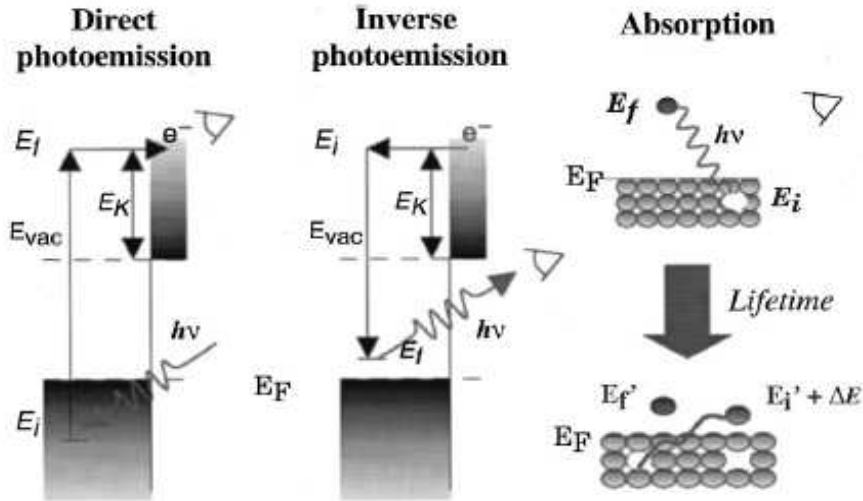}
\caption{Rappresentazione schematica delle eccitazioni coinvolte in differenti tipi di spettroscopia:
         fotoemissione diretta, fotoemissione inversa e assorbimento.}
\label{fig:Onida}
\rule{\linewidth}{0.2 mm}
\end{center}
\end{figure}

Le eccitazioni elettroniche in particolare sono alla base degli spettri di eccitazione
che vengono pi\`{u} comunemente studiati.
Negli ultimi vent'anni ci sono stati numerosi miglioramenti delle tecniche
sperimentali utilizzate per lo studio delle eccitazioni elettroniche in materia condensata;
d'altra parte molti degli aspetti fisici legati a tali esperimenti richiedono, per essere
descritti, opportuni modelli teorici.
Appare dunque interessante sviluppare tali modelli e verificarne la validit\`{a}
utilizzando i dati sperimentali pi\`{u} recenti.
Lo scopo \`{e} quello di disporre di strumenti in grado di prevedere i risultati
degli esperimenti prima di eseguirli. Infatti l'approccio teorico
offre il vantaggio di essere meno dispendioso, sia in termini economici che in termini 
di risorse umane e di tempo.

In ogni esperimento di spettroscopia un campione viene perturbato e si misura la risposta
del sistema a tale perturbazione, quindi in generale non \`{e} sufficiente calcolare le
propriet\`{a} di stato fondamentale del sistema per interpretare e prevedere i risultati
degli esperimenti. La foto-emissione diretta ed inversa e l'assorbimento possono essere
considerati come il prototipo dei fenomeni che si vorrebbero descrivere in questo contesto
(fig. \ref{fig:Onida}). Il lavoro della presente tesi \`{e} focalizzato sullo studio dello
spettro di assorbimento di sistemi isolati che presentino effetti legati allo spin.

La tesi si apre, Capitolo 1, con un'introduzione alle equazioni che sono alla base dello studio di
un sistema fisico tramite il formalismo della meccanica quantistica e vengono illustrate
quali siano le maggiori difficolt\`{a} che si incontrano nel tentativo di risolverle.
Sono quindi esposte nei due capitoli successivi le due teorie che vengono utilizzate per superare
tali difficolt\`{a} nella descrizione dello stato fondamentale del nostro sistema: la Density
Functional Theory (DFT), Capitolo 2, e la Many Boduy perturbation Theory (MBPT), Capitolo 3.
Nel Capitolo 4 vengono quindi illustrate le approssimazione introdotte nell'ambito della DFT, che
costituisce la teoria che utilizzeremo per i calcoli eseguiti nell'ambito di questa tesi.

La descrizione dello stato fondamentale del sistema \`{e} il punto di partenza per poterne descrivere
lo spettro di eccitazione ma, come gi\`{a} detto, non \`{e} sufficiente a comprendere
la fisica che sta alla base degli esperimenti di spettroscopia. Nel Capitolo 5 della tesi vengono
quindi esposte le estensioni delle due teorie presentate necessarie per lo studio delle energie di
eccitazione di un sistema ed in particolare l'equazione di Bethe-Salpeter, estensione della MBPT, e
la Time Dependet Density Functional Theory (TDDFT), estensione della DFT. Come per la parte di stato
fondamentale le approssimazioni necessarie per l'utilizzo della TDDFT vengono esposte nella parte finale
del capitolo.

Nella seconda parte della tesi ci concentriamo invece in modo pi\`{u} dettagliato su quello che \`{e}
il formalismo di Casida per la TDDFT, esposto nel Capitolo 6,
formalismo atto allo studio di sistemi isolati e che \`{e} in
particolare legato alla descrizione degli effetti di spin, presenti sia nello studio di sistemi a spin
perfettamente equilibrato, sia nello studio di sistemi con stato fondamentale spin polarizzato, polarizzazione
collineare. Nel capitolo vengono in particolare messe in luce le differenze tra i due tipi di sistemi
nell'ambito dell'approssimazione eseguita.

\begin{figure}[t]
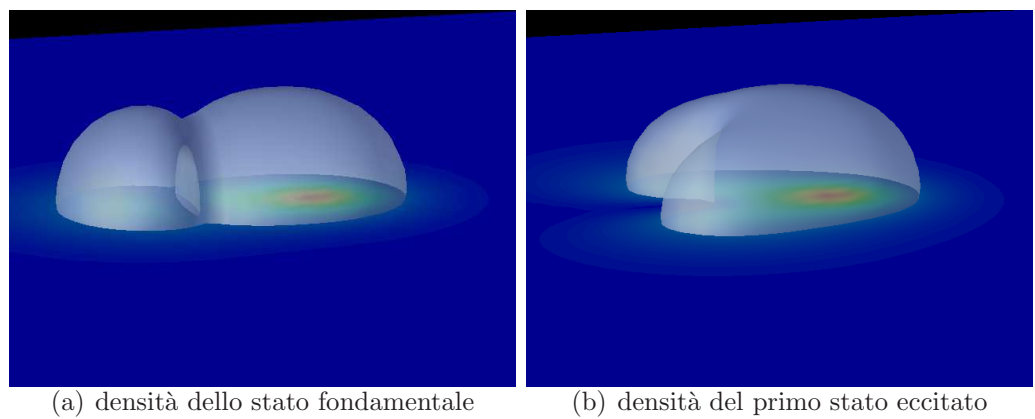
 
\begin{center}
%\vspace{-1 cm}
%\rule{\linewidth}{0.5mm}
\vspace{0.2 mm}
\subfigure[densit\`{a} dello stato fondamentale]{\includegraphics[width=5 cm, angle=-90]{immagini/densita_tot_ac30_ec10.epsf}}
\subfigure[densit\`{a} del primo stato eccitato]{\includegraphics[width=5 cm, angle=-90]{immagini/densita_ecc_ac30_ec10.epsf}}
\caption{Molecola test del BeH. %\\
         Raffronto tra la densit\`{a} dello stato fondamentale e quella del primo stato eccitato.
         Le immagini sono state costruite a partire dall'interpretazione delle funzioni d'onda di KS.
         Solo l'immagine dello stato fondamentale potrebbe, in base a giustificazione
         formale, essere interpretata come la densit\`{a} reale del sistema. Possiamo comunque osservare come
         la densit\`{a} elettronica attorno all'atomo di sinistra, atomo di idrogeno, sia diminuita  
         poich\`{e} la nube elettronica \`{e} ora distribuita su di uno stato eccitato. Quelle rappresentate
         in figura sono due isosuperfici a valore costante della densit\`{a}; i colori sul piano indicano invece
         il valore della densit\`{a} sul piano stesso, il rosso indica le zone di massima densit\`{a}, il blu le
         zone a densit\`{a} pi\`{u} bassa.}
\vspace{0.2 cm}
\rule{\linewidth}{0.2 mm}
\vspace{-1 cm}
\label{fig:BeH_view} 
\end{center}
\end{figure}

\begin{figure}[t]
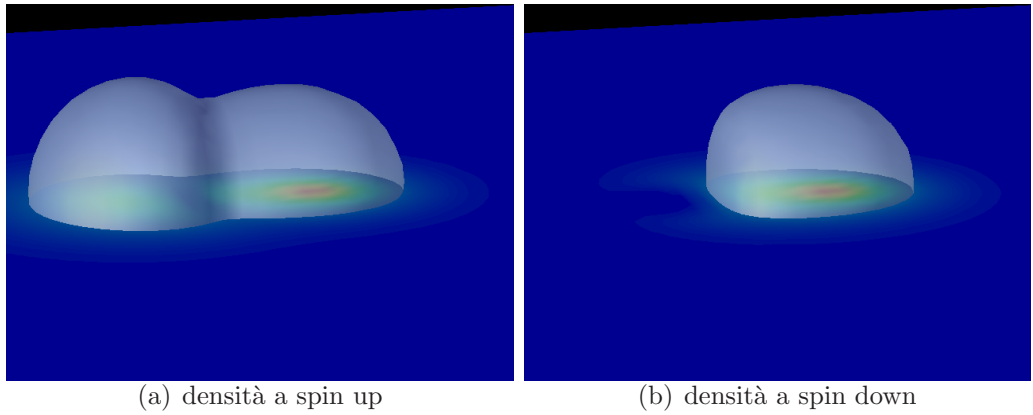
 
\begin{center}
%\vspace{-1 cm}
%\rule{\linewidth}{0.5mm}
\vspace{0.2 mm}
\subfigure[densit\`{a} a spin up]{\includegraphics[width=5 cm, angle=-90]{immagini/densita_up_ac30_ec10.epsf}}
\subfigure[densit\`{a} a spin down]{\includegraphics[width=5 cm, angle=-90]{immagini/densita_down_ac30_ec10.epsf}}
\caption{Molecola test del BeH. %\\
         Raffronto tra la componente a spin up e quella a spin down della densit\`{a} dello stato fondamentale.}
\vspace{0.2 cm}
\rule{\linewidth}{0.2 mm}
\vspace{-1 cm}
\label{fig:BeH_view2} 
\end{center}
\end{figure}

Il lavoro della presente tesi \`{e} stato in particolare mirato allo studio di sistemi con stato
fondamentale spin-polarizzato.
Le modifiche necessarie per lo studio di sistemi spin-polarizzati sono state implementate
all'interno del software ABINIT \cite{Abinit}, che in precedenza permetteva di studiare al livello
TDDFT solo sistemi non polarizzati. \`{E} stata poi scelta la molecola di BeH (fig. \ref{fig:BeH_view},
\ref{fig:BeH_view2}) per testare l'implementazione
fatta confrontando i risultati da noi ottenuti con altri calcoli teorici e con dati sperimentali,
Capitolo 7.

%% file: convenzioni.tex
\newpage
\section*{Convenzioni utilizzate nel capitolo 1} 
% come realizzare una pagina rotata di 90 gradi
% questa però non verra visualizzata nel file dvi
% ma solo nel pdf. Ciao
%\begin{landscape}
%\begin{table}
%\begin{center}
%\begin{tabular}{c c c c c}
%pippo & pippo & pippo & pippo & pippo \\
%pippo & pippo & pippo & pippo & pippo
%\end{tabular}
%\end{center}
%\end{table}
%\end{landscape}
%\end{document}

%Capitolo 1: \\
n: numero di elettroni \\
N: numero di ioni \\
m: massa elettrone \\
M: massa ione \\
e: carica dell'elettrone \\ 
$\mathbf{r}$: coordinata elettrone \\
$\mathbf{R}$: coordinata ione \\
p,q: indice elettroni \\
P,Q: indice ioni \\
io: abbreviazione per ione \\
el: abbreviazione per elettrone \\
T: energia cinetica \\
V: potenziale esterno \\
H: Hamiltoniana \\
i: $\sqrt{-1}$ \\
$\hbar$: costante di Plank ridotta \\
$\Psi$: funzione d'onda molti corpi \\

\begin{comment}
\\
Capitoli 2 e 3: \\
$\rho$: densit\`{a} del sistema \\
$E$: energia del sistema \\
$v_H$, $E_H$: potenziale ed energia di Hartree \\
$v_{xc}$, $E_{xc}$: potenziale ed energia di scambio e correlazione \\
$w$: interazione \\
pedice s: sistema non interagente \\
pedice I: rappresentazione di interazione \\
pedice S: rappresentazione di Schr\"{o}eginger \\
pedice H: rappresentazione di Heisinberg \\
indice J: autostati molti corpi dell'hamiltoniana completa \\
indici i,j,h,k: autostati a particella singola dell'hamilniana non interagente  \\
$h_0=T+V^{ext}$: Hamiltoniana di particella singola \\
$\Phi_0$: stato fondamentale a molti corpi del sistema non interagente \\
$\psi$: funzione d'onda particella singola \\
%$\phi$: funzione d'onda di quasi particella \\
$\varphi$: potenziale esterno applicato \\
$\varepsilon$, $\eta$: costanti arbitrarie positive \\
$\epsilon$: autovalore energia di particella singola e costante dielettrica \\
$\hat{\psi}, \hat{\psi}^\dag$ operatori di campo \\
$\hat{S}, \hat{S}_\epsilon$ operatore di scatering \\
$E$: autovalore energia stato a molticorpi \\
%$\xi$: autovalore energia di quasi-particella \\
T: grandezza T-ordinamento
\\
Capitolo 5: \\
$f_{Hxc}$, $K_{Hxc}$: Kernel della TDDFT e dell'equazione di Bethe Salpeter \\
apice R: grandezza ritardata
\end{comment}

%% file: many_body.tex
\section{Lo studio dei sistemi a molti corpi}
Studiare le propriet\`{a} di una molecola o di un materiale qualsiasi
scegliendo come punto di partenza la meccanica quantistica significa
cercare di risolvere l'equazione che regola il comportamento di tutte
le particelle che costituiscono tale sistema.
Si tratta pi\`{u} in specifico dell'equazione di Schro\"{e}dinger
\index{prova} per il
moto accoppiato di atomi ed elettroni, che, definite le seguenti grandezze
\begin{eqnarray*}
&&\hat{T}_{el}=-\sum_{p=1}^n \frac{\hbar^2 \nabla^2_{(p)}}{2m}                
\hspace{2 cm} \hat{T}_{ion}=-\sum_{P=1}^N \frac{\hbar^2 \nabla^2_{(P)}}{2M}              \\
&&\hat{V}_{el-el}=\sum_{p \neq q}\frac{e^2}{\left| \mathbf{r}_p - \mathbf{r}_q \right|}     
\hspace{2 cm}
\hat{V}_{ion-ion}=\sum_{P \neq Q}\frac{e^2}{\left| \mathbf{R}_P - \mathbf{R}_Q \right|}        \\
&&\hat{V}_{ion-el}=
  -\sum_{p=1}^n \sum_{P=1}^N \frac{e^2}{\left| \mathbf{R}_P - \mathbf{r}_p \right| }        \\
&&\Psi = \Psi \left( \mathbf{R}_1, \ldots, \mathbf{R}_N, \mathbf{r}_1, \ldots, \mathbf{r}_n, t \right)
\hspace{0.5 cm} \text{,}
\end{eqnarray*}
risulta:
\begin{equation} \label{Schroedinger}
\left( \hat{T}_{el}+\hat{T}_{ion}+\hat{V}_{el-el}+\hat{V}_{ion-ion}+\hat{V}_{ion-el} \right) \Psi =
i \hbar \frac{\partial \Psi}{\partial t} \hspace{0.5 cm} \text{.}
\end{equation}

Cos\`{i} formulata tale
equazione rappresenta un problema di difficolt\`{a} non risolvibile,
perlomeno con i mezzi matematici che abbiamo oggi a disposizione. Questo per
diversi motivi: innanzi tutto perch\'{e} \`{e} un'equazione differenziale alle
derivate parziali e perch\'{e} deve descrivere il comportamenti di un numero di
particelle interagenti che pu\`{o} essere dell'ordine del numero di Avogadro
($10^{23}$ particelle!).
Di questo tipo di equazione
esiste una soluzione esatta, almeno in 3 dimensioni, solo per lo studio di
una singola particella nel vuoto o immersa in potenziali che presentino particolari
simmetrie o infine per lo studio di due particelle interagenti. In altre parole per
quanto riguarda la fisica della materia siamo in grado di trovare una soluzione esatta
a questa equazione solo per quanto concerne lo studio di un elettrone libero o per
l'elettrone dell'atomo di idrogeno (l'unico elemento che presenta un solo elettrone).

L'avvento dei computer ha offerto una possibilit\`{a} alternativa al calcolo analitico
per la soluzione del problema. D'altra parte, come mostrato da Walter Kohn nelle sue
\emph{Nobel Lecture} \cite{Kohn_Nobel} anche con il supporto di software numerici la
possibilit\`{a} di studiare la funzione d'onda molti corpi resta limitata a sistemi che
presentano un numero di particelle ridotto. Cercando di studiare sistemi pi\`{u}
complessi ci si trova infatti di fronte a quello che Kohn definisce un ``Exponential
Wall'' (Muro Esponenziale) e che di fatto limita le attuali possibilit\`{a}, nel caso
di soluzione diretta dell'equazione, a sistemi contenenti circa 10 elettroni.

Per poter utilizzare la meccanica quantistica per i nostri scopi dovremo dunque ricorrere
a nuovi strumenti matematici che ci permettano di trasformare l'equazione scritta
e quindi introdurre diverse
approssimazioni cercando in qualche modo di mantenere il controllo, durante i calcoli,
di tutte le approssimazioni eseguite. La teoria che illustreremo all'interno di
questo lavoro costituir\`{a} proprio quell'insieme di approssimazioni e teoremi che ci
permetteranno di giungere, partendo dall'equazione (\ref{Schroedinger}),
ad un problema risolvibile.

Come vedremo, un ruolo chiave nella teoria verr\`{a} rivestito da equazioni di
Schro\"{e}dinger a particella singola che costituiscono, come gi\`{a} detto,
l'unico caso in cui siamo in grado di ottenere una soluzione.
Un ruolo chiave sar\`{a} svolto inoltre dalle grandezze integrate, in particolare la
densit\`{a} del sistema e la funzione di Green.
Le equazioni che troveremo saranno di tipo auto-consistente. Ci\`{o} significa che per
studiarle un ruolo dominante sar\`{a} svolto dai calcoli numerici,
poich\'{e} solo grazie all'avvento dei moderni computer siamo in grado di risolvere
tali equazioni che, altrimenti, rimarrebbero non risolvibili cos\`{i} come
l'equazione iniziale.

La teoria non verr\`{a} presentata secondo l'evoluzione storica che ha avuto,
piuttosto secondo quello che \`{e} lo schema logico che ho utilizzato per studiarli
e comprenderli.
 
Le due teorie pi\`{u} importanti che andremo ad illustrare nei prossimi capitoli
saranno la teoria della funzione di Green, che indicheremo come MBPT (Many-Body
Perturbation Theory), e la teoria del funzionale densit\`{a}, che indicheremo come
DFT (Density Functional Theory).

I teoremi di cui ci serviremo per lo sviluppo della teoria saranno: il teorema di
Hohenberg-Kohn e il teorema di Kohn e Sham per la DFT, e il loro equivalente per
l'estensione al caso dipendente dal tempo (TD-DFT); il teorema di Gell-Mann e Low
e il teorema di Wick per la MBPT.

Le prime due approssimazioni che introdurremo saranno l'approssimazione di
Born-Oppenheimer, che ci servir\`{a} per disaccoppiare il moto degli elettroni da
quello degli ioni e l'approssimazione di potenziale statico per il campo di
Coulomb (approssimazione che abbiamo in realt\`{a} gi\`{a} introdotto scrivendo
l'equazione di Schro\"{e}dinger), poich\'{e} considerando i ritardi anche le equazioni
della fisica classica diventano troppo complesse, immaginate quindi quelle della
meccanica quantistica.
Le altre approssimazioni principali di cui parleremo saranno la LDA (Local Density
Approximation) per la DFT e la A-LDA (Adiabatic-LDA) per il caso dipendente
dal tempo; l'approssimazione GW (chiariremo inseguito il significato) per la
formulazione Many-Body.

Le equazioni pi\`{u} importanti che verranno nominate per le due teorie saranno infine
le equazioni di Kohn e Sham per la DFT ed il loro equivalente per il caso dipendente
dal tempo (TDDFT); le equazioni di Hedin e l'equazione di Bethe-Salpeter per la MBPT

\section{L'approssimazione di Born-Oppenheimer}
L'approssimazione di Born-Oppenheimer consiste nel considerare disaccoppiati
il moto degli elettroni e il moto dei nuclei che costituiscono il nostro sistema
partendo dalla considerazione che essi presentano masse molto differenti
e quindi evolvono su di una scala di tempo molto differente.
Dal punto di vista matematico ci\`{o} significa che possiamo idealmente separare
la nostra equazione in due parti, la prima per il moto degli elettroni, nella quale
le posizioni degli atomi vengono considerate dei parametri e la seconda che possiamo
risolvere per trovare la posizione degli atomi una volta che abbiamo ottenuto la
funzione d'onda degli elettroni. Per il momento ci concentreremo solamente sull'equazione
del moto degli elettroni:

\begin{equation}
\left(\hat{T}_{el} +\hat{V}_{el-el}+\hat{V}_{ion-el}\right) \Psi=
i \hbar \frac{\partial}{\partial t} \Psi  \hspace{0.5 cm} \text{.}
\end{equation}

La ragione che ci induce a concentrarci, per il momento, sulla risoluzione dell'equazione per gli
elettroni nasce dal dal fatto che gran parte delle propriet\`{a} del sistema dipenderanno
proprio dalla soluzione di questa equazione, mentre per quanto riguarda gli ioni ci
baster\`{a} semplicemente conoscere la loro posizione di equilibrio (e non la funzione
d'onda) principalmente per determinare i parametri dell'equazione elettronica. Se
volessimo calcolare anche il moto degli ioni questo potrebbe ad esempio
essere fatto in approssimazione armonica per studiare le propriet\`{a} vibrazionali del sistema.
All'interno di questa tesi
comunque ci occuperemo solo dello studio delle eccitazioni elettroniche e dunque gli ioni,
una volta trovata la posizione di equilibrio, saranno considerati fermi.

\section{Il ruolo dell'interazione, le grandezze integrate e le funzioni d'onda
a particella singola}
L'equazione resta comunque un problema formidabile.
La maggiore difficolt\`{a} che deve essere affrontato \`{e} dovuta al fatto che il
nostro sistema presenta un numero di particelle molto elevato e che queste sono tra
loro interagenti. La conseguenza matematica \`{e} dunque quella di avere un'equazione
che presenta un numero molto elevato di variabili che devono essere studiate
contemporaneamente.
Entrambe le teorie che abbiamo citato in precedenza, la DFT e la teoria MBPT, sono state
elaborate appunto per superare tale problema.

Il problema del numero elevato di variabili viene superato da entrambe le teorie
grazie all'utilizzo delle ``grandezze integrate'', cio\`{e} di funzioni in cui tutte
le variabili, tranne poche, sono state integrate. Si tratta della densit\`{a}
per la DFT e della funzione di Green per la MBPT.
L'idea \`{e} quella di non voler necessariamente calcolare la funzione d'onda dell'intero
sistema ma di calcolare quantit\`{a} pi\`{u} maneggevoli, che siano per\`{o}
in grado di descrivere tutte le propriet\`{a} del sistema.

Anche in questo modo per\`{o} tali teorie restano inutilizzabili poich\'{e}
la presenza dell'interazione render\`{a} impossibile anche il calcolo
di queste due grandezze. Nell'affrontare questo problema le due teorie
presentano un approccio completamente differente.

La DFT riesce in modo elegante a spostare il ruolo dell'interazione all'interno di
un potenziale esterno che agisca su un sistema di particelle non interagenti.
A partire da questo, che sottolineiamo in principio non ha senso fisico,
vengono poi ricavate le grandezze fisiche ed in particolare la 
densit\`{a} e l'energia dello stato fondamentale. Il pregio
maggiore di questa teoria resta la semplicit\`{a} delle equazioni che, a 
partire da essa, possono essere costruite. In particolare grazie a questo tipo
di approccio si riesce a superare il problema del ``Muro Esponenziale'' di cui
abbiamo parlato in precedenza \cite{Kohn_Nobel}.
Il limite maggiore invece \`{e} costituito
dal fatto che la forma di tale potenziale non \`{e} noto e che dunque la teoria 
necessiti di un input esterno per poter essere utilizzata.

La MBPT affronta invece il problema cercando di giungere alla soluzione attraverso
un approccio di tipo perturbativo a partire dalla soluzione del sistema non interagente.
Il punto di partenza di tale approccio \`{e} quello di considerare l'interazione
come la grandezza perturbante del sistema.
La teoria viene poi riscritta sotto forma di un set di equazioni che descrivono le grandezze
fisiche caratteristiche del sistema.
In questa formulazione si partire
dalla soluzione del sistema non interagente e si calcolno via via le grandezze fisiche del
sistema in modo iterativo. Cos\`{i} facendo ad ogni iterazione si aggiungono le correzioni
alla soluzione del sistema non interagente dovute all'interazione.
Questo teoria offre il vantaggio, dal punto di vista dell'autore, di
essere pi\`{u} facilmente interpretabile grazie anche ai diagrammi di Feynman.
Lo svantaggio \`{e} in primo
luogo quello di richiedere strumenti matematici molto pi\`{u} raffinati e di presentare
calcoli molto pi\`{u} complessi della DFT.

Concludiamo questo capitolo anticipando che ci\`{o} che viene fatto in pratica \`{e}
utilizzare equazioni che sono il risultato dell'incrocio delle due teorie, spesso
andando al di l\`{a} di ci\`{o} che \`{e} stato formalmente giustificato. Ci\`{o} di
cui parleremo in questa tesi infatti si avvicina spesso a quelli che sono le 
frontiere della ricerca e a volte i modelli utilizzato trovano giustificazione solo a
posteriori nel fatto che sono in grado prevedere in modo corretto i risultati sperimentali.

%% file: dft.tex
\begin{comment}
References:
\\ - Hohenberg and Kohn, Phys. Rev. B 136, B864 (1964)
\\ - Kohn and Sham, Phys. Rev. 140, A1133 (1965)
\end{comment}

%%%%%%%%%%%%%%%%%%%%%%%%%%%%%%%%%%%%%%%%%%
%The Hohenberg-Kohn theorem
%%%%%%%%%%%%%%%%%%%%%%%%%%%%%%%%%%%%%%%%%%
\section{Il teorema di Hohenberg-Kohn}
\subsection{Dimostrazione del teorema}
In questo capitolo presenteremo quelle che sono le basi della
teoria del funzionale densit\`{a}. Il punto di partenza di tale teoria
\`{e} il teorema di Hohenberg-Kohn, il quale afferma che, dato un sistema
fisico di particelle in cui l'interazione \`{e} fissata, esiste una relazione
biunivoca tra il potenziale esterno applicato a tale sistema, la funzione
d'onda dello stato fondamentale e la densit\`{a} di particelle dello stato
fondamentale
\begin{equation}
\hat{V}^{ext} \Longleftrightarrow 
\Psi_0 \Longleftrightarrow
\rho_0
\nonumber \hspace{0.5 cm} \text{.}
\end{equation}

Dimostriamo il teorema scegliendo come modello un sistema di particelle descritto
dalla seguente Hamiltoniana
\begin{equation}
\hat{H}=\hat{T}+\hat{V}^{ext}+\hat{w} \hspace{0.5 cm} \text{;}
\nonumber
\end{equation}
dove $\hat{T}$ rappresenta l'operatore energia cinetica, $\hat{w}$ \`{e} l'interazione
e $\hat{V}^{ext}$ il potenziale esterno. Immaginiamo ora che l'interazione sia fissata,
mentre che il potenziale esterno possa essere libero, con la sola condizione per\`{o}
che lo stato fondamentale del sistema risulti non degenere.
Delle relazioni tra potenziale esterno, funzione d'onda dello stato fondamentale e
densit\`{a} dello stato fondamentale due sono, in principio, banali: la funzione
d'onda pu\`{o} essere infatti ricavata a partire dal potenziale esterno risolvendo
l'equazione di Schro\"{e}dinger%
\footnote{Affermiamo qui che la relazione \`{e} banale poich\`{e} nella dimostrazione
del teorema siamo semplicemente interessati alla sua esistenza in principio.
Resta invece non banale utilizzare in pratica tale relazione}
 e la densit\`{a} dalla funzione d'onda
\begin{equation}
\rho_0=\langle \Psi_0| \hat{\rho} |\Psi_0\rangle \text{.}
\end{equation}
Restano dunque da dimostrare le due relazioni inverse.

Partiamo dalla prima delle due implicazioni. Dimostreremo per assurdo che fissare
per il nostro sistema uno stato fondamentale significa fissare il potenziale esterno
a meno di una costante additiva, che per\`{o} non riveste alcuna importanza
dal punto di vista fisico poich\'{e} l'energia di un sistema pu\`{o} sempre essere
stabilita a meno di una costante.
Ipotizziamo dunque che esistano due differenti potenziali esterni che diano come
soluzione lo stesso stato fondamentale
\begin{eqnarray}
(\hat{T}+\hat{w}+\hat{V}^{ext}) |\Psi_0 \rangle = E_0 |\Psi_0 \rangle   \nonumber \\
(\hat{T}+\hat{w}+\hat{V}'^{ext}) |\Psi'_0 \rangle = E'_0 |\Psi'_0 \rangle
\hspace{0.5 cm} \text{,}
\end{eqnarray}
con $|\Psi_0 \rangle = |\Psi'_0 \rangle $. \\
Se ora sottraiamo la prima equazione alla seconda otteniamo
\begin{equation}
(\hat{V}^{ext}-\hat{V}'^{ext}) |\Psi_0 \rangle = (E_0 - E'_0) |\Psi_0 \rangle
\nonumber
\hspace{0.5 cm} \text{.}
\end{equation}
Applicando ad entrambe le equazioni l'operatore $\langle \Psi_0|$ e assumendo
che la funzione d'onda sia normalizzata ad uno otteniamo
\begin{equation}
{V}^{ext}(\mathbf{x})-{V}'^{ext}(\mathbf{x})= (E_0 - E'_0)
\hspace{0.5 cm} \text{,}
\end{equation}
il che dimostra appunto, assumendo che le funzioni d'onda siano eventualmente uguali
a zero solo in regioni dello spazio di misura nulla, che i due
potenziali possono differire al pi\`{u} di una semplice costante additiva.

Dimostriamo ora la seconda implicazione, vale a dire che fissata la densit\`{a}
risulta stabilita anche la funzione d'onda. A questo scopo dimostriamo che
due funzioni d'onda differenti non possono generare la stessa densit\`{a},
in altre parole che fissato $|\Psi_0 \rangle \neq |\Psi'_0 \rangle$ risulter\`{a}
$\rho_0 \neq \rho'_0$.
Per la dimostrazione utilizziamo il fatto che $|\Psi_0 \rangle$ \`{e} lo stato fondamentale
del sistema e dunque lo stato a minima energia:
\begin{equation}
E_0 = \langle \Psi_0  | \hat{H}  | \Psi_0  \rangle <
\langle \Psi'_0 | \hat{H} | \Psi'_0 \rangle
\nonumber
\hspace{0.5 cm} \text{,}
\end{equation}
dove $\hat{H}=\hat{T}+\hat{w}+\hat{V}^{ext}$ \`{e} l'Hamiltoniana associata al
potenziale esterno che ha $|\Psi_0 \rangle$ come ground state.
Riscriviamo ora l'equazione come
\begin{equation} \label{passaggio-dimostrazione}
\langle \Psi'_0 | \hat{H}'-\hat{V}'^{ext}+\hat{V}^{ext} | \Psi'_0 \rangle =
E'_0 + \int d^3 \mathbf{r}\ \rho'(\mathbf{r}) \big[V^{ext}(\mathbf{r})-V'^{ext}(\mathbf{r})\big] 
\hspace{0.5 cm} \text{,}
\end{equation}
dove $\hat{H}'=\hat{T}+\hat{w}+\hat{V}'^{ext}$ e abbiamo utilizzato il fatto che
l'operatore potenziale \`{e} un operatore moltiplicativo in spazio $(\mathbf{r})$.
Otteniamo cos\`{i} la seguente relazione
\begin{equation}
E_0 < E'_0 + \int d^3 \mathbf{r}\ \rho'(\mathbf{r})
\big[V^{ext}(\mathbf{r})-V'^{ext}(\mathbf{r})\big]
\hspace{0.5 cm} \text{.}
\end{equation}

Ripetendo lo stesso ragionamento partendo dalla funzione d'onda $| \Psi'_0 \rangle$
otteniamo l'equazione
\begin{equation}
E'_0 < E_0 + \int d^3 \mathbf{r}\ \rho(\mathbf{r})
\big[V'^{ext}(\mathbf{r})-V^{ext}(\mathbf{r})\big]
\nonumber
\hspace{0.5 cm} \text{.}
\end{equation}
Se infine ipotizziamo che $\rho(\mathbf{r}) = \rho'(\mathbf{r})$ e sommiamo le due
equazioni otteniamo
\begin{equation}
\nonumber
E_0 + E'_0 < E_0 + E'_0
\hspace{0.5 cm} \text{,}
\end{equation}
il che \`{e} ovviamente impossibile.

Ci\`{o} che abbiamo sin qui dimostrato \`{e} in realt\`{a} che le seguenti mappe
\begin{eqnarray}
F: \hat{V}^{ext} &\rightarrow& \Psi_0 \\ \nonumber
G: \Psi_0      &\rightarrow& \rho_0
\hspace{0.5 cm} \text{,}
\end{eqnarray}
sono iniettive. Affinch\'{e} esista invece una corrispondenza
biunivoca tra funzioni d'onda, densit\`{a} e potenziali abbiamo bisogno che esse siano
anche suriettive. La questione potrebbe apparire superflua poich\'{e} se vogliamo
utilizzare tale teoria per descrivere sistemi fisici reali (e se crediamo nella
validit\`{a} dell'equazione di Schro\"{e}dinger) \`{e} ovvio che utilizzeremo funzioni
d'onda che sono associate ad un qualche potenziale esterno. Allo stesso modo rester\`{a}
dunque ovvio che utilizzeremo, per la densit\`{a}, funzioni che abbiano un significato
fisico e che dunque debbano essere ottenibili a partire da una qualche funzione d'onda.
In realt\`{a}, poich\'{e} proprio la densit\`{a} costituir\`{a}, come vedremo, il punto
di partenza di tale teoria sar\`{a} ragionevole chiedersi quale debba essere lo spazio
delle funzioni da cui sia necessario partire nello svolgere i calcoli. Resta dunque non
banale la suriettivit\`{a} della seconda delle due mappe rispetto allo spazio delle funzioni.

Lo studio di tale problema necessita la formulazione di ulteriori teoremi ed una
discussione di questo fatto esula dagli scopi di questo lavoro.
Altro punto di cui non ci occuperemo all'interno di questa tesi \`{e} quello dell'estensione
di tale teorema a sistemi che presentano uno stato fondamentale degenere. Per chi fosse
interessato alla questione \`{e} possibile trovare una discussione approfondita del problema
nella referenza \cite{Testo_DFT}.

%******************************************************************************************
\subsection{Lo scopo del teorema di Hohenberg Kohn e il ruolo della densit\`{a}}
Come abbiamo gi\`{a} accennato in precedenza la densit\`{a} del sistema vuole essere la
grandezza da cui partire per studiare un sistema a molti corpi nell'ambito della teoria
del funzionale densit\`{a}. Si tratta infatti di una grandezza che, a differenza della
funzione d'onda, \`{e} funzione di una sola variabile e che dunque pu\`{o} essere pi\`{u}
facilmente controllabile. L'idea per poter utilizzar il teorema appena dimostrato \`{e}
quella di formulare un principio di minimo partendo proprio da questa grandezza.
Utilizzando infatti la mappa inversa della funzione $G$ possiamo costruire, in linea
teorica, una funzione d'onda a partire da una data densit\`{a}. Variando la densit\`{a} nello
spazio delle funzioni ad una variabile ci muoveremo nello spazio delle funzioni d'onda individuando
tutte quelle che sono lo stato fondamentale di un qualche sistema descritto da un'Hamiltoniana
del tipo
\begin{equation}
\hat{H}[\rho] = \hat{T} + \hat{w} + \hat{V}^{ext}[\rho]
\nonumber
\hspace{0.5 cm} \text{,}
\end{equation}
dove abbiamo indicato che il potenziale esterno sar\`{a} il funzionale della densit\`{a}
stabilito dall'inversione della funzione $F \cdot G$. Se ora costruiamo la seguente grandezza,
\begin{equation}
E = \langle \Psi[\rho] | \hat{H} | \Psi[\rho] \rangle
\nonumber
\hspace{0.5 cm} \text{,}
\end{equation}
dove per\`{o} questa volta l'Hamiltoniana \`{e} quella che descrive il nostro sistema fisico
e in cui dunque il potenziale esterno \`{e} fissato,
otteniamo un funzionale il cui minimo corrisponder\`{a} con il prendere la funzione d'onda
dello stato fondamentale e dunque la funzione d'onda e la
densit\`{a} che, all'interno dello schema di Hohenberg-Kohn sono legate all'Hamiltoniana
di partenza e che hanno significato fisico.

Abbiamo dunque raggiunto uno dei nostri scopi, cio\`{e} quello di costruire una teoria
che possa partire semplicemente dalla densit\`{a} del sistema piuttosto che dalla sua
funzione d'onda e quindi dalla soluzione dell'equazione di Schro\"{e}dinger a molti
corpi. Per trovare infatti lo stato fondamentale del nostro sistema non dovremo far altro
che minimizzare, rispetto alla densit\`{a} il funzionale appena costruito.
Il problema di questo approccio consiste per\`{o} nel fatto che il funzionale passa di
fatto ancora tramite la costruzione della funzione d'onda molti corpi e che il teorema di
Hohenberg-Kohn non fornisce alcun metodo pratico per costruire tale funzione.
Il problema della costruzione della funzione d'onda resta in ultima analisi legato a quello
del fatto che le particelle sono interagenti. Il modello che riesce a superare questo
ostacolo verr\`{a} spiegato nel prossimo paragrafo ed \`{e} stato formulato un anno dopo
il teorema appena dimostrato. Come abbiamo detto nell'introduzione, sebbene la grandezza
fondamentale per questa teoria resti la densit\`{a} del sistema, all'interno di questo
modello assumeranno un ruolo fondamentale le funzioni d'onda di particella singola che
sono le uniche funzioni d'onda che siamo in grado di calcolare
a partire dall'equazione di Schro\"{e}dinger. Sottolineiamo per\`{o} che tali funzioni
d'onda non avranno di fatto nessun significato fisico ma saranno semplicemente un artifizio
utilizzato per raggiungere i nostri scopi.

%***********************************************************************************************
%***********************************************************************************************
%***********************************************************************************************
\section{Lo schema di Kohn e Sham}
\subsection{Perch\`{e} scegliere di lavorare con un sistema non interagente}
Lo schema di Kohn e Sham nasce dall'idea di utilizzare come supporto
per i nostri scopi un sistema di particelle non interagenti, in modo che,
nei calcoli che dovremo fare, ci troveremo a lavorare con equazioni di
Schro\"{e}dinger a particella singola che saremo quindi in grado di risolvere.
La scelta di lavorare con le funzioni d'onda
di un sistema non interagente, anzich\'{e} direttamente con la densit\`{a}
del sistema, sembrerebbe una complicazione della teoria sin qui presentata
poich\'{e} in questo modo ci
troveremo a lavorare con grandezze a molte variabili e dunque dovremo nuovamente
affrontare il problea del numero elevato di particelle. D'altra parte questa scelta
offre il vantaggio di essere un punto di partenza migliore per introdurre delle
approssimazioni che consentano di utilizzare in pratica la teoria presentata.
In particolare il successo di questa scelta \`{e}
in gran parte dovuto al successo dell'approssimazione LDA di cui parleremo nel
capitolo 4 di questa tesi.

Anche in questo tipo di approccio la densit\`{a} resta comunque la grandezza
fondamentale dello schema e il punto di partenza della teoria.

\subsection{Le equazioni di Kohn e Sham}
L'idea \`{e} dunque quella di utilizzare il teorema di Hohenberg-Kohn per un
sistema ausiliario in cui l'interazione \`{e} nulla. In questo modo otterremo
a partire da una densit\`{a} di prova una funzione d'onda di particelle
indipendenti $\Psi_s[\rho]$ soluzione di un'Hamiltoniana con potenziale
esterno $V_s^{ext}[\rho]$
entrambi costruiti a partire dalle mappe $F_s$ e $G_s$ per un sistema
non interagente.
\begin{equation} \label{KS-equation}
\left(\hat{T}+\hat{v}_s[\rho]\right)\Psi_s[\rho]=E\Psi_s[\rho]
\hspace{0.5 cm} \text{.}
\end{equation}
L'apice ``s'' indica appunto il fatto che facciamo riferimento
ad un sistema in cui $\hat{w}=0$.

Vogliamo ora provare a riscrivere il funzionale della densit\`{a} da cui ricaviamo
l'energia del sistema nel suo minimo partendo dalla funzione d'onda $\Psi_s[\rho]$.
L'idea per tale funzionale \`{e} quella di fare il valor medio dell'Hamiltoniana
non interagente rispetto alla nuova funzione d'onda e di inserire poi dei
termini correttivi per tener conto dell'interazione e della differenza tra la
funzione d'onda reale e quella non interagente.
\begin{equation}
\begin{split}
E_s[\rho] =& \langle \Psi_s[\rho]
                     | \hat{T} + \hat{V}^{ext} | \Psi_s[\rho] \rangle
                     + \int \int 
                     \frac{\rho(\mathbf{r}) \rho(\mathbf{r}')}
                                       {\left| \mathbf{r}-\mathbf{r}' \right|}
                      d^3\mathbf{r}\ d^3 \mathbf{r}'
                     + E_{xc}[\rho]  \\    
                  =& T_s[\rho] + U^{ext}[\rho] + E_H[\rho]
                     +E_{xc}[\rho]
\hspace{0.5 cm} \text{,}
\end{split}
\end{equation}
Cerchiamo di analizzare il funzionale scelto.
Il primo termine \`{e} costituito dall'energia cinetica delle particelle non interagenti
($T_s[\rho]$) poich\'{e} la speranza \`{e} quella che non differisca di molto
dall'energia cinetica del sistema reale. Il secondo termine \`{e} costituito dall'energia
dovuta al potenziale esterno ($U^{ext}$) e poich\'{e} questa dipende solo dalla densit\`{a}
sar\`{a} identica a quella del sistema interagente. Il terzo termine \`{e} costituito
dall'energia di Hartree ($E_H[\rho]$) cio\`{e} l'energia dovuto a quello che
\`{e} i campo medio degli elettroni. \`{E} stato inserito poich\'{e} \`{e} l'unica parte
dell'interazione che sappiamo scrivere in modo esatto sotto forma di funzionale della densit\`{a}
e di cui quindi sappiamo calcolare l'energia; anch'esso dipende unicamente dalla densit\`{a}
e corrisponde all'energia di campo medio del sistema interagente. L'ultimo termine
infine viene definito come la differenza tra il funzionale costruito per il sistema interagente
e i termini sin qui considerati del funzionale appena costruito. Esso in particolare tiene conto
della differenza tra l'energia cinetica dei due funzionali e di quella dovuta tra
l'energia di Hartree e l'energia dovuta all'interazione.
\begin{equation}
E_{xc}[\rho]= \left( T[\rho] - T_s[\rho] \right) +
\left( E_w[\rho] - E_H[\rho] \right)
\nonumber
\hspace{0.5 cm} \text{,}
\end{equation}
dove con $E_w$ abbiamo indicato il valor medio dell'interazione $\hat{w}$ rispetto alla
funzione d'onda del sistema interagente. Scritta in particolare l'espressione dell'interazione
come
\begin{equation}
\hat{w}=\int\int d^3\mathbf{r}d^3\mathbf{r'}\ \frac{1}{|\mathbf{r-r'}|}
\mathbf{\hat{\psi}^{\dag}(r)\hat{\psi}^{\dag}(r')\hat{\psi}(r')\hat{\psi}(r)}
\hspace{0.5 cm} \text{,}
\end{equation}
possiamo ora definire la ``funzione di correlazione a coppie''
\begin{equation} \label{correlazione-coppie}
g(\mathbf{r,r'},[\rho])=\mathbf{\frac{\langle\hat{\rho}(r)\hat{\rho}(r')\rangle}
{\rho(r)\rho(r')}-\frac{\delta(r-r')}{\rho(r)}}
\hspace{0.5 cm} \text{,}
\end{equation}
e scrivere in modo esplicito il valor medio $\langle w(\mathbf{r,r'})\rangle$
\begin{equation}
E_{w}[\rho]=\frac{1}{2}\int d^3\mathbf{r}\int d^3\mathbf{r'}\ \frac{\rho(\mathbf{r})\rho(\mathbf{r'})}
{|\mathbf{r-r'}|}g(\mathbf{r,r'},[\rho])
\hspace{0.5 cm} \text{,}
\end{equation}
il che ci d\`{a} un'idea della differenza di questo rispetto all'energia di Hartree.

Poich\'{e} il funzionale cos\`{i} costruito \`{e} identico a quello del sistema interagente
se ora cerchiamo il minimo rispetto alla densit\`{a} otterremo, come in precedenza,
la densit\`{a} e l'energia di stato fondamentale del nostro sistema. Ci\`{o} che
invece non avr\`{a} significato fisico sar\`{a} la funzione d'onda.

Possiamo ora sfruttare il fatto che stiamo studiando un sistema di particelle non
interagenti e scrivere la funzione d'onda come il prodotto di funzioni d'onda di
particella singola. In questo modo in particolare la densit\`{a}
del sistema non sar\`{a} altro che la somma dei moduli quadri delle funzioni d'onda.
Pensando ora al nostro funzionale come a un funzionale composto scegliamo di minimizzarlo
rispetto alle funzioni d'onda anzich\'{e} rispetto alla densit\`{a}. Poich\'{e} per\`{o}
le funzioni d'onda rispetto a cui vogliamo minimizzare il nostro funzionale non
possono variare liberamente nello spazio delle funzioni d'onda (che \`{e} il prodotto
tensoriale degli spazi di Hilbert di particella singola) ma devono essere ortonormali
dovremo cercare un minimo vincolato. Utilizzando la teoria dei moltiplicatori di
Lagrange l'equazione che dovremo risolvere sar\`{a}:
\begin{equation}
\sum_{i=1}^n \frac{\delta}{\delta \psi_i^*}
             \left(E[\rho(\psi_1,...,\psi_n)] +
             \sum_{h,k} \left( \delta_{h,k} - \lambda_{h,k} \int \psi_h(\mathbf{r})
             \psi_k(\mathbf{r}) d^3\mathbf{r} \right) \right) |{\psi}_i\rangle = 0
\hspace{0.1 cm} \text{.}
\end{equation}
Il fatto che le funzioni d'onda sono ortogonali ci permette ora di pensare
all'equazione appena scritta come ad un set di $n$ equazioni differenti.
Scriviamo dunque in modo esplicito tali equazioni, poich\'{e} esse costituiscono
in effetti il set di equazioni a particella singola che siamo in grado di risolvere
e che cercavamo. A questo scopo utilizziamo la seguente relazione tra
la derivata funzionale rispetto alle funzioni d'onda e quella rispetto alla densit\`{a}
\begin{equation}
\frac{\delta}{\delta \psi_i^{\ast}} = \frac{\partial \rho}{\partial \psi_i^{\ast}}
       \frac{\delta}{\delta \rho}
        = \psi_i \frac{\delta}{\delta \rho}
\hspace{0.5 cm} \text{,}
\end{equation}
in modo da poter finalmente scrivere l'equazione a particella singola nel seguente modo:
\begin{eqnarray}
\frac{\delta}{\delta \psi_i^{\ast}} T_s + \psi_i\frac{\delta}{\delta \rho}
    \left( E_H + U^{ext} + E_{xc} \right) = \lambda_i \psi_i   \nonumber   \\
\left( \hat{T}+ \hat{v}_H[\rho]+ \hat{V}^{ext}+
       \hat{v}_{xc}[\rho] \right) \psi_i = \lambda_i \psi_i
\hspace{0.5 cm} \text{.}
\end{eqnarray}

Siamo dunque giunti alla fine delle nostre fatiche, almeno per quanto concerne la formulazione
della teoria che dovremo poi utilizzare. Cerchiamo di analizzare i termini di questa
equazione e capire come possiamo pensare di utilizzarla ai nostri scopi.
Come volevamo \`{e} un'equazione a particella singola, abbiamo un termine per
l'energia cinetica, un termine per il campo medio, il potenziale esterno e un potenziale
che tiene conto del fatto che il risultato della nostra equazione deve essere un set
di funzioni d'onda che generi una densit\`{a} identica a quella di un set di particelle
reali poste all'interno dello stesso potenziale esterno presente nella nostra equazione.
Otteniamo dunque che, se sommiamo in un unico termine il potenziale esterno,
il potenziale di Hartree
e il termine di scambio e correlazione otteniamo esattamente il potenziale $\hat{v}_s$ 
introdotto nell'equazione (\ref{KS-equation}).

%************************************************************************************************
%************************************************************************************************
%************************************************************************************************
\section{Alcune considerazioni}
\subsection{Lo spin}
Finora abbiamo parlato di DFT prendendo in considerazione la densit\`{a} e abbiamo mostrato
come l'energia dello stato fondamentale del sistema possa essere espressa come funzionale di
questa grandezza. Tutte le dimostrazioni sin qui eseguite restano di fatto valide se il potenziale
esterno pu\`{o} essere espresso nella forma scelta per la dimostrazione del teorema, ovvero
nel caso in cui non siano presenti campi magnetici. Se volessimo includere gli effetti di un
campo magnetico, perlomeno l'accopiamento del campo con il momento magnetico dovuto allo spin
\begin{equation}
\hat{V}^{ext}= \int d^3\mathbf{r}\ \left[V^{ext}(\mathbf{r})\hat{\rho}(\mathbf{r})-
                                \mathbf{B}(\mathbf{r}) \cdot \hat{\mathbf{m}}(\mathbf{r})
                                  \right]
\end{equation}
dovremmo riscirivere il passaggio (\ref{passaggio-dimostrazione}) nel seguente modo
\begin{multline}
\langle \Psi'_0 | \hat{H}'-\hat{V}'^{ext}+\hat{V}^{ext} | \Psi'_0 \rangle =
E'_0 \\
+ \int d^3 \mathbf{r}\ \rho'(\mathbf{r}) \big[V^{ext}(\mathbf{r})-V'^{ext}(\mathbf{r})\big]-
\mathbf{m}[\rho'](\mathbf{r}) \cdot \big[\mathbf{B}(\mathbf{r})-\mathbf{B}'(\mathbf{r})\big]
\hspace{0.5 cm} \text{,}
\end{multline}
il che porterebbe per\`{o} ad avere un potenziale che agisce in modo non locale sulla funzione
d'onda nello schema di KS \cite{Testo_DFT}. 

Quello che dunque viene
solitamente fatto \`{e} sviluppare la teoria scrivendo un funzionale della densit\'{a} e della
magnetizzazione (o della matrice densit\`{a} di spin, il che \`{e} equivalente) in
modo da poter lavorare con dei campi locali. Non riportiamo qui la dimostrazione di
questa estensione dei teoremi visti; nel corso del nostro lavoro faremo comunque uso
di questo funzionale
\begin{equation}
E = E[\rho_{\alpha\beta}]
\hspace{0.5 cm} \text{,}
\end{equation}
poich\'{e} tale teoria viene utilizzata anche per lo studio di sistemi che presentano uno stato
fondamentale spin-polarizzato.

Poich\'{e} in realt\`{a} all'interno di questo lavoro ci limiteremo a studiare sistemi collineari
e perturbazioni con campi magnetici diretti secondo l'asse $z$ sar\`{a} sufficiente studiare il
seguente funzionale
\begin{equation}
E = E[\rho_{\uparrow},\rho_{\downarrow}]
\end{equation}
in cui l'energia \`{e} cio\`{e} funzionale delle sole componenti diagonali della matrice densit\`{a}
di spin. Vedremo pi\`{u} in dettaglio quale sia il significato di tale scelta nel capitolo 4.

Se infine volessimo inserire anche l'accoppiamento del campo magnetico con la corrente elettronica
dovremmo sviluppare ulteriormente la teoria. Ci\`{o} esula per\`{o} dagli scopi di questo lavoro.

%************************************************************************************************
\subsection{La teoria costruita}
Abbiamo costruito una teoria che possa essere utilizzata per studiare un sistema di particelle
interagenti. Come volevamo siamo riusciti a semplificare l'equazione di Schro\"{e}dinger in modo
da dover studiare solo equazioni a particella singola, partendo da una grandezza integrata
(la densit\`{a}).
Le quantit\`{a} che siamo in grado di calcolare sono l'energia dello stato fondamentale del
sistema e la sua densit\`{a}. Non possiamo invece dire niente, almeno secondo giustificazione
formale, della funzione d'onda. 

La cosa pi\`{u} notevole del nostro risultato credo consista nel fatto che nei passaggi sin qui
svolti non abbiamo introdotto nessuna approssimazione. Le uniche approssimazioni fin qui
utilizzate sono quella di Born-Oppenheimer (di cui abbiamo gi\`{a} parlato) e il fatto che
stiamo utilizzando un'interazione di tipo statico. Ovviamente sarebbe tutto troppo bello se
potessimo accontentarci di queste due approssimazioni. La nostra teoria presenta infatti un
elemento oscuro che \`{e} appunto il potenziale di scambio e correlazione (o se preferite
l'energia di scambio e correlazione). Per una discussione sulla questione rimandiamo
nuovamente al capitolo 4 del presente lavoro.

Una considerazione a parte merita la questione dello studio dello spettro di eccitazione del sistema.
Fino ad ora infatti ci siamo limitati a parlare di energia di stato fondamentale e abbiamo sviluppato
un metodo pratico, lo schema di Kohn e Sham, per poter calcolare tale grandezza. Ritornando al teorema
di Hohenberg-Kohn per\`{o} possiamo osservare come, poich\'{e} il potenziale esterno pu\`{o} essere
scritto come funzionale della densit\`{a} a partire da questo, tramite la soluzione dell'equazione di
Schro\"{e}dinger potremmo, in principio, riuscire a calcolare tutti gli stati eccitati del nostro
sistema. In pratica avremmo bisogno di sviluppare un metodo, equivalente allo schema di Kohn e
Sham, per il computo delle energie di eccitazione del sistema. Nel capitolo sullo studio dello spettro
di eccitazione vedremo come tale metodo pratico \`{e} in effetti costituito dalla Time-Dependent DFT
(TDDFT) per quel che riguarda lo studio delle energie di eccitazione per stati che presentano
lo stesso numero di particelle dello stato fondamentale. Non ci occuperemo invece dello studio 
delle energie di eccitazione di stati con un numero di particelle differenti dal quello del stato
fondamentale. Per chi fosse interessato al problema rimandiamo come sempre alla referenza \cite{Testo_DFT}
e all'articolo \cite{Onida}.

%% file: mbpt.tex
\section{Introduzione}
In questo capitolo cercheremo di introdurvi a quella che \`{e} la
``Many Body Perturbation Theory'' o se preferite ``Teoria della funzione di Green''.
Per la comprensione del capitolo sar\`{a} richiesta, oltre che la conoscenza
della meccanica quantistica secondo la formulazione di Schro\"{e}dinger anche
il formalismo matematico della seconda quantizzazione e dunque la definizione
di operatore di campo. Cercheremo di delineare tutti quei passaggi che portano
alla costruzione delle formule matematiche che verranno poi utilizzate all'interno
di questo lavoro e, per quanto possibile di riportare anche una dimostrazione dei
teoremi che via via citeremo. Poich\'{e} d'altro canto
questa teoria \`{e} di fatto molto pi\`{u} elaborata della DFT non ci sar\`{a}
possibile presentare in modo semplice e lineare tutti i suoi risultati.
Saremo quindi spesso costretti a rimandare il lettore ad altri libri specialistici
o a pubblicazioni sull'argomento.
In particolare all'interno di questa tesi faremo spesso riferimento
al testo di Fetter e Walecka \cite{Testo_MB}, ``Quantum Theory of
Many-Particle Systems'' che costituisce il libro su cui ho studiato tale teoria.

%********************************************************************************
%********************************************************************************
%********************************************************************************
\section{La funzione di Green}
\subsection{Definizione}
La MBPT cerca
di descrivere il nostro sistema di particelle interagenti a partire da una grandezza
integrata, la funzione di Green, cos\`{i} come la DFT partiva dalla densit\`{a} del
sistema. Se per\`{o} \`{e} certo chiaro a tutti cosa sia la densit\`{a} \`{e} sicuramente
pi\`{u} difficile spiegare cosa sia la funzione di Green. A dire il vero esistono perlomeno
due differenti definizioni di questa grandezza e, non avendo ben chiaro quale sia il suo
significato fisico, risulta ancora pi\`{u} complicato capire perch\'{e} le differenti
definizioni possano portare alla stessa grandezza. La scelta che faremo all'interno di
questo lavoro sar\`{a} quella di presentare innanzi tutto la funzione di Green secondo la
definizione che \`{e} alla base della MBPT. A partire da questa cercheremo di capire
quale possa essere il suo significato fisico e quale sia il motivo per cui sia stata
scelta tale grandezza per descrivere il nostro sistema. Il riferimento all'altro modo in
cui \`{e} possibile definire tale funzione verr\`{a} fatto unicamente in un secondo momento
per aiutarci nella comprensione del significato fisico di tale grandezza.

La funzione di Green \`{e} dunque definita nel seguente modo:
\begin{equation} \label{Green-def}
\begin{split}
iG_{\sigma_1\sigma_2}(\mathbf{x}_1,t_1;\mathbf{x}_2,t_2) =& 
\frac{\langle \Psi_0 |T \left[ \hat{\psi}_{\hat{H},\sigma_1}
      (\mathbf{x}_1,t_1) \hat{\psi}^{\dag}_{\hat{H},\sigma_2}(\mathbf{x}_2,t_2) \right] |
      \Psi_0 \rangle}
      {\langle \Psi_0 | \Psi_0 \rangle}  \\
iG(1,2) =& \frac{\langle \Psi_0 |T \left[ \hat{\psi}_{\hat{H}}(1)
          \hat{\psi}^{\dag}_{\hat{H}}(2) \right] |
          \Psi_0 \rangle}{\langle \Psi_0 | \Psi_0 \rangle}
\hspace{0.5 cm} \text{.}
\end{split}
\end{equation}
Nella seconda linea abbiamo introdotto la convenzione $1=(\mathbf{x}_1,t_1,\sigma_1)$ di
cui faremo uso a volte all'interno di questa tesi.
Questa grandezza \`{e} in particolare la funzione di Green $T$-ordinata a particella singola.
\`{E} dunque definita come il
valor medio, rispetto allo stato fondamentale $|\Psi_0 \rangle$ di un sistema descritto
da un'Hamiltoniana indipendente dal tempo.
Possiamo immaginare sia la stessa Hamiltoniana che abbiamo utilizzato nel presentare la DFT e
dunque l'Hamiltoniana
del nostro problema many-body introdotto nel primo capitolo di questa tesi
\begin{equation}
\hat{H}= \hat{T} + \hat{w} + \hat{V}^{ext}
\hspace{0.5 cm} \text{,}
\end{equation}
dove ancora $\hat{T}$ rappresenta l'operatore energia cinetica, $\hat{w}$ l'interazione
e $\hat{V}^{ext}$ il potenziale esterno del sistema, nel nostro caso quello degli ioni.
Il valor medio \`{e} fatto sugli operatori di campo in descrizione di Heisenberg
(vedi paragrafo successivo).
I due operatori sono sotto l'operatore di $T$-ordinamento, o ordinamento temporale,
che \`{e} definito nel seguente modo%
\footnote{Pi\`{u} in generale il $T$-ordinamento di N operatori di campo \`{e} definito come
$T[\hat{\psi}(t_3)\hat{\psi}(t_1)\hat{\psi}(t_2)]=(-1)^P \hat{\psi}(t_1)\hat{\psi}(t_2)
\hat{\psi}(t_3)$ con $P$ uguale al numero di permutazioni eseguite e con
$t_1 > t_2 > t_3$}
\begin{eqnarray}
T \left[ \hat{\psi}_{\hat{H}}(1) \hat{\psi}^{\dag}_{\hat{H}}(2) \right] =
          \theta (t_1-t_2) \hat{\psi}_
          {\hat{H}}(1) \hat{\psi}^{\dag}_{\hat{H}}(2) - 
          \theta (t_2-t_1) \hat{\psi}^{\dag}_{\hat{H}}(2) \hat{\psi}_{\hat{H}}(1)
\hspace{0.2 cm} \text{.}
\end{eqnarray}

In altre parole i due operatori vengono $T$-ordinati facendo agire prima quello il cui
argomento temporale \`{e} precedente. Il segno meno tra i 2 operatori \`{e} dovuto al
fatto che stiamo studiando un sistema di fermioni e che quindi gli operatori di campo
sono definiti secondo regole di anti-commutazione%
\footnote{La teoria della funzione di Green \`{e} in realt\`{a} pi\`{u} generale e pu\`{o}
essere utilizzata anche per lo studio di bosoni. All'interno di questa tesi ci limiteremo per\`{o}
a studiare la funzione di Green fermionica}.

%**************************************************************************************
\subsection{Rappresentazione di Schro\"{e}dinger, Heisenberg e interazione}
Nella formulazione pi\`{u} semplice della meccanica quantistica, che chiamiamo
rappresentazione di Schro\"{e}dinger, l'evoluzione temporale del sistema \`{e}
descritta dall'evoluzione delle funzioni d'onda o stati del sistema.
Come noto infatti sono le funzioni d'onda a descrivere tutte le propriet\`{a} del
sistema. Quando ci si trova per\`{o} a dover studiare sistemi con un numero molto
elevato di particelle, come abbiamo gi\`{a} detto, le funzioni d'onda diventano
oggetti difficili da controllare e si deve dunque ricorrere ad altro.
Il formalismo matematico che viene utilizzato per questo scopo \`{e} appunto
quello della seconda quantizzazione. In esso parte centrale viene svolta dagli
operatori e in particolare dagli operatori di campo, l'idea \`{e} dunque quella
di spostare anche l'evoluzione temporale del sistema su questi oggetti.
Vediamo come.

Nella formulazione di Schro\"{e}dinger le funzioni d'onda sono rappresentate dalle
soluzioni dell'equazione
\begin{equation}
i \hbar \frac{\partial}{\partial t} |\Psi_S(t)\rangle = \hat{H} 
   | \Psi_S(t)\rangle
\hspace{0.5 cm} \text{,}
\end{equation}
dove il pedice $S$ indica che \`{e} la rappresentazione di Schro\"{e}dinger.
L'operatore evoluzione temporale in questa descrizione risulta:
\begin{equation}
e^{-i\frac{\hat{H}t}{\hbar}}
\hspace{0.5 cm} \text{,}
\end{equation} 
Se ora vogliamo una rappresentazione in cui l'evoluzione temporale
venga tolta dagli stati (che chiamiamo rappresentazione di Heisenberg) possiamo
definire
\begin{equation} \label{Heisenberg-def}
|\Psi_H\rangle = e^{i\frac{\hat{H}t}{\hbar}} |\Psi_H(t)\rangle 
\hspace{0.5 cm} \text{,}
\end{equation}
da cui si vede facilmente che $|\Psi_H\rangle $ \`{e} indipendente dal tempo:
\begin{equation}
i \hbar \frac{\partial}{\partial t} |\Psi_H\rangle = 0
\hspace{0.5 cm} \text{.}
\end{equation}
Se ora vogliamo imporre che le due formulazioni siano equivalenti dobbiamo richiedere
che i valori medi degli operatori di Heisenberg rispetto a tali stati siano uguali
ai valori medi degli operatori di Schro\"{e}dinger rispetto agli stati di Schro\"{e}dinger.
\begin{equation}
\langle \Psi_H | \hat{O}_H | \Psi_H \rangle = \langle \Psi_S | \hat{O} | \Psi_S \rangle
\hspace{0.5 cm} \text{,}
\end{equation}
da cui utilizzando l'inversa della formula (\ref{Heisenberg-def}) si vede facilmente che
\begin{equation}
\hat{O}_H = e^{i\frac{\hat{H}t}{\hbar}} \hat{O} e^{-i\frac{\hat{H}t}{\hbar}}
\hspace{0.5 cm} \text{,}
\end{equation}
dove gli operatori senza pedice sono quelli in descrizione di Schro\"{e}dinger.

La maggior difficolt\`{a} del problema many-body \`{e} costituita dal fatto che le
particelle siano tra loro interagenti.
Un punto di partenza migliore per superare tale difficolt\`{a} sarebbe quindi costituito
da una descrizione che separa l'evoluzione temporale dovuta all'interazione da quella dovuta
alla parte restante dell'Hamiltoniana. Vediamo come.

Immaginiamo la nostra Hamiltoniana suddivisa in due parti, entrambe indipendenti dal tempo,
\begin{equation}
\hat{H}=\hat{H}_0 + \hat{H}_1
\hspace{0.5 cm} \text{,}
\end{equation}
e definiamo lo stato come
\begin{equation} \label{interazione-def}
| \Psi_I(t) \rangle = e^{i\frac{\hat{H}_0 t}{\hbar}} | \Psi_S(t) \rangle 
\hspace{0.5 cm} \text{,}
\end{equation}
il pedice $I$ sta per ``interazione''.
Vediamo ora qual'\`{e} la dipendenza dal tempo egli stati cos\`{i} costruiti
\begin{equation} \label{eq-interazione}
\begin{split}
i \hbar \frac{\partial}{\partial t} |\Psi_I(t)\rangle =&
-\hat{H}_0 e^{i\frac{\hat{H}_0 t}{\hbar}} |\Psi_S(t)\rangle +
    e^{i\frac{\hat{H}_0 t}{\hbar}}[\hat{H}_0+\hat{H}_1] |\Psi_S(t)\rangle  \\
              =&\ e^{i\frac{\hat{H}_0 t}{\hbar}} \hat{H}_1 e^{-i\frac{\hat{H}_0 t}{\hbar}}
                  |\Psi_S(t)\rangle  
\hspace{0.5 cm} \text{,}
\end{split}
\end{equation}
dove abbiamo usato l'inversa della (\ref{interazione-def}) e il fatto che
$e^{i\frac{\hat{H}_0 t}{\hbar}}$ commuta con $\hat{H}_0$.
Lo stato \`{e} dunque ancora dipendente dal tempo ma si vede che la sua dipendenza sar\`{a}
dovuta alla seconda parte dell'Hamiltoniana, moltiplicata per i due esponenziali.
Richiedendo come in precedenza che anche per questa formulazione risulti
\begin{equation}
\langle \Psi_I | \hat{O}_I | \Psi_I \rangle = \langle \Psi_S | \hat{O} | \Psi_S \rangle 
\hspace{0.5 cm} \text{,}
\end{equation}
otteniamo
\begin{equation}
\hat{O}_I = e^{i\frac{\hat{H}_0t}{\hbar}} \hat{O} e^{-i\frac{\hat{H}_0t}{\hbar}}
\hspace{0.5 cm} \text{.}
\end{equation}
Dunque l'evoluzione dello stato \`{e} dovuta semplicemente all'operatore $\hat{H}_1$
in descrizione di interazione mentre l'operatore evolve secondo il solo operatore $\hat{H}_0$.
Se ora scegliamo $\hat{H}_0=\hat{T}+\hat{V}^{ext}$ e
$\hat{H}_1=\hat{w}$ abbiamo ottenuto il nostro scopo: l'evoluzione libera \`{e} stata
spostata sugli operatori mentre l'evoluzione dovuta all'interazione resta sugli stati
e in particolare pu\`{o} essere racchiusa nell'operatore $\hat{U}(t,t_0)$ che diventa il
termine rilevante della teoria.

Vediamo dunque come esprimere proprio tale operatore studiando l'evoluzione di uno stato
\begin{equation} \label{ev-interazione1}
\begin{split}
|\Psi_I(t) \rangle =&\ e^{i\frac{\hat{H}_0t}{\hbar}} |\Psi_S(t)\rangle
                    = e^{i\frac{\hat{H}_0t}{\hbar}} e^{i\frac{\hat{H}(t-t_0)}{\hbar}} 
                                                       |\Psi_S(t_0)\rangle      \\
                   =&\ e^{i\frac{\hat{H}_0t}{\hbar}} e^{i\frac{\hat{H}(t-t_0)}{\hbar}} 
                       e^{-i\frac{\hat{H}_0t_0}{\hbar}} |\Psi_I(t_0)\rangle
\hspace{0.5 cm} \text{.}
\end{split}
\end{equation}
\`{E} possibile infine mostrare come \cite{Testo_MB} l'operatore di evoluzione temporale cos\`{i}
scritto pu\`{o} essere ricondotto ad un'espressione del tutto simile a quella dell'operatore
che descrive l'evoluzione degli stati di Schro\"{e}dinger, con la sola differenza che in questo
caso l'Hamiltoniana che descrive l'evoluzione \`{e} dipendente dal tempo, poich\'{e} l'operatore
$\hat{H}_1$ \`{e} in descrizione di interazione e che quindi ``tutto \`{e} pi\`{u} complicato''.
Il risultato che viene dimostrato in Ref.\cite{Testo_MB} \`{e} il seguente:
\begin{equation} \label{ev-interazione}
\begin{split}
\hat{U}(t,t_0)=& \sum_{n=0}^{+\infty} \left(\frac{i}{\hbar}\right)^n \frac{1}{n!}
                \int_{t_0}^{t}dt_1 \ldots \int_{t_0}^{t}dt_n\ \hat{T}
                [\hat{H}_1(t_1)] \ldots \hat{H}_1(t_n)]   \\
              =&\ \hat{T} \left[ e^{-\frac{i}{\hbar}\int_{t_0}^{t}\hat{H}_1(t')dt' } \right]
\hspace{0.5 cm} \text{.}
\end{split}
\end{equation}
L'ultima riga \`{e} un'espressione formale. In particolare tale espressione descrive
l'evoluzione di un qualsiasi stato in presenza di un'Hamiltoniana dipendente dal tempo, come
in effetti \`{e} lo stato in descrizione di interazione essendo soluzione della 
(\ref{eq-interazione}).

%*******************************************************************************************
\subsection{Perch\'{e} la funzione di Green}
La scelta di utilizzare la funzione di Green come grandezza integrata da cui
partire a studiare il nostro sistema pu\`{o} essere motivata da diverse ragioni;
uno dei motivi principali resta per\`{o} il fatto che a partire
da essa \`{e} possibile ricavare il valore di aspettazione rispetto al
ground state di un qualsiasi operatore a particella singola e che \`{e} inoltre
possibile calcolare l'energia totale del sistema.

La prima delle due affermazioni \`{e} facilmente verificabile. Consideriamo
un generico operatore $\hat{A}$ a particella singola (per comodit\`{a} di notazione
scegliamo un operatore indipendente dal tempo, per operatori dipendenti dal
tempo la dimostrazione sarebbe identica) e scriviamone la sua
rappresentazione in spazio $(\mathbf{x},\sigma)$ prima in rappresentazione di
Schro\"{e}dinger, quindi in rappresentazione di Heisenberg:
\begin{equation}
\begin{split}
\hat{A}   &= \sum_{\sigma_1,\sigma_2}\int d^3\mathbf{x}_1 d^3\mathbf{x}_2\
            A_{\alpha,\beta}(\mathbf{x}_1,\mathbf{x}_2)
            \hat{\psi}_{\sigma_1}^{\dag}(\mathbf{x}_1) \hat{\psi}_{\sigma_2}(\mathbf{x}_2)
            \\
\hat{A}_H(t) &= \sum_{\sigma_1,\sigma_2}\int d^3\mathbf{x}_1 d^3\mathbf{x}_2\
            A_{\sigma_1,\sigma_2}(\mathbf{x}_1,\mathbf{x}_2)
            \hat{\psi}_{\hat{H},\sigma_1}^{\dag}(\mathbf{x}_1,t)
            \hat{\psi}_{\hat{H},\sigma_2}(\mathbf{x}_2,t)
\hspace{0.5 cm} \text{.}
\end{split}
\end{equation}
Se ora facciamo il valor medio rispetto allo stato fondamentale in descrizione di
Heisenberg della seconda riga e
inseriamo una variabile $t_2 \rightarrow t^+$ otteniamo (e scegliendo $t_1=t$)
\begin{equation}
\langle \Psi_0 | \hat{A}_H | \Psi_0 \rangle =
   -i \sum_{\sigma_1,\sigma_2} \int d^3\mathbf{x}_1 d^3\mathbf{x}_2 \lim_{t_2 \rightarrow t^+}
   A_{\sigma_1,\sigma_2}(\mathbf{x}_1,\mathbf{x}_2) G_{\sigma_1,\sigma_2}(\mathbf{x}_1t,\mathbf{x}_2t_2)
\hspace{0.5 cm} \text{.}
\end{equation} 
Il simbolo $t^+$ indica un tempo che \`{e} un infinitesimo successivo all'istante
$t$ ed \`{e} stato utilizzato per garantire che la funzione di Green sia ben definita.
Se infine vogliamo studiare un operatore diagonale in spazio $\mathbf{x}$ dovremo
banalmente inserire un limite $\mathbf{x}_2 \rightarrow \mathbf{x}$ in modo analogo a quanto
abbiamo fatto per l'istante T.
Il computo dell'energia totale del sistema risulta invece pi\`{u} complicata poich\'{e}
per ottenere tale grandezza dobbiamo calcolare anche il valor medio di un operatore a
due particelle quale \`{e} l'interazione che, in linea di principio, richiederebbe
l'utilizzo di una funzione di Green a due particelle. L'ostacolo pu\`{o} essere per\`{o}
aggirato utilizzando l'equazione di Schro\"{e}dinger; riportiamo il risultato che \`{e}
possibile trovare sulla referenza \cite{Testo_MB}:
\begin{equation}
\langle \hat{w} \rangle = -\frac{i}{2} \sum_{\sigma} \int d^3\mathbf{x}
            \lim_{t_2 \rightarrow t^+} \lim_{\mathbf{x}_2 \rightarrow \mathbf{x}}
            \left[ i\hbar\frac{\partial}{\partial t} - \hat{h}_0(\mathbf{x}) \right]
            G_{\sigma,\sigma}(\mathbf{x}t,\mathbf{x}_2t_2)
\hspace{0.5 cm} \text{,}
\end{equation}
dove $\hat{h}_0=\hat{T}+\hat{V}^{ext}$.
Da questa l'energia totale del sistema nel suo stato fondamentale pu\`{o} essere facilmente
ottenuta come la somma dei valori medi degli operatori che compongono la nostra Hamiltoniana.

%***************************************************************************************
%***************************************************************************************
%***************************************************************************************
\section{Il teorema di Gell-Mann e Low}
\subsection{Introduzione: l'accensione adiabatica}
Abbiamo gi\`{a} anticipato che delle tre rappresentazioni di cui abbiamo parlato in
precedenza quella che utilizzeremo sar\`{a} la rappresentazione d'interazione.
L'idea \`{e} quella di utilizzare infatti tale rappresentazione per legare in qualche
modo la funzione d'onda soluzione dell'Hamiltoniana completa a quella soluzione
dell'Hamiltoniana non interagente facendo evolvere quest'ultima in descrizione di
interazione.

Partiamo dalla seguente Hamiltoniana:
\begin{equation}
\hat{H}(t)=\hat{H}_0 + e^{-\varepsilon |t|} \hat{H}_1
\hspace{0.5 cm} \text{,}
\end{equation}
dove nel nostro caso $\hat{H}_1$ \`{e} l'interazione e $\hat{H}_0 = \hat{T}+\hat{V}^{ext}$
\`{e} l'Hamiltoniana del sistema non interagente, $\varepsilon$ \`{e} una costante
positiva arbitraria. Notiamo che per $t \gg \varepsilon^{-1}$ risulta $\hat{H}(t) \simeq 
\hat{H}_0$ e che per $t=0$ invece $\hat{H}(0)=\hat{H}_0 + \hat{H}_1$.
Utilizziamo ora la descrizione di interazione%
\footnote{\`{E} possibile dimostrare che la
(\ref{ev-interazione}) \`{e} valida anche nel caso in cui l'Hamiltoniana
dipenda dal tempo} e scriviamo l'evoluzione di uno stato dal tempo $t_0$ al tempo $t$:
\begin{equation}
| \Psi_I(t) \rangle = \hat{U}_{\varepsilon}(t,t_0) | \Psi_I(t_0)\rangle
\hspace{0.5 cm} \text{,}
\end{equation}
dove per ora lo stato $| \Psi_I(t_0) \rangle$ non \`{e} specificato.
Utilizziamo quindi nuovamente la definizione (\ref{interazione-def}) e scegliamo come stato
di partenza per la descrizione di Schro\"{e}dinger lo stato fondamentale del sistema non
interagente. Infine scegliamo come istante iniziale un tempo $t_0\ll \epsilon^{-1}$ in modo
che a tale istante l'evoluzione temporale dello stato fondamentale del sistema sia (quasi)
quella libera e otteniamo
\begin{equation}
| \Psi_I(t_0) \rangle = e^{\frac{i \hat{H}_0}{\hbar} t_0} | \Psi_S(t_0)\rangle
\simeq | \Phi_0 \rangle
\hspace{0.5 cm} \text{,}
\end{equation}
cio\`{e} l'autostato a energia minima del sistema non interagente (potremmo partire da un autostato
qualsiasi, o anche da una qualsiasi combinazione lineare di tali autostati, ma poich\'{e} siamo
interessati ala ricerca dello stato fondamentale del sistema...).
Se ora facciamo il limite per $t_0 \to -\infty$ le relazioni scritte diventano esatte ed otteniamo:
\begin{equation}
| \Psi_I(t) \rangle = \hat{U}_{\epsilon}(t,- \infty) | \Phi_0\rangle
\hspace{0.5 cm} \text{,}
\end{equation}
siamo cio\`{e} in grado di scrivere uno stato in descrizione di interazione in un istante
$t$ arbitrario. Ci\`{o} che ci chiediamo ora \`{e} cosa accade se facciamo il limite per
$\varepsilon \to 0$, poich\'{e} siamo interessati a far sparire la dipendenza da $\varepsilon$
(non ha senso che i nostri risultati dipendano da una costante arbitraria) e poich\'{e} in
questo caso l'Hamiltoniana si riduce appunto a quella che vogliamo studiare%
\footnote{Parte essenziale della teoria \`{e} che i due processi di limite vengano
eseguiti nell'ordine che abbiamo indicato, e cio\`{e} prima $t_0 \rightarrow +\infty$ e
poi $\varepsilon \rightarrow 0$. L'inversione dei due limiti porterebbe ad una
formulazione del tutto inutile.}.

La risposta a tale domanda \`{e} tutt'altro che banale e costituisce il risultato
del teorema di Gell-Mann e Low.

%*******************************************************************************************
\subsection{Enunciato}
L'enunciato del teorema di Gell-Mann e Low \`{e} dunque: \\
Se la seguente quantit\`{a} esiste a tutti gli ordini in teoria perturbativa
\begin{equation}
\lim_{\varepsilon\rightarrow 0} \frac{\hat{U}_{\varepsilon}(0,-\infty)|\Phi_0\rangle}
    {\langle \Phi_0 | \hat{U}_{\varepsilon}(0,-\infty)|\Phi_0 \rangle}
\hspace{0.5 cm} \text{,}
\end{equation}
\`{e} un autostato dell'Hamiltoniana $\hat{H}=\hat{H}_0+\hat{H}_1$

Per la dimostrazione di tale teorema rimandiamo alla bibliografia \cite{Testo_MB}.
Segnaliamo per\`{o}
come in effetti il teorema non parla di stato fondamentale dell'Hamiltoniana ma di semplice
autostato. Gi\`{a} nel paragrafo precedente abbiamo accennato al fatto che in effetti tutto
ci\`{o} che \`{e} stato fatto finora vale per un qualsiasi autostato del sistema. Il problema
diventa dunque come individuare lo stato fondamentale. In precedenza abbiamo suggerito di
scegliere come stato di partenza lo stato fondamentale del sistema non interagente ($\Phi_0$
appunto). Il fatto che lo stato fondamentale non interagente porti allo stato fondamentale
interagente non \`{e} in realt\`{a} garantito in alcun modo dal teorema; questo punto
rappresenta anzi una delle maggiori difficolt\`{a} della MBPT ed \`{e} dovuta alla scelta
di considerare l'interazione come una perturbazione del sistema.

%*******************************************************************************************
%*******************************************************************************************
%*******************************************************************************************
\section{Interpretazione fisica della funzione di Green}
\subsection{Analizziamo l'espressione}
Cerchiamo ora di dare un'interpretazione fisica alla funzione di Green analizzandone
l'espressione in descrizione di interazione. Partiamo dunque dal considerare uno stato
$|\Psi_I(t')\rangle$ a cui aggiungiamo una particella nel punto $(\mathbf{x}'t')$ in modo
da ottenere lo stato $\psi_I^{\dag}(\mathbf{x}'t')|\Psi_I(t')\rangle$. Facciamo quindi evolvere
lo stato dal tempo $t'$ al tempo $t$ utilizzando l'operatore di evoluzione temporale in
questa descrizione (\ref{ev-interazione})
$\hat{U}(t,t')\psi_I^{\dag}(\mathbf{x}'t')|\Psi_I(t')\rangle$.

Ora per un istante qualsiasi $t>t'$ consideriamo qual'\`{e} la sovrapposizione dello stato
cos\`{i} ottenuto rispetto allo stato che si ottiene creando una particella nel punto
$(\mathbf{x}t)$, $\psi_I^{\dag}(\mathbf{x}t)|\Psi_I(t)\rangle$.
Il risultato sar\`{a} appunto la funzione di Green per $t>t'$. Per $t<t'$ \`{e} possibile
fare un ragionamento analogo a quello sin qui svolto, considerando la creazione di una buca
all'istante $t$ fatta poi evolvere fino all'istante $t'$.\footnote{Questa situazione pu\`{o}
anche essere interpretata come una particella creata al tempo $t'$ che evolve indietro nel
tempo. L'interpretazione \`{e} dovuta a Feynman in un suo famoso articolo}

La funzione di Green descrive quindi il propagarsi di una particella (o di una buca)
creata all'interno del gas elettronico e che quindi interagisce con esso. Pi\`{u} in particolare
essa ci d\`{a} la probabilit\`{a} di trovare una particella (o una buca), creata nel punto
$(\mathbf{x}'t')$, nel punto $(\mathbf{x}t)$.

%*******************************************************************************************
\subsection{L'equazione del moto e confronto con la definizione classica di funzione di Green}
Per scrivere l'equazione del moto della funzione di Green non dobbiamo far altro che applicare
l'operatore di derivazione rispetto al tempo alla definizione (\ref{Green-def}), svolgere quindi
in modo esplicito la derivazione dei vari termini (la funzione $\theta(t)$ e l'operatore di campo),
utilizzare l'equazione del moto per gli operatori in descrizione di Heisenberg
\begin{equation}
\frac{\partial \hat{\psi}_H^{\dag}(t)}{\partial t}= -i \hbar[\hat{H},\hat{\psi}_H^{\dag}(t)]
\end{equation}
e infine calcolare il commutatore dell'Hamiltoniana con l'operatore di campo. Non riportiamo
qui la dimostrazione che, come sempre, pu\`{o} essere trovata nella Ref.\cite{Testo_MB} e
scriviamo direttamente il risultato:
\begin{equation} \label{moto1}
\left[i\hbar\frac{\partial}{\delta t} - \hat{h}_0(\mathbf{x}_1)\right]G(1,2)=\hbar\delta(1,2)
 -i\hbar\ w(1,3)G(1,3;2,3^+)
\hspace{0.5 cm} \text{.}
\end{equation}
La notazione qui introdotta \`{e} l'estensione di quella gi\`{a} utilizzata in precedenza:
$3^+=(\mathbf{x}_3,\sigma_3,t_3^+)$.
Se eliminiamo l'ultimo termine di sinistra dell'equazione ritroviamo quella che \`{e} la definizione
di funzione di Green utilizzata nell'elettromagnetismo classico per la soluzione delle equazioni delle
onde e che pu\`{o} essere facilmente interpretata come il campo generato da una sorgente di tipo
deltiforme (nello spazio e nel tempo). Il termine che compare in pi\`{u} non rappresenta nient'altro
che una ``correzione'' a tale definizione. Tale correzione \`{e} dovuta al fatto che il campo fermionico
(a differenza di quello elettromagnetico classico) ``interagisce con se stesso''. Alla luce di questa
analogia possiamo capire l'interpretazione che in precedenza abbiamo fatto della funzione di Green.
Il creare una particella nel punto $(\mathbf{x}'t')$ corrisponde al porre una sorgente puntiforme nello
spazio e nel tempo che poi si propaga all'interno del gas elettronico.

Ora l'analogia si arresta qui poich\'{e} nei due casi le funzioni di Green vengono utilizzate in modo
differente.

Per quanto riguarda l'elettromagnetismo
classico viene utilizzato il principio di sovrapposizione per ottenere il campo in presenza di una certa
densit\`{a} di carica attraverso l'integrale di convoluzione della funzione di Green con tale densit\`{a};
oltre che alla densit\`{a} sorgente la soluzione dell'equazione viene poi definita imponendo le condizioni
al contorno nello spazio e nel tempo.

Nel caso fermionico non ha pi\`{u} senso parlare di sorgente del campo (o perlomeno non \`{e} in questo
modo che si procede). Ci\`{o} a cui siamo interessati \`{e} la funzione di Green stessa che descrive
il modo in cui il campo si propaga mentre interagisce con se stesso. Ancora una volta la difficolt\`{a}
principale nasce dalla presenza dell'interazione e quindi del termine in pi\`{u} che in precedenza non
compariva. Vediamo infatti come la funzione di Green a una particella, a causa della presenza
dell'interazione sia legata a quella a due particelle; allo steso modo per una funzione di Green a $n$
particelle si ricava un'equazione del moto che conterr\`{a} una funzione di Green a $n+1$ particelle.
Per superare il problema dovremo dunque cercare di esprimere in modo diverso l'interazione che come
sempre \`{e} la causa di tutte le nostre difficolt\`{a}. Vedremo nei paragrafi successivi come ci\`{o}
sar\`{a} fatto.

%********************************************************************************************
%********************************************************************************************
%********************************************************************************************
\section{La rappresentazione di Lehmann}
Prima di procedere con la presentazione della teoria many-body ci vogliamo soffermare sull'esporre
quella che \`{e} la rappresentazione di Lehmann della funzione di Green. In questa sezione
le funzioni d'onda vanno intese in rappresentazione di Heisenberg.
L'idea \`{e} semplicemente quella di partire dalla funzione di Green e di inserire tra la
coppia di operatori di campo l'operatore identit\`{a} scritto come:
\begin{equation}
\hat{I}=\sum_{J} |\Psi_J \rangle \langle \Psi_J|
\hspace{0.5 cm} \text{,}
\end{equation}
dove $|\Psi_J\rangle$ sono gli autostati dell'Hamiltoniana completa%
\footnote{Abbiamo preso la funzione d'onda dello stato fondamentale normalizzata}.
\begin{multline} \label{Lehmann1}
iG(1,2)=\sum_{J\neq0}\langle\Psi_0|\hat{\psi}_H^{\dag}(1)|\Psi_J\rangle
            \langle\Psi_J|\hat{\psi}_H(2)|\Psi_0\rangle \Theta(t_1-t_2)-  \\
            \langle\Psi_0|\hat{\psi}_H(2)|\Psi_J\rangle
            \langle\Psi_J|\hat{\psi}_H^{\dag}(1)|\Psi_0\rangle\Theta(t_2-t_1)
\hspace{0.2 cm} \text{.}
\end{multline}
Notiamo che nella sommatoria non compare il termine per $J=0$ poich\'{e} il valor medio sullo
stato fondamentale di un operatore di campo \`{e} nullo dato che l'Hamiltoniana scelta
commuta con l'operatore numero di particelle. Ora, sfruttiamo il fatto
che le funzioni d'onda sono autostati dell'Hamiltoniana completa otteniamo:
\begin{multline*}
iG(1,2)=\sum_J\langle\Psi_0|\hat{\psi}_{\sigma_1}^{\dag}(\mathbf{x}_1)|\Psi_J\rangle
            \langle\Psi_J|\hat{\psi}_{\sigma_2}(\mathbf{x}_2)|\Psi_0\rangle \Theta(t_1-t_2)
             e^{-i\hbar^{-1}(E_J-E_0)(t_1-t_2)}-  \\
            \langle\Psi_0|\hat{\psi}_{\sigma_2}(\mathbf{x}_2)|\Psi_J\rangle
            \langle\Psi_J|\hat{\psi}^{\dag}_{\sigma_1}(\mathbf{x}_1)|\Psi_0\rangle
            \Theta(t_2-t_1)e^{-i\hbar^{-1}(E_J-E_0)(t_2-t_1)}
\hspace{0.1 cm} \text{.}
\end{multline*}
Scrivendo infine la funzione $\Theta$ come
\begin{equation}
\Theta(t-t')= \frac{1}{2\pi i}\lim_{\eta\rightarrow 0} \int_{-\infty}^{+\infty}
              \frac{e^{-i \omega(t-t')}}{\omega+i\eta} d\omega
\end{equation}
e applicando la trasformata di Fourier otteniamo:
\begin{multline} \label{Lehmann_Green}
G_{\sigma_1\sigma_2}(\mathbf{x}_1,\mathbf{x}_2,\omega)=\sum_J
                   \frac{\langle\Psi_0|
                          \hat{\psi}^{\dag}(\mathbf{x}_1)|\Psi_J\rangle
                          \langle\Psi_J|\hat{\psi}(\mathbf{x}_2)|\Psi_0\rangle}
                        {\omega-\omega_J+i\eta}+    \\
                   -\frac{\langle\Psi_0|\hat{\psi}(\mathbf{x}_2)|\Psi_J\rangle
                          \langle\Psi_J|\hat{\psi}^{\dag}(\mathbf{x}_1)|\Psi_0\rangle}
                        {\omega+\omega_J-i\eta}
\hspace{0.2 cm} \text{,}
\end{multline}
dove abbiamo definito $\hbar\omega_J=(E_J-E_O)$.
Questo modo di scrivere la funzione di Green nello spazio delle frequenze ci permette di 
individuarne i poli come la differenza tra l'energia dello stato fondamentale e quella di
un qualsiasi stato che presenti una particella in pi\`{u} o in meno di tale stato. Il fatto
che consideriamo solo gli stati che abbiano una particella in pi\`{u} o in meno
dello stato eccitato dipende dal fatto che stiamo considerando gli elementi di matrice degli
operatori di campo che creano o distruggono una particella.

La funzione di Green potrebbe costituire quindi, in principio, un metodo
per ottenere le eccitazioni del sistema, per stati con $N\pm 1$ particelle,
a partire dalla DFT e in particolare dal teorema di Hohenberg-Kohn, dato che la funzione
di Green \`{e} calcolata come valor medio rispetto allo stato fondamentale del sistema. 
D'altra parte, ancora una volta, il teorema di Hohenberg-Kohn non fornisce un metodo
pratico per la costruzione della funzione d'onda e dunque per il calcolo dello spettro
di eccitazione del sistema. Lo schema di Kohn e Sham non \`{e} infatti qui applicabile
dato che l'espressione necessita della funzione d'onda molti corpi del sistema.
In effetti non \`{e} a tutt'oggi ancora stato sviluppato un metodo pratico, per la DFT,
che permetta di accedere alle energie di eccitazione che con $N\pm 1$ particelle.
Per ulteriori approfondimenti sull'argomento si veda la referenza \cite{Testo_DFT}.

%********************************************************************************************
%********************************************************************************************
%********************************************************************************************
\section{Il teorema di Wick}
\subsection{Introduzione al teorema}
Ora che abbiamo trovato un modo di esprimere lo stato fondamentale del
sistema interagente a partire dallo stato fondamentale del sistema in cui
$\hat{w}=0$ vogliamo utilizzare questo risultato per calcolare la funzione di
Green.
Dalla definizione di funzione di Green otteniamo:
\begin{multline}
\frac{\langle \Psi_0 | T\left[ \hat{\psi}^{\dag}_H(1)
      \hat{\psi}_{H}(2)  \right] |\Psi_0 \rangle}
     {\langle \Psi_0 |\Psi_0 \rangle} =  \\
\lim_{\varepsilon \rightarrow 0} \frac{D}{D}
\frac{\langle \Phi_0 | \hat{U}_{\varepsilon}(+\infty,0) T \left[ \hat{\psi}^
        {\dag}_{H}(1) \hat{\psi}_{H}(2) \right]
        \hat{U}_{\varepsilon}(0,-\infty) |\Phi_0 \rangle}
     {\langle \Phi_0 | \hat{U}_{\varepsilon}(+\infty,0)\hat{U}_{\varepsilon}(0,-\infty)
        | \Phi_0 \rangle}
\hspace{0.5 cm} \text{.}
\end{multline}
Definiamo quindi l'operatore $\hat{S}_{\varepsilon}=\hat{U}_{\varepsilon}(+\infty,-\infty)$
e riscriviamo gli operatori di campo in rappresentazione di interazione e otteniamo:
\begin{equation}
\lim_{\varepsilon \rightarrow 0}
\frac{\langle \Phi_0 | T \left[ \hat{U}_{\varepsilon}(+\infty,t) \hat{\psi}^
        {\dag}_I(1) \hat{U}_{\varepsilon}(t_1,t_2) \hat{\psi}_I(2)
        \hat{U}_{\varepsilon}(t_2,-\infty) \right] |\Phi_0 \rangle}
     {\langle \Phi_0 | \hat{S}_{\varepsilon} | \Phi_0 \rangle}
\hspace{0.5 cm} \text{,}
\end{equation}
dove gli operatori unitari $\hat{U}_{\varepsilon}(+\infty,0)$ e
$\hat{U}_{\varepsilon}(0,-\infty)$ sono passati all'interno del T-ordinamento
poich\'{e} indipendenti dal tempo. Scrivendo infine esplicitamente tutti
gli operatori di evoluzione temporale e facendo alcuni ``giochetti con gli indici''
otteniamo quello che pu\`{o} essere definito ``The most useful result of
quantum field theory'' \cite{Testo_MB}:
\begin{equation} \label{the-most}
\begin{split}
iG_{\alpha\beta}(\mathbf{x}t_x,\mathbf{y}t_y) =& 
 \sum_{n=0}^{+\infty} \left(\frac{-i}{\hbar}\right)^n \frac{1}{n!}
 \int_{-\infty}^{+\infty}dt_1 \ldots dt_n   \\
 &  \phantom{\sum_{n=0}^{+\infty} \left(\frac{-i}{\hbar}\right)^n}
        \frac{\langle \Phi_0 | \hat{T} \left[\hat{H}_1(t_1) \ldots \hat{H}_1(t_n)
        \hat{\psi}^{\dag}_{I,\alpha}(\mathbf{x},t_x) \hat{\psi}_{I,\beta}
        (\mathbf{y},t_y) \right] | \Phi_0 \rangle}
      {\langle \Phi_0 | \hat{S} | \Phi_0 \rangle}    \\
iG(1,2) =& \frac{\langle \Phi_0 | \hat{T} \left[\hat{S}
        \hat{\psi}^{\dag}_I(1) \hat{\psi}_I(2) \right] | \Phi_0 \rangle}
      {\langle \Phi_0 | \hat{S} | \Phi_0 \rangle}
\hspace{0.5 cm} \text{,}
\end{split}
\end{equation}
dove nella prima riga, avendo scritto in modo esplicito l'operatore di scattering $\hat{S}$
abbiamo usato una notazione differente per i due estremi non integrati della funzione di Green
$(\mathbf{x},\alpha,t_x)$ e $(\mathbf{y},\beta,t_y)$ per evidenziarli.
Notiamo in particolare che nell'ultima espressione \`{e} stato svolto il limite
per $\varepsilon \rightarrow 0$, la cui convergenza \`{e}
garantita dal fatto che gli operatori di evoluzione temporale compaiono
sia al numeratore che al numeratore,
in analogia a quanto accade nella dimostrazione del teorema di Gell-Mann e Low,
cancellando la divergenza della fase nell'espressione. Risulta quindi
$\hat{S}= \lim_{\varepsilon\rightarrow 0}\hat{S}_{\varepsilon}$.

Scopo del teorema di Wick \`{e} esattamente quello di valutare il $T$-ordinamento
di un numero arbitrario di operatori di campo e dunque di valutare in modo esplicito
l'ultima formula scritta.

%****************************************************************************************
\subsection{Enunciato}
Prima di scrivere l'enunciato del teorema dobbiamo introdurre due nuovi operatori,
il primo \`{e} l'$N$-ordinamenti (o ordinamento normale) di operatori di campo, mentre
il secondo \`{e} la contrazione di coppie di operatori di campo.

L'ordinamento normale del prodotto di un numero arbitrario di operatori di campo
non \`{e} altro che un riordinamento di tali operatori in cui tutti gli operatori
di distruzione, ovvero quegli operatori che applicati allo stato fondamentale di
un sistema danno come risultato zero, vengono spostati a destra. In altre parole
\begin{equation}
N \left[ \hat{\psi}_1 \hat{\psi}_2^\dag \hat{\psi}_3 \hat{\psi}_4^\dag \right] =
(-1)^P \hat{\psi}_2^\dag \hat{\psi}_4^\dag \hat{\psi}_1 \hat{\psi}_2
\hspace{0.5 cm} \text{,}
\end{equation}
dove con $\psi$ e $\psi^\dag$ abbiamo indicato appunto se un operatore \`{e} di distruzione
o di creazione. $P$ indica il numero di permutazioni eseguite per ottenere la
formula finale in questo esempio.

La contrazione di una coppia di operatori di campo \`{e} invece definita come la
differenza tra il $T$-ordinamento e l'ordinamento normale di tale coppia di operatori:
\begin{equation}
\overbracket[0.4 pt]{\hat{\psi}\hat{\psi}}=T[\hat{\psi}\hat{\psi}]-N[\hat{\psi}\hat{\psi}]
\hspace{0.5 cm} \text{.}
\end{equation}

Poniamo l'accento sul fatto che sia il T-ordinamento che l'N-ordinamento sono definiti
a partire dagli operatori di campo e che possono essere facilmente estesi a qualsiasi
operatore poich\'{e} un qualsiasi operatore pu\`{o} essere scritto utilizzando gli
operatori di campo (anche se nel caso del T-ordinamento si incontrano dei problemi
dovuti al fatto che esso non \`{e} ben definito per operatori a tempi uguali).
L'operatore di contrazione \`{e} invece definito solo per coppie di operatori di campo,
la definizione non \`{e} valida per una coppia qualsiasi di operatori.

Definiti questi operatori il teorema di Wick pu\`{o} essere enunciato nella seguente maniera: \\
\\
``Il $T$-ordinamento di un numero qualsiasi di operatori di campo \`{e} uguale alla somma
dell'ordinamento normale di tali operatori, pi\`{u} la somma su tutte le possibili coppie
dell'ordinamento normale di tali operatori in cui una coppia \`{e} per\`{o} ``contratta'',
pi\`{u} la somma su tutte le possibili doppie coppie di operatori
dell'ordinamento normale di tali operatori in cui due coppie sono per\`{o} ``contratte'',
pi\`{u}...'' continuando cos\`{i} fino al termine in cui tutti gli operatori, tranne al
pi\`{u} uno, sono contratti. \\

Tale teorema afferma in altre parole che
\begin{equation}
T[\hat{\psi}\hat{\psi}\hat{\psi}\hat{\psi}..] =
N[\hat{\psi}\hat{\psi}\hat{\psi}\hat{\psi}..]+
\sum_{\overbracket[0.4 pt]{\phantom{ab}}}N[\overbracket[0.4 pt]{\hat{\psi}\hat{\psi}\hat{\psi}}\hat{\psi}..]+
\sum_{\overbracket[0.4 pt]{\phantom{ab}}\overbracket[0.4 pt]{\phantom{ab}}}
N[\overbracket[0.4 pt]{\hat{\psi}\hat{\psi}}\overbracket[0.4 pt]{\hat{\psi}\hat{\psi}}...]+...
\hspace{0.5 cm} \text{.}
\end{equation}

L'idea del teorema \`{e} quella di spostare via via verso destra tutti gli operatori 
di annichilazione che poi, applicati allo stato fondamentale del sistema non interagente
danno appunto risultato nullo. In questo modo restano dei termini dovuti al fatto che la
commutazione di due operatori \`{e} in generale differente da zero.
Poich\'{e} anche i termini in cui compaiono le somme sulle contrazioni sono $N$-ordinati gli
unici termini non nulli, una volta applicato il valor medio sullo stato fondamentale, 
saranno quelli in cui tutti gli operatori sono contratti a coppie. \`{E} dunque evidente che
un espressione in cui compare un numero dispari di operatori di campo sar\`{a} identicamente
nulla.

In definitiva ci\`{o} che resta da calcolare \`{e} il prodotto di coppie di operatori di campo
``contratte''. Per calcolare il valore di una contrazione partiamo dalla considerazione
che la contrazione di una coppia qualsiasi di operatori \`{e} un numero e non un operatore.
Questa affermazione pu\`{o} essere dimostrata osservando
che la differenza tra il $T$-ordinamento e l'ordinamento normale di una coppia di operatori di
campo o \`{e} nulla o \`{e} pari al commutatore di tali operatori che sono entrambi numeri
\cite{Testo_MB}.
Se ora applichiamo l'operazione di valor medio sullo stato fondamentale a entrambi i termini
della definizione di operatore di contrazione otteniamo facilmente che:
\begin{equation}
\begin{split}
&\overbracket[0.4 pt]{\hat{\psi}^{\dag}(1)\hat{\psi}}(2)= ig(1,2) \\
&\overbracket[0.4 pt]{\hat{\psi}(1)\hat{\psi}^{\dag}}(2)= -ig(1,2) \\
&\overbracket[0.4 pt]{\hat{\psi}(1)\hat{\psi}}(2)=0 \\
&\overbracket[0.4 pt]{\hat{\psi}^{\dag}(1)\hat{\psi}^{\dag}}(2)=0
\hspace{0.5 cm} \text{,}
\end{split}
\end{equation}
dove $g(1,2)$ la funzione di Green del sistema non interagente, ovvero l'ordine
zero dell'espressione ricavata nella sezione precedente. Essa \`{e} infatti il valor medio
rispetto allo stato fondamentale del sistema non interagente di una coppia di operatori di
campo. \`{E} implicito che tutto il formalismo \`{e} stato sviluppato per operatori in
descrizione di interazione, poich\'{e} il nostro scopo era quello di calcolare il valor
medio di una serie di operatori in descrizione di interazione.

%***********************************************************************************************
%***********************************************************************************************
%***********************************************************************************************
\section{Analizziamo la funzione di Green con il teorema di Wick}
\subsection{La prima equazione di Hedin}
Abbiamo finalmente a disposizione tutti gli strumenti matematici necessari per studiare la
funzione di Green. Partiamo innanzi tutto dall'espressione (\ref{the-most}) scrivendo
in modo esplicito il termine:
\begin{equation}
\hat{H}_1(t_1)=\sum_{\sigma_1,\sigma_2}
\int d^3\mathbf{z}_1 d^3\mathbf{z}_2\ w(\mathbf{z}_1,\mathbf{z}_2)\delta(t_1-t_2)
\hat{\psi}^{\dag}(\mathbf{z}_1t_1^+)\hat{\psi}^{\dag}(\mathbf{z}_2t_2^+)
\hat{\psi}(\mathbf{z}_2t_2)\hat{\psi}(\mathbf{z}_1t_1)
\end{equation}
A questo punto le cose diventano ``abbastanza semplici'' poich\'{e} non dobbiamo far altro che
contrarre in tutti i modi possibili gli operatori di campo che compaiono ad ogni ordine di tale
espressione, ricordando che la contrazione di una coppia da come risultato la funzione di Green libera.
Nell'eseguire le contrazioni partiamo sempre dall'operatore di campo
$\hat{\psi}^{\dag}_{\alpha}(\mathbf{x}t_x)$
e terminiamo sempre con il termine $\hat{\psi}^{\dag}_{\beta}(\mathbf{y}t_y)$ poich\'{e} sono evidentemente
gli estremi della funzione di Green completa e sono le uniche due grandezze non integrate nell'espressione.
Il risultato sar\`{a} il prodotto di convoluzione di una serie di funzioni di $G(1,2)$ e di $w(1,2)$
che possono essere rappresentati con i diagrammi di Feynman.

Tralasciamo i dettagli di questo procedimento e ci limitiamo a segnalare che svolgendo lo sviluppo
sia per il numeratore che per il denominatore \`{e} possibile dimostrare che tutti i termini che compaiono
al denominatore cancellano in modo esatto alcuni termini che compaiono al numeratore ed in particolare
vengono cancellati tutti quei termini che contengono diagrammi di vuoto, dove i diagrammi di vuoto sono
definiti da una serie di contrazioni che parte e termina su due punti dello spazio
$N=(\mathbf{z}_N,\sigma_N,t_N)$ e $N'=(\mathbf{z}_{N'},\sigma_{N'},t_{N'})$ entrambi integrati.

Resta dunque intuitivo che a tutti gli ordini in teoria perturbativa potremo scrivere il risultato
come una prima funzione di Green che parte dal punto $(\mathbf{x},\alpha,t_x)$ ed arriva nel punto
$1=(\mathbf{z}_1,\sigma_1,t_1)$, pi\`{u} una serie di termini pi\`{u} un'ultima funzione di Green
che parte da un punto $N=(\mathbf{z}_N,\sigma_N,t_N)$
e che termina nel punto $(\mathbf{y},\beta,t_y)$. Raccogliendo su tutti gli ordini queste due funzioni di Green
e definendo ``ci\`{o} che rimane'' come la self energia del sistema otteniamo la seguente equazione
\begin{equation} \label{Green1}
G(1,2) = g(1,2) + g(1,3) \Sigma_H(3,4) g(4,2)   
\hspace{0.5 cm} \text{,}
\end{equation}
dove stiamo integrando sulle variabili ripetute, il pedice alla self energia ci ricorda che
il potenziale di Hartree \`{e} in essa incluso. L'ultimo passo che ci manca per scrivere l'equazione
di Dyson \`{e} per la funzione di Green, che costituisce la prima equazione di Hedin, \`{e} quello di
studiare il termine appena definito. 

Studiando i diagrammi di Feynman \`{e} possibile mostrare che la Self-energia cos\`{i} definita pu\`{o}
essere scritta come una serie di infiniti termini identici che si ripetono e quindi
\begin{equation}
\begin{split}
\Sigma_H(1,2)=&\ \Sigma_H^{\star}(1,2)+\Sigma_H^{\star}(1,3)g(3,4)\Sigma_H^{\star}(4,2)+  \\
            &\phantom{\Sigma_H^{\star}(1,3)}
             \Sigma_H^{\star}(1,3)g(3,4)\Sigma_H^{\star}(4,5)g(5,6)\Sigma_H^{\star}(6,2)...  \\
           =&\ \Sigma_H^{\star}(1,2)+\Sigma_H^{\star}(1,3)g(3,4)\Sigma_H(4,2)
\hspace{0.5 cm} \text{.}
\end{split}
\end{equation}
Per la definizione di $\Sigma_H^{\star}(1,2)$ rimandiamo alla referenza \cite{Testo_MB}.
Inserendo l'equazione appena scritta all'interno della (\ref{Green1}) e usando l'equazione stessa
otteniamo finalmente:
\begin{equation} \label{Green2}
G(1,2) = g(1,2) + g(1,3) \Sigma_H^{\star}(3,4) G(4,2)   
\hspace{0.5 cm} \text{.}
\end{equation}

\begin{comment}
%***********************************************************************************************
%***********************************************************************************************
%***********************************************************************************************
\section{Funzioni d'onda di quasi particella}
\subsection{Equazione del moto con la self-energia}
\end{comment}

\section{Equazione del moto con la self-energia}
Ora che abbiamo costruito l'equazione di Dyson per la funzione di Green possiamo cercare di 
guardare da un nuovo punto di vista quale sia il significato fisico delle grandezze sin qui
definite. A questo scopo ci serviremo, oltre che di tale equazione, della rappresentazione di
Lehmann della funzione di Green, per confrontare la funzione di Green libera e quella interagente.

Iniziamo con l'osservare che la funzione di Green libera, cio\`{e} soluzione dell'equazione
\begin{equation} \label{moto_libero}
\left[i\hbar\frac{\partial}{\delta t} - \hat{h}_0(\mathbf{x}_1)\right]g(1,2)=\hbar\delta(1,2)
\hspace{0.5 cm} \text{,}
\end{equation}
pu\`{o} essere facilmente scritta in rappresentazione di Lehmann come
\begin{equation}
g_{\sigma_1\sigma_2}(\mathbf{x}_1,\mathbf{x}_2,\omega)=\sum_I
                   \frac{\psi_{i\sigma_1}(\mathbf{x}_1)\psi_{i\sigma_2}^*(\mathbf{x}_2)}
                        {\omega-\omega_i+i\eta}-
                   \frac{\psi_{i\sigma_2}^*(\mathbf{x}_2)\psi_{i\sigma_1}(\mathbf{x}_1)}
                        {\omega+\omega_i+i\eta}
\hspace{0.5 cm} \text{,}
\end{equation}
dove $\hbar\omega_i=\epsilon_i-\epsilon_0$ e $\epsilon_i$ son gli autovalori di singola particella
dell'Hamiltoniana $\hat{h}_0$. 
Partendo dall'espressione (\ref{Lehmann_Green}) abbiamo utilizzato come base le funzioni d'onda
dell'Hamiltoniana non interagente e abbiamo quindi sfruttato il fatto che esse sono il prodotto
di funzioni d'onda a particella singola. Vediamo quindi che la funzione di Green non interagente
\`{e} facilmente calcolabile (poich\'{e} \`{e} possibile calcolare le funzioni d'onda di particelle
non interagenti) ed \`{e} pi\`{u} immediata la sua interpretazione fisica: i suoi poli sono
le auto-energie di singola particella
del sistema e a numeratore troviamo il prodotto delle funzioni d'onda e quindi la matrice densit\`{a}
del sistema.

Ci\`{o} che vogliamo fare ora \`{e} cercare di ottenere una formulazione simile per la funzione di Green
completa. A questo scopo applichiamo l'operatore
$\displaystyle{i\hbar\frac{\partial}{\delta t} - \hat{h}_0(\mathbf{x}_1)}$
all'equazione di Dyson della funzione di Green completa (\ref{Green2}) ottenendo:
\begin{equation}
\left[i\hbar\frac{\partial}{\delta t} - \hat{h}_0(\mathbf{x}_1)\right]G(1,2)=\
   \hbar\delta(1,2)+\hbar\Sigma_H^{\star}(1,3)G(3,2) \hspace{0.5 cm} \text{,}   \\
\end{equation}
dunque
\begin{equation} \label{moto2}
\left[(i\hbar\frac{\partial}{\delta t} - \hat{h}_0(\mathbf{x}_3))\delta(1,3)-\hbar\Sigma_H^{\star}(1,3)\right]
G(3,2)=\ \hbar\delta(1,2)
\hspace{0.5 cm} \text{,}
\end{equation}
dove abbiamo utilizzato l'equazione (\ref{moto_libero}).

\begin{comment}
%************************************************************************************************
\subsection{Dalle ampiezze di Lehmann alle funzioni di quasi particella}
\end{comment}
%Innanzitutto

\`{E} possibile dimostrare \cite{Onida,Tesi_dot_Fabien} che
se inseriamo l'espressione di Lehmann (\ref{Lehmann_Green}) nella (\ref{moto2})
e riscriviamo il numeratore definendo le ampiezze di Lehmann come:
\begin{equation}
f_{J\sigma}(\mathbf{x})=
\begin{cases}
\langle E_0|\hat{\psi}_{\sigma}(\mathbf{x})|E_J\rangle & \text{if } \omega_J \geq \mu \\
\langle E_J|\hat{\psi}_{\sigma}(\mathbf{x})|E_0\rangle & \text{if } \omega_J < \mu
\hspace{0.5 cm} \text{,}
\end{cases}
\end{equation}
otteniamo un'equazione del tutto simile a quella per il caso non interagente,
\begin{equation}
\left[(\hbar\omega_J - \hat{h}_0(\mathbf{x}_3))\delta(1,3)-
      \hbar\Sigma_H^{\star}(1,3;\omega_J)\right]f_{J\sigma_3}(\mathbf{x}_3)=0
\hspace{0.5 cm} \text{,}
\end{equation}
con la differenza che
ora anzich\'{e} il prodotto delle funzioni d'onda di particella singola abbiamo il prodotto delle
funzioni $f_{J\sigma}(\mathbf{x})$ e che questa volta compare la self energia del sistema.
$\hbar\omega_J=(E_J-E_0)$ risulta essere l'autostato di queste funzioni rispetto
all'equazione di Schro\"{e}dinger appena scritta e coincide con i poli della funzione di Green.
Nelle ultime due equazioni \`{e} stata inoltre introdotta la notazione $1=(\mathbf{x}_1,\sigma_1)$
in cui, rispetto alla notazione abituale, non \`{e} pi\`{u} inclusa la variabile temporale.

\section{Le equazioni di Hedin}
Lo studio della funzione di Green tramite lo sviluppo della self-energia a ordini successivi
sin qui presentato porta alla formulazione di equazioni che non possono
essere risolte se non nel caso del gas uniforme di elettroni poich\'{e} si ottengono delle
espressioni molto complesse dal punto di vista matematico. Anche in questo caso per\`{o} la
soluzione pu\`{o} essere trovata solo per i primi ordini perturbativi e di nuovo \`{e} tutt'altro
che banale. Un ulteriore problema \`{e} costituito dal fatto che abbiamo scelto di scrivere
come correzioni perturbative delle quantit\`{a} che non siamo sicuri siano trascurabili o piccole
rispetto ad altri. Ci\`{o} si traduce in quella che pu\`{o} essere considerata una delle maggiori
difficolt\`{a} di tutte le teorie di campo, quale \`{e} la MBPT, e che \`{e} costituita dalla presenza
di termini di ordine elevato che, anzich\'{e} che essere trascurabili, risultano essere divergenti.
Un'esempio \`{e} costituito dal secondo ordine dell'energia del gas uniforme appunto e pi\`{u}
in particolare dal termine che viene solitamente indicato \cite{Testo_MB} come $E_2^{ring}$.

Un approccio alternativo a quello sin qui introdotto, sempre basato sulle grandezze della funzione di
Green e della self-energia \`{e} quello delle cinque equazioni di Hedin. L'idea \`{e} quella
di trovare delle quantit\`{a} che insieme alla funzione di Green e alla self-energia formino un set di
equazioni chiuso. Tale set di equazioni ancora una volta risulta essere troppo complesso per essere
risolto in modo esatto ma ci sono tentativi di risoluzione, attraverso l'uso dei calcolatori,
in modo iterativo.
L'idea per superare il problema dei termini divergenti che compaiono nella MBPT \`{e} quella di non
scegliere come ``grandezze fondamentali'' quelle libere, $g(1,2)$ e $w(1,2)$, ma le grandezze 
``rivestite'' $G(1,2)$ e $W(1,2)$.

Prima di procedere con la formulazione del set di equazioni di Hedin segnaliamo che il punto di partenza sar\`{a}
l'espressione gi\`{a} scritta per la funzione di Green grazie all'ausilio del teorema di Gell-Mann e Low. In
questo caso per\`{o} noteremo come, a differenza si quanto sin qui fatto, l'operatore $S(+\infty,-\infty)$ sia
costruito a partire da un potenziale esterno anzich\'{e} dall'interazione $\hat{w}$. Questo poich\'{e}
utilizzeremo il teorema di Gell-Mann e Low pensando l'Hamiltoniana separata in due parti con $\hat{H}_0$
costituito dall'Hamiltoniana completa (interazione inclusa) e $\hat{H}_1=\hat{\varphi}$, che rappresenta
un potenziale esterno applicato al sistema.
Ovviamente la funzione di Green cos\`{i} definita sar\`{a} differente da quella sinora definita, ritroveremo
la funzione di Green sin qui studiata semplicemente ponendo il potenziale esterno pari a zero.

%*********************************************************************************************************************
\subsection{Introduzione alle equazioni}
Partiamo come annunciato con lo scrivere la funzione di G+reen secondo la formula (\ref{the-most})
\begin{equation}
iG_{\varphi}(1,2) = \frac{\langle \Psi_0 | \hat{T} \left[\hat{S}
        \hat{\psi}^{\dag}(1) \hat{\psi}(2) \right] | \Psi_0 \rangle}
               {\langle \Psi_0 | \hat{S} | \Psi_0 \rangle}
\hspace{0.5 cm} \text{,}
\end{equation}
dove l'operatore $\hat{S}$ \`{e} definito come
\begin{equation}
T[\hat{S}] = T \left[ \exp\left(-\frac{i}{\hbar}\
           \hat{\psi}^{\dag}(1^+) \hat{\psi}(1) \varphi(1) \right) \right]
\end{equation}
e lo stato rispetto a cui facciamo il valor medio \`{e} questa volta lo stato fondamentale
dell'Hamiltoniana completa.

Eseguiamo ora alcuni ``giochetti'' che ci permettono di scrivere una relazione tra la funzione di
Green a particella singola e la funzione di Green a due particelle; scriviamo innanzitutto la variazione
della funzione di Green cos\`{i} definita rispetto al potenziale esterno $\hat{\varphi}$
\begin{equation} \label{delta-G}
i\delta G_{\varphi}(1,2) =\frac{\langle \Psi_0 | T \left[\delta \hat{S}
        \hat{\psi}^{\dag}(1) \hat{\psi}(2) \right] | \Psi_0 \rangle}
               {\langle \Psi_0 | \hat{S} | \Psi_0 \rangle} -
                iG_{\varphi}(1,2)\frac{\langle \Psi_0 | T[\delta\hat{S}] | \Psi_0 \rangle}
               {\langle \Psi_0 | \hat{S} | \Psi_0 \rangle}
\hspace{0.5 cm} \text{,}
\end{equation}
dove abbiamo utilizzato
\begin{equation}
T[\delta\hat{S}]= -\frac{i}{\hbar}T \left[ \hat{S}\left(\
          \hat{\psi}^{\dag}(1^+) \hat{\psi}(1) \delta \varphi(1) \right) \right]
\hspace{0.5 cm} \text{,}
\end{equation}
quindi utilizziamo la definizione di funzione di Green a due particelle ottenedo
\begin{equation} \label{two-particels}
\frac{\delta G_{\varphi}(1,2)}{\delta \varphi(3)} = -\hbar^{-1}G_{\varphi}(1,3;2,3^+)
+\hbar^{-1}G_{\varphi}(1,2)G_{\varphi}(3,3^+)
\hspace{0.5 cm} \text{.}
\end{equation}

Possiamo finalmente valutare la nostra espressione nel caso in cui il potenziale esterno sia uguale a zero
in modo da avere delle grandezze confrontabili con i risultati sin qui ottenuti per via perturbativa%
\footnote{D'ora in poi tutte le funzioni di Green saranno intese valutate a potenziale esterno nullo e quando
indicheremo una variazione rispetto al potenziale sottintenderemo di fare la variazione e poi valutare
l'espressione a campo nullo}.
In particolare sostituendo l'espressione ricavata per la funzione di Green a due particelle nell'equazione
del moto per la funzione di Green (\ref{moto1}) otteniamo:
\begin{multline}
\left[i\hbar\frac{\delta}{\delta t}-\hat{h}_0(\mathbf{x}_1)\right] G(1,2)
      +i\ w(1,3)G(3,3^+)G(1,2) \\
 -i\hbar\ w(1,3)\frac{\delta G(1,2)}{\delta \varphi(3)}= \hbar\delta(1,2)
\hspace{0.5 cm} \text{.}
\end{multline}
Utilizziamo ora la seguente identit\`{a} matematica per riscrivere l'equazione ottenuta:
\begin{equation} \label{rel-Green}
\frac{\delta G(1,2)}{\delta \varphi(3)} =
-\hbar^{-1}\ G(1,4)\frac{\delta G^{-1}(4,5)}{\delta \varphi(3)}G(5,2)
\end{equation}
e confrontandone il risultato con l'equazione (\ref{moto2}) otteniamo
\begin{equation}
\begin{split}
&\hbar\Sigma_H^{\star}(1,2)= -i\delta(1,2)\ w(1,3)G(3,3^+)+ \\
&\phantom{\hbar\Sigma^{\star}(1,2)= -i\delta(1,2)}
-i\ w(1,3)G(1,4)\frac{\delta G^{-1}(4,2)}{\delta \varphi(3)} \\
&\phantom{\hbar\Sigma^{\star}(1,2)} =\delta(1,2)v_H(1) + i\ w(1,3)G(1,4)\Gamma(4,2;3)
\hspace{0.5 cm} \text{.}
\end{split}
\end{equation}
Il primo termine \`{e} evidentemente il potenziale di Hartree, mentre il secondo termine costituisce
un'espressione in forma funzionale (e non pi\`{u} perturbativa quindi) dei termini della self-energia oltre
il potenziale di Hartree. Nell'ultima linea abbiamo gi\`{a} definito la funzione di vertice come
\begin{equation}
\Gamma(4,2;3)=-\frac{\delta G^{-1}(4,2)}{\delta \varphi(3)}
\hspace{0.5 cm} \text{.}
\end{equation}
Potremmo ora procedere con il
costruire un'equazione di Dyson per la funzione vertice e ottenere cos\`{i} un set di 3 equazioni con
tre funzioni incognite: il propagatore, la self-energia e il vertice (l'interazione libera \`{e} nota).
Ci\`{o} che invece si preferisce fare \`{e} partire dall'interazione ``rivestita'' anzich\'{e} da quella
libera e costruire il set di cinque equazioni che dimostreremo nel prossimo paragrafo poich\'{e}, come
gi\`{a} detto, l'utilizzo dell'interazione libera porta dei problemi di divergenza nella teoria.

%**********************************************************************************************************
\subsection{Dimostrazione delle equazioni}
In questo paragrafo presenteremo la teoria cos\`{i} come l'abbiamo trovata in letteratura. Poich\'{e}
d'altra parte crediamo che questo modo di procedere (sebbene elegante dal punto di vista matematico) 
impedisca in qualche modo di vedere la fisica che dietro di essa \`{e} nascosta, nel paragrafo successivo
cercheremo di svolgere quelle considerazioni a nostro avviso necessarie per capire il significato delle
grandezze che compaiono. Come altrove nell'esposizione della MBPT saremo dunque costretti a presentare
la teoria in modo non lineare, crediamo d'altronde che sia questo il modo migliore di guardare ad essa e che
entrambi i livelli di analisi siano necessari per poterla comprendere.

Iniziamo dunque con il definire un ``potenziale esterno efficace'' pari alla somma della perturbazione
introdotta pi\`{u} il potenziale di Hartree
\begin{equation} \label{pot-eff}
V(1)=\varphi(1)+v_H(1)
\hspace{0.5 cm} \text{.}
\end{equation}
Definiamo quindi la funzione dielettrica come (vedremo pi\`{u} avanti il perch\'{e} di questa scelta)
\begin{equation}
\begin{split}
\epsilon^{-1}(1,2)=&\ \frac{\delta V(1)}{\delta \varphi(2)} \\
                  =&\ \delta(1,2)+ \ w(1,3)\Pi(3,2)
\hspace{0.5 cm} \text{,}
\end{split}
\end{equation}
dove per la seconda linea abbiamo utilizzato la definizione classica di funzione di risposta,
ricordando che la densit\`{a} \`{e} pari alla parte diagonale della funzione di Green
\begin{equation} \label{pi-def}
\Pi(1,2)=-i\frac{\delta G(1,1^+)}{\delta \varphi(2)}
\hspace{0.5 cm} \text{.}
\end{equation}
Otteniamo quindi l'interazione efficace come
\begin{equation} \label{interazione}
\begin{split}
W(1,2) =& \ \epsilon^{-1}(1,3) w(3,2)       \\
       =&\ w(1,2) + \ w(1,3)\Pi(3,4)w(4,2)
\hspace{0.5 cm} \text{.}
\end{split}
\end{equation}
Utilizziamo ora le regole di derivazione a catena nell'equazione scritta per la funzione risposta
(\ref{pi-def}), o polarizzazione del sistema, e scriviamo:
\begin{equation} \label{polarizzazione-def}
\begin{split}
\Pi(1,2)=&\ -i\ \frac{\delta G(1,1^+)}{\delta V(3)} \frac{\delta V(3)}{\delta \varphi(2)} \\
        =&\ \tilde{\Pi}(1,2)+ \tilde{\Pi}(1,3)w(3,4)\Pi(4,2)
\hspace{0.5 cm} \text{,}
\end{split}
\end{equation}
dove abbiamo definito la polarizzazione ridotta come la derivata della densit\`{a} rispetto al potenziale
efficace; in questo modo abbiamo ``messo in evidenza'' il ruolo del campo di Hartree. Inserendo l'ultima
equazione scritta nella (\ref{interazione}) otteniamo
\begin{equation}
W(1,2)= w(1,2)+ \ w(1,3)\Pi(3,4)W(4,2)
\end{equation}

Utilizzando ora nuovamente la relazione (\ref{rel-Green}) insieme alle due definizioni di
funzione risposta e funzione risposta ridotta otteniamo
\begin{equation} \label{Hedin-polarizzazione}
\begin{split}
\hbar\Pi(1,2)         =& -i \ G(1,3)G(4,1^+)\Gamma(2;3,4)          \\
\hbar\tilde{\Pi}(1,2) =& -i\ G(1,3)G(4,1^+)\tilde{\Gamma}(2;3,4)
\hspace{0.5 cm} \text{.}
\end{split}
\end{equation}
Le due grandezze differiscono solo per il fatto che una \`{e} considerata rispetto al potenziale
libero mentre l'altra rispetto al potenziale efficace. Allo stesso modo differiscono la funzione
$\Gamma$ e la funzione $\tilde{\Gamma}$ che abbiamo qui introdotto.

Se ora riprendiamo l'equazione per la self-energia scritta nel paragrafo precedente e utilizziamo
la regola di derivazione a catena a la definizione di interazione efficace appena introdotta
otteniamo\footnote{Indichiamo qui con $\Sigma^{\star}$ solo il secondo termine della self energia,
ovvero la self-energia meno il potenziale di Hartree $\Sigma_H^{\star}-V_H$, per questa ragione
non comparir\`{a} pi\`{u}, d'ora in poi, il pedice $H$ accanto all'operatore della self-energia.}
\begin{equation}
\begin{split}
\Sigma^{\star}(1,2)=& -i\ w(1^+,3)\frac{\delta V(5)}{\delta \varphi(3)}
                        \frac{\delta G^{-1}(1,4)}{\delta V(5)} G(4,2)    \nonumber \\
                   =&\ i\ G(1,4) W(3,1^+) {\Gamma}(4,2;3)
\hspace{0.5 cm} \text{.}
\end{split}
\end{equation}

Siamo cos\`{i} quasi giunti alla fine dei nostri sforzi. Ci manca ora solo di ricavare l'equazione per
il vertice. Questa pu\`{o} essere ricavata partendo dalla definizione di funzione di vertice,
utilizzando l'inverso dell'equazione del propagatore
\begin{equation}
g^{-1}(1,2)=G_{\varphi}^{-1}(1,2)+\delta(1,2)(V_H(1)+\varphi(1))+\Sigma^{\star}(1,2)
\hspace{0.5 cm} \text{,}
\end{equation}
dove abbiamo esplicitato nuovamente il potenziale esterno per maggior chiarezza,
nell'equazione per il vertice si deve poi immaginare, come sempre,
di mettere il potenziale pari a zero.
Utilizziamo ora le regole di derivazione a catena insieme con la relazione (\ref{rel-Green}). In questo
modo si ottiene (scriviamo qui il risultato per la funzione ridotta, ma si potrebbe fare lo stesso
con la funzione $\Gamma$ completa)
\begin{equation} \label{vertice}
\begin{split}
\tilde{\Gamma}(1;2,3)&=\delta(1,2)\delta(1,3)+\frac{\Sigma^{\star}(2,3)}{\delta V(1)} \\
             &=\delta(1,2)\delta(1,3)+ \\
&\phantom{\delta(1,2)\delta()1,3}\frac{1}{\hbar}\ \tilde{\Gamma}(1;4,5)G(6,4)G(7,5)
                \frac{\delta \Sigma^{\star}(2,3)}{\delta G(6,7)}
\hspace{0.5 cm} \text{.}
\end{split}
\end{equation}

Scriviamo dunque le equazioni che abbiamo sin qui ricavato, introducendo per la scrittura finale
la funzione di Green di Hartree in modo da utilizzare solo il secondo termine della come gi\`{a}
fatto in precedenza.
\begin{equation}
g_H^{-1}(1,2)=g^{-1}(1,2)+V_H(1,2)
\hspace{0.5 cm} \text{.}
\end{equation}
Le equazioni di Hedin risultano quindi cos\`{i} scritte:
\begin{flalign}
\label{Hedin's equations}
&G(1,2) = g_H(1,2) + g_H(1,3) \Sigma^{\star}(3,4) G(4,2)   \\
&W(1,2) = w(1,2) + w(1,3) \tilde{\Pi}(3,4) W(4,2)   \\
&\hbar \tilde{\Pi}(1,2) = -i \tilde{\Gamma}(1;3,4) G(2,3) G(4,2)     \\
&\hbar \Sigma^{\star}(1,2) = i \Gamma(4;1,3) G(3,2) W(2,4)  \\
&\tilde{\Gamma}(1;2,3)=\delta(1,2) \delta(1,3)+   \nonumber \\
&\phantom{\Gamma(1;2,3)=\delta(1,2)}\hbar^{-1}\tilde{\Gamma}(1;4,5) G(6,4) G(5,7)
              \frac{\delta \Sigma^{\star}(2,3)}{\delta G(6,7)} 
\hspace{0.5 cm} \text{.}
\end{flalign}

%*********************************************************************************************************
\subsection{La fisica dietro le equazioni di Hedin}
Prima di procedere oltre con l'esposizione della teoria sin qui esposta vorrei soffermarmi ad osservare
le grandezze che abbiamo definito e cercare di capire quale sia il loro significato fisico. 
A questo scopo faremo uso, in modo decisivo delle regole di cambio base in seconda quantizzazione
per un operatore e del concetto stesso di operatore che abbiamo precisato in appendice.

La prima
domanda che mi sono posto dopo aver studiato questa teoria \`{e} perch\'{e} la polarizzazione (e di
conseguenza la funzione dielettrica) e l'interazione ``rivestita'' cos\`{i} definite dovrebbero coincidere
con la loro definizione nell'elettromagnetismo classico.
A questo scopo osserviamo pi\`{u} con calma le grandezze che compaiono. Innanzi tutto abbiamo un sistema di
particelle che interagiscono e l'operatore che rappresenta l'interazione \`{e} il seguente
\begin{equation}
\hat{w}=\int \int d^3\mathbf{x} d^3\mathbf{y}\ \hat{\psi}^{\dag}(\mathbf{x})\hat{\psi}^{\dag}(\mathbf{y})
                  w(\mathbf{x},\mathbf{y})\hat{\psi}(\mathbf{y})\hat{\psi}(\mathbf{x})
\hspace{0.5 cm} \text{,}
\end{equation}
che \`{e} evidentemente un operatore a due particelle, dove
$w(\mathbf{x},\mathbf{y})=\frac{1}{|\mathbf{x}-\mathbf{y}|}$.
Il campo medio \`{e} il potenziale di Hartree ed \`{e} rappresentato dall'operatore ad una particella
\begin{equation}
\hat{v}_H=\int \int d^3\mathbf{x} d^3\mathbf{y}\ \hat{\psi}^{\dag}(\mathbf{x})\frac{\rho(\mathbf{y})}
                        {|\mathbf{x}-\mathbf{y}|}\hat{\psi}(\mathbf{x})
\hspace{0.5 cm} \text{.}
\end{equation}
Ipotizziamo infine, cos\`{i} come facciamo nella costruzione delle equazioni di Hedin, che sia presente
un potenziale esterno applicato a tale sistema:
\begin{equation}
\hat{\varphi}=\int d^3\mathbf{x}\ \varphi(\mathbf{x})\hat{\psi}^{\dag}(\mathbf{x})\hat{\psi}(\mathbf{x})
\hspace{0.5 cm} \text{.}
\end{equation}
Immaginiamo ora che il sistema si trovi nel suo stato fondamentale e di applicare una variazione del
campo esterno $\delta \hat{\varphi}(\mathbf{x})$
il campo di cui risentir\`{a} una particella qualsiasi del sistema sar\`{a} dovuto alla somma del campo
applicato pi\`{u} la variazione dell'interazione con tutte le altre particelle dovuta alla variazione
della funzione d'onda delle stesse a causa del campo applicato.
\begin{equation}
\delta\hat{V}^{eff}=\delta\hat{\varphi}+ \delta \hat{w}
\hspace{0.5 cm} \text{.}
\end{equation}
Dove l'operatore $\delta\hat{w}$ deve rappresentare in qualche modo il campo dovuto alla variazione delle
funzioni d'onda delle particelle%
\footnote{Formalmente non ha senso parlare di variazione di un operatore poich\'{e} esso \'{e} definito
in modo unico e non dipende dalla funzione d'onda. Quello che viene fatto in pratica \`{e} cercare di costruire
un operatore che possa tener conto della variazione delle funzioni d'onda del sistema e che risulti essere
un operatore a particella singola. Nella presente teoria si riesce a includere solo una parte di questo effetto,
il campo medio; se ad esempio guardiamo alla DFT in questa teoria tale effetto \`{e} viene incluso in modo satto
per il sistema non interagente come variazione del potenziale di scambio e correlazione. Il prezzo da pagare \`{e}
per\'{o} che in questo caso la funzione d'onda molti-corpi non sar\`{a} quella esatta.}.

Cio\`{e} che nelle equazioni di Hedin \`{e} approssimare tale operatore con la variazione del campo di Hartree.
Possiamo guardare all'approssimazione fatta come ad un modo di trascurare gli effetti quantistici di scambio
e correlazione e tenere solo il campo medio.
Se ora scriviamo esplicitamente la variazione del potenziale di Hartree
\begin{equation}
\delta \hat{v}_H=\int \int d^3\mathbf{x} d^3\mathbf{y}\ \hat{\psi}^{\dag}(\mathbf{x})\frac{\delta\rho(\mathbf{y})}
                                      {|\mathbf{x}-\mathbf{y}|}\hat{\psi}(\mathbf{x})
\hspace{0.5 cm} \text{,}
\end{equation}
vediamo facilmente come la funzione di risposta, che lega la variazione di densit\`{a} al
potenziale esterno all'ordine lineare rispetto alla densit\`{a}, possa essere anche utilizzata per definire
l'interazione rivestita e quindi la funzione dielettrica. Scriviamo in modo esplicito quanto affermato: 
\begin{equation}
\delta \hat{V}^{eff} \simeq \delta\hat{\varphi} +
\ w(1,2)\chi(2,3)\delta\hat{\varphi}(3)
\hspace{0.5 cm} \text{,}
\end{equation}
da cui si capisce il perch\'{e} della funzione dielettrica definita come la variazione del potenziale
efficace rispetto al potenziale esterno.

Abbiamo parlato di approssimazione per l'interazione e abbiamo chiarito in che senso l'interazione \`{e}
stata considerata in modo approssimato. Questo non ci deve per\`{o} in alcun modo far pensare che la
soluzione delle equazioni di Hedin sia in qualche modo approssimata! Tali equazioni sono infatti state
costruite per permetterci di calcolare la funzione di Green e la sua self-energia e sono queste le due
grandezze a cui dobbiamo guardare. L'interazione che compare (e con essa la polarizzazione) sono solo un
mezzo per ottenere questo scopo. L'idea di utilizzare un'interazione cos\`{i} fatta nasce dalla necessit\`{a}
di una grandezza fisica che possa essere un punto di partenza migliore dell'interazione libera, ma che non
deve necessariamente essere esatta. Gli effetti quantistici che non compaiono in tale grandezza saranno
quindi inclusi altrove ed in particolare questo ci porta a guardare alla funzione di vertice.

L'interpretazione fisica di questa grandezza appare meno immediata delle altre (perlomeno all'autore di
questa tesi) ma pu\`{o} essere in qualche modo delineata perlomeno in modo intuitivo. L'idea \`{e} che
stiamo studiando un sistema in cui compaiono due grandezza, il campo fermionico degli elettroni e
l'interazione. Per entrambe abbiamo definito ``un propagatore'' (possiamo immaginare l'interazione libera
come il propagatore del fotone, sebbene in realt\`{a} il campo venga trattato qui in modo classico) e la
sua self-energia (per l'interazione essa sar\`{a} costituita dalla polarizzazione). Appare quindi naturale
la presenza di un termine che ``leghi'' tali ``campi'' e che, come detto, dovr\`{a} in qualche modo supplire
al fatto che stiamo studiando il problema in modo semi-classico (o perlomeno con l'approssimazione appena
indicata per quanto concerne l'interazione). Per l'interpretazione della funzione di vertice come inclusione
degli effetti quantistici trascurati
nella definizione dell'interazione rivestita rimandiamo alla bibliografia \cite{Testo_MB,Fabien_kernel}.

%% file: lda.tex
\section{Introduzione}
Fino ad ora abbiamo presentato la formulazione esatta di due teorie per lo stato
fondamentale di un sistema di elettroni interagenti. Se per\`{o} vogliamo
applicarle per lo studio di sistemi fisici dobbiamo necessariamente introdurre alcune
approssimazioni. La necessit\`{a} di eseguire tali approssimazioni nasce in modo differente
per le due teorie: nel caso della DFT la necessit\`{a} nasce dal
fatto che non \`{e} noto il potenziale di scambio e correlazione e che dunque dovremo in
qualche modo trovare una buona approssimazione per tale potenziale. Per la MBPT il problema
nasce invece dal fatto che i calcoli che compaiono sono troppo complessi per poter essere
risolti in modo esatto.

Le due approssimazioni che solitamente vengono utilizzate sono l'approssimazione locale o
``Local Density Approximation'' (LDA) per la DFT e l'approssimazione ``GW'' per la MBPT.
La prima consiste nello scrivere il funzionale dell'energia del sistema a partire dal
funzionale del gas omogeneo che pu\`{o} essere calcolato a partire dalla MBPT e da calcoli
Montecarlo, la seconda nel risolvere le equazioni di Hedin (\ref{Hedin's equations})
approssimando la funzione $\Gamma$ con l'identit\`{a} e dunque la Self-Energia con l'espressione
$\Sigma=GW$, da cui il nome dell'approssimazione.

Nei prossimi paragrafi ci concentreremo soprattutto sulle approssimazioni necessarie all'utilizzo
della DFT poich\`{e} i calcoli che faremo all'interno di questa tesi saranno svolti nell'ambito
di tale teoria. La teoria MBPT rester\`{a} comunque sempre presente sia come uno dei possibili
punti di partenza per la costruzione di approssimazioni per i funzionali del potenziale e dell'energia
della DFT sia, come vedremo nell'ultimo capitolo di questa tesi, come possibile punto di partenza
per correzioni alle approssimazioni svolte nei calcoli da noi eseguiti.

%***************************************************************************************************
%***************************************************************************************************
%***************************************************************************************************
\section{L'approssimazione locale}
\subsection{Approssimazione LDA}
L'approssimazione locale costituisce una notevole semplificazione della teoria presentata e si basa
sull'assunzione che la densit\`{a} di energia in un punto dato del nostro sistema dipenda solamente
dalla densit\`{a} del gas elettronico in quel punto, da cui il nome di approssimazione locale.
Facendo questo presupposto \`{e} possibile approssimare l'espressione dell'energia di scambio e
correlazione come funzionale della densit\`{a} nel seguente modo
\begin{equation}
E_{xc}[\rho]= \int d^3\mathbf{r}\ \rho(\mathbf{r})\epsilon_{xc}[\rho(\mathbf{r})]
       \simeq \int d^3\mathbf{r}\ \rho(\mathbf{r})\epsilon^{hom}_{xc}(\rho(\mathbf{r}))
\hspace{0.5 cm} \text{,}
\end{equation}
dove $\epsilon^{hom}_{xc}(\rho)$ \`{e} una semplice funzione di una variabile che pu\`{o} essere ricavata
dall'energia del sistema con densit\`{a} uniforme $\rho$
\begin{equation}
\epsilon^{hom}_{xc}(\rho)=
\frac{E^{hom}_{xc}(\rho)}{V\rho} %\Bigg{|}_{\rho_0=\rho(\mathbf{r})}
\hspace{0.5 cm} \text{,}
\end{equation}
dove $V$ \`{e} il volume del sistema.
%dove evidentemente l'energia del sistema omogeneo \`{e} una semplice funzione della densit\`{a}
%in quanto quest'ultima non \`{e} una funzione in questo caso ma una costante.
Il potenziale di scambio e correlazione in questo modo risulta
\begin{equation}
v_{xc}[\rho](\mathbf{r})=\frac{\delta E_{xc}[\rho]}{\delta \rho(\mathbf{r})}
      \simeq \epsilon^{hom}_{xc}(\rho(\mathbf{r}))+
      \rho(\mathbf{r})\frac{d \epsilon^{hom}_{xc}(\mathbf{r})}{d \rho(\mathbf{r})}
\hspace{0.5 cm} \text{.}
\end{equation}

Per esprimere il funzionale dobbiamo quindi scrivere qual'\`{e} l'energia del sistema omogeneo in
funzione della densit\`{a}. A questo scopo scegliamo innanzi tutto di separare l'energia di scambio
e correlazione in due parti, un termine per l'energia di scambio ed un termine per l'energia di
correlazione appunto. Il primo termine di approssimazione locale pu\`{o} essere ottenuto
a partire dall'espressione dell'energia di scambio dell'interazione che sappiamo pu\`{o} essere scritto
in modo esatto in funzione della matrice densit\`{a} di Kohn e Sham del sistema \cite{Testo_DFT}.
\begin{equation}
E^{KS}_x[\rho]=-\frac{e^2}{4}\int d^3\mathbf{r} \int d^3 \mathbf{r}'
\frac{|\gamma_s(\mathbf{r},\mathbf{r}')|^2}{|\mathbf{r}-\mathbf{r}'|}
\hspace{0.5 cm} \text{.}
\end{equation}
dove $\gamma_s(\mathbf{r},\mathbf{r}')=\langle \Psi_0|\hat{\psi}^{\dag}(\mathbf{r})
\hat{\psi}(\mathbf{r'})|\Psi_0\rangle$ \`{e} appunto la matrice densit\`{a} del sistema%
\footnote{Vediamo come la matrice densit\`{a} differisca dalla funzione di Green poich\'{e}
quest'ultima \`{e} definita cpon gli operatori in rappresentazione di Heisenberg e quindi \`{e}
dipendente dal tempo.}.
Sviluppando infatti all'ordine lineare la matrice densit\`{a} rispetto alla densit\`{a} del
sistema si ottiene \cite{Testo_DFT}
\begin{equation} \label{xLDA1}
\epsilon_x^{KS} = -d_0 \rho_0^{1/3}
\hspace{7 mm}\mbox{ con}\hspace{7 mm}
d_0=\frac{3e^2}{4}\left(\frac{3}{\pi}\right)^{1/3}
\hspace{0.5 cm} \text{.}
\end{equation}
Definendo la quantit\`{a} $r_s^3=\frac{3}{4\pi \rho}$ (cio\`{e} il raggio della sfera che contiene
una quantit\`{a} unitaria di materia), possiamo riscrivere l'espressione come \cite{Kohn_Nobel}:
\begin{equation} \label{xLDA}
\epsilon_x = -\frac{0.458\ e^2}{r_s}
\hspace{0.5 cm} \text{.}
\end{equation}
Per la parte di correlazione una delle
prime parametrizzazioni proposte in questa approssimazione \`{e} invece la seguente \cite{Kohn_Nobel}:
\begin{equation} \label{cLDA}
\epsilon_c = -\frac{0.44\ e^2}{r_s+7.8\ a_0}
\hspace{0.5 cm} \text{,}
\end{equation}
dove $a_0$ \`{e} il raggio di Bohr. Abbiamo dunque
\begin{equation}
\epsilon_{xc}= -\frac{0.458\ e^2}{r_s}-\frac{0.44\ e^2}{r_s+7.8\ a_0}
\hspace{0.5 cm} \text{.}
\end{equation}

Molte altre parametrizzazioni pi\`{u} sofisticate sono comunemente note \cite{Testo_MB} e sono ottenute
da fits di calcoli ab-initio virtualmente esatti.

%************************************************************************************************************
\subsection{Approssimazione LSDA}
Fino ad ora abbiamo parlato di parametrizzazioni per il funzionale $E_{xc}[\rho]$; all'interno di questa tesi
faremo per\`{o} principalmente uso del funzionale $E_{xc}[\rho_{\uparrow},\rho_{\downarrow}]$ per lo studio di
sistemi spin polarizzati. Il funzionale $E_{xc}[\rho]$, e dunque anche la sua approssimazione LDA, tratta in
modo simmetrico la componente $\rho_{\uparrow}$ e la componente $\rho_{downarrow}$
della densit\`{a} e dunque se provassimo a costruire da questa i potenziali
di scambio e correlazione $v^{\uparrow}_{xc}[\rho]$ e $v^{\downarrow}_{xc}[\rho]$ otterremmo
\begin{equation}
\begin{split}
v^{\uparrow}_{xc}[\rho]&=\ \frac{\delta E_{xc}[\rho_{\uparrow}+\rho_{\downarrow}]}
           {\delta \rho_{\uparrow}}=\frac{\delta E_{xc}[\rho]}{\delta \rho}
           \frac{\delta \rho}{\delta \rho_{\uparrow}}=v_{xc}[\rho]  \\
v^{\downarrow}_{xc}[\rho]&=\ \frac{\delta E_{xc}[\rho_{\uparrow}+\rho_{\downarrow}]}
           {\delta \rho_{\downarrow}}=\frac{\delta E_{xc}[\rho]}{\delta \rho}
           \frac{\delta \rho}{\delta \rho_{\downarrow}}=v_{xc}[\rho]
\hspace{0.5 cm} \text{.}
\end{split}
\end{equation}
Appare dunque evidente che per sistemi con magnetizzazione di spin sia necessario introdurre uno
schema differente per approssimare il funzionale. L'idea \`{e} quella di partire dall'energia del
sistema omogeneo completamente polarizzato, dove per completamente polarizzato intendiamo che
la grandezza
\begin{equation}
\zeta=\frac{\rho_{\uparrow}-\rho_{\downarrow}}{\rho}
\end{equation}
sia pari ad $1$, il che significa che tutti gli elettroni hanno spin orientato nella stessa direzione.
Calcolata l'energia del sistema omogeneo imponendo $\zeta=1$, si trovano, in approssimazione LDA,
le seguenti parametrizzazioni
\begin{equation}
\begin{split}
e_x(\rho,\zeta=1)&=\ 2^{1/3}\ e_x(\rho,\zeta=0) \\
e_c(\rho,\zeta=1)&=\ 0.5\ e_c(2^{4/9}\rho,\zeta=0)
\hspace{0.5 cm} \text{,}
\end{split}
\end{equation}
dove con $e_{x/c}(\rho,\zeta=0)$ abbiamo indicato la densit\`{a} di energia LDA.

L'approssimazione per sistemi spin polarizzati%
\footnote{o per lo studio di sistemi in presenza di campo magnetico},
che chiamiamo ``Local Spin Density Approximation'' (LSDA), viene costruita utilizzando
\begin{equation} \label{LSDA}
e_{xc}(\rho,\zeta)= e_{xc}(\rho,\zeta=0)+\big(e_{xc}(\rho,\zeta=1)-e_{xc}(\rho,\zeta=0)\big)g(\zeta)
\hspace{0.5 cm} \text{,}
\end{equation}
dove $g(\zeta)$ \`{e} una funzione di interpolazione che vale
\begin{equation}
g(\zeta)=\frac{(1+\zeta)^{4/3}+(1-\zeta)^{4/3}-2}{2(2^{1/3}-1)}
\hspace{0.5 cm} \text{.}
\end{equation}

%*******************************************************************************************************
%*******************************************************************************************************
%*******************************************************************************************************
\section{Perch\'{e} l'approssimazione LDA funziona cos\`{i} bene}
\subsection{Introduzione}
L'approssimazione eseguita (LDA) \`{e} basata sull'idea che il funzionale dell'energia di scambio
e correlazione non differisca di molto, nella forma, da quello del sistema omogeneo. Ci si aspetterebbe
dunque che un approccio di questo tipo possa esser ragionevole per sistemi in cui la densit\`{a}
varia lentamente. In realt\`{a} applicazioni di questa approssimazione sono state utilizzate, con ottimi
risultati, anche in sistemi che sono molto lontani dall'avere un comportamento della densit\`{a}
lentamente variabile, come possono essere ad esempio le molecole che studieremo all'interno di questa
tesi.

Il successo ottenuto pu\`{o} comunque essere spiegato dal fatto che l'approssimazione
locale soddisfa una regola di somma che riguarda ``la densit\`{a} di buca di scambio e correlazione''
e che la stessa densit\`{a} di buca in approssimazione LDA, pur essendo di molto differente da quella
reale, presenta una media sferica molto simile a quest'ultima. Nei prossimi paragrafi spiegheremo
dunque pi\`{u} in dettaglio questi punti, partendo dal teorema di connessione adiabatica e utilizzando
nella dimostrazione il teorema di Hellmann Feynman \cite{Payne}.

Concludiamo questa introduzione segnalando per\`{o} che pur essendo un ottimo punto di partenza
anche per lo studio di sistemi isolati l'approssimazione LDA ha l'inconveniente di non presentare il
corretto andamento asintotico del potenziale di scambio e correlazione; questo
inconveniente sar\`{a} una delle fonti di problemi nel calcolo dello spettro di eccitazione di 
alcune molecole.

%**************************************************************************************************
\subsection{Il teorema di connessione adiabatica}
Partiamo dal considerare un'Hamiltoniana $\hat{H}_{\lambda}$ cos\`{i} costruita
\begin{multline}
\hat{H}_{\lambda}=\hat{T}+\sum_{\sigma}\int d^3\mathbf{r}\ v_{\lambda}(\mathbf{r})
\hat{\psi}^{\dag}_{\sigma}(\mathbf{r})\hat{\psi}_{\sigma}(\mathbf{r})  \\
+\frac{\lambda}{2}\sum_{\sigma\sigma'}\int\int d^3\mathbf{r}d^3\mathbf{r}'\
w(\mathbf{r},\mathbf{r}')
\hat{\psi}^{\dag}_{\sigma}(\mathbf{r})\hat{\psi}^{\dag}_{\sigma'}(\mathbf{r'})
\hat{\psi}_{\sigma}(\mathbf{r})\hat{\psi}_{\sigma'}(\mathbf{r}')
\hspace{0.5 cm} \text{,}
\end{multline}
e per il valore $\lambda=1$ scegliamo di prendere $v_1=V^{ext}$, il potenziale esterno applicato
al nostro sistema fisico.

Ora sfruttiamo il teorema di Hohenberg-Kohn per il sistema descritto dalla seguente Hamiltoniana
e fissiamo idealmente il potenziale $v_{\lambda}$ per tutti i valori di $\lambda$ differenti da $1$
in modo che risulti
\begin{equation}
\rho_{\lambda}(\mathbf{r})=\sum_{\alpha}\langle\Psi^0_{\lambda}|\hat{\psi}^{\dag}_{\sigma}(\mathbf{r})
\hat{\psi}_{\sigma}(\mathbf{r})|\Psi^0_{\lambda}\rangle=
\rho_1(\mathbf{r})=\rho_o(\mathbf{r})
\hspace{0.5 cm} \text{,}
\end{equation}
In questo modo il potenziale $v_0$ \`{e} evidentemente $v_0=v_{KS}=V^{ext}+v_H+v_{xc}$.

Utilizziamo ora il teorema di Hellmann-Feynman, che corrisponde alla prima riga della
(\ref{Hellmann-Feynman}), e scriviamo la variazione
dell'energia dello stato fondamentale del sistema, intesa come il valor medio rispetto
alla funzione d'onda $|\Psi^0_{\lambda}\rangle$ dell'Hamiltoniana $\hat{H}_{\lambda}$, come
\begin{equation} \label{Hellmann-Feynman}
\begin{split}
\frac{dE(\lambda)}{d \lambda}=&\ \left\langle \Psi^0_{\lambda} \bigg|
\frac{\partial \hat{H}_{\lambda}}{\partial \lambda} \bigg| \Psi^0_{\lambda} \right\rangle \\
       =&\ \frac{1}{2} \int\int d^3\mathbf{r}d^3\mathbf{r}'\ w(\mathbf{r},\mathbf{r}')
\big\langle\Psi^0_{\lambda}|\hat{\rho}(\mathbf{r})\hat{\rho}(\mathbf{r}')-\delta(\mathbf{r-r'})
\hat{\rho}(\mathbf{r})|\Psi^0_{\lambda}\big\rangle \\
        &\phantom{\frac{1}{2} \int\int d^3\mathbf{r}d^3\mathbf{r}'\ w(\mathbf{r},\mathbf{r}')
\big\langle\Psi^0_{\lambda}|\hat{\rho}(\mathbf{r})\hat{\rho}(\mathbf{r}')}
+\frac{\partial}{\partial \lambda}\int d^3\mathbf{r}\ v_{\lambda}(\mathbf{r})\rho(\mathbf{r})
\hspace{0.5 cm} \text{.}
\end{split}
\end{equation}

L'energia dello stato fondamentale del sistema fisico, che \`{e} quella dell'Hamiltoniana
a $\lambda=1$, pu\`{o} ora essere scritta come
\begin{multline}
E(1)=E(0)+\int_0^1 d\lambda \frac{dE(\lambda)}{d\lambda}=T_{s}[\rho]+\int d^3\mathbf{r}\
V^{ext}(\mathbf{r})\rho(\mathbf{r}) \\
+\frac{1}{2}\int \int d^3\mathbf{r}d^3\mathbf{r}'\ w(\mathbf{r},\mathbf{r}')
\rho(\mathbf{r})\rho(\mathbf{r}')+E_{xc}[\rho]
\hspace{0.5 cm} \text{,}
\end{multline}
dove abbiamo definito
\begin{multline}
E_{xc}=\frac{1}{2}\int_0^1 d\lambda \int\int d^3\mathbf{r}d^3\mathbf{r}'\
w(\mathbf{r,r'}) \\
\left[\big\langle\Psi^0_{\lambda}|\left(\hat{\rho}(\mathbf{r})-\rho(\mathbf{r})\right)
\left(\hat{\rho}(\mathbf{r'})-\rho(\mathbf{r}')\right)|\Psi^0_{\lambda}\big\rangle
-\rho(\mathbf{r})\delta(\mathbf{r-r'})\right]
\hspace{0.5 cm} \text{.}
\end{multline}
Definendo ora la funzione di correlazione a coppie $g_{\lambda}$ come estensione della
(\ref{correlazione-coppie}) nel caso in cui la funzione d'onda dipenda da $\lambda$ 
otteniamo
\begin{equation}
E_{xc}=\frac{1}{2} \int\int d^3\mathbf{r}d^3\mathbf{r}'\
w(\mathbf{r,r'})\rho(\mathbf{r})\rho(\mathbf{r'})
\int_0^1 d\lambda [g_{\lambda}(\mathbf{r,r',[\rho]})-1]
\hspace{0.5 cm} \text{.}
\end{equation}

%*****************************************************************************************
\subsection{La buca di scambio e correlazione}
Per capire il successo dell'approssimazione LDA ci resta ora da capire cosa significhi
eseguire tale approssimazione per la funzione $g_{\lambda}$

A questo scopo definiamo un'ulteriore grandezza
\begin{equation}
\varrho_{xc}=\rho(\mathbf{r'})
\left[\int_0^1 d\lambda [g_{\lambda}(\mathbf{r,r',[\rho]})-1]\right]
\hspace{0.5 cm} \text{,}
\end{equation}
che pu\`{o} essere interpretata come la densit\`{a} di una buca di
scambio e correlazione \cite{Testo_DFT}. Inserendo questa espressione nel funzionale per l'energia di scambio
e correlazione \`{e} dunque possibile interpretare tale energia come un'interazione 
particella-buca di scambio e correlazione.

Ora \`{e} possibile mostrare che, espandendo tale densit\`{a} in termini di multipolo rispetto
alla distanza $\mathbf{r-r'}$ solo il termine di monopolo, cio\`{e} il termine con andamento
di tipo sferico, d\`{a} contributo all'energia di scambio e correlazione%
\footnote{Questo \`{e} vero solo nel caso di un'interazione che dipende soltanto da
$|\mathbf{r-r'}|$.}
e che \`{e} sempre il
termine di monopolo il solo a dare contributo alla regola di somma
\begin{equation}
\int d^3\mathbf{r'}\varrho_{xc}(\mathbf{r,r'})=-1
\hspace{0.5 cm} \text{.}
\end{equation}

\begin{figure}[!ht] 
\begin{center}
\subfigure[Buca di scambio e correlazione]
{\includegraphics[width=14 cm]{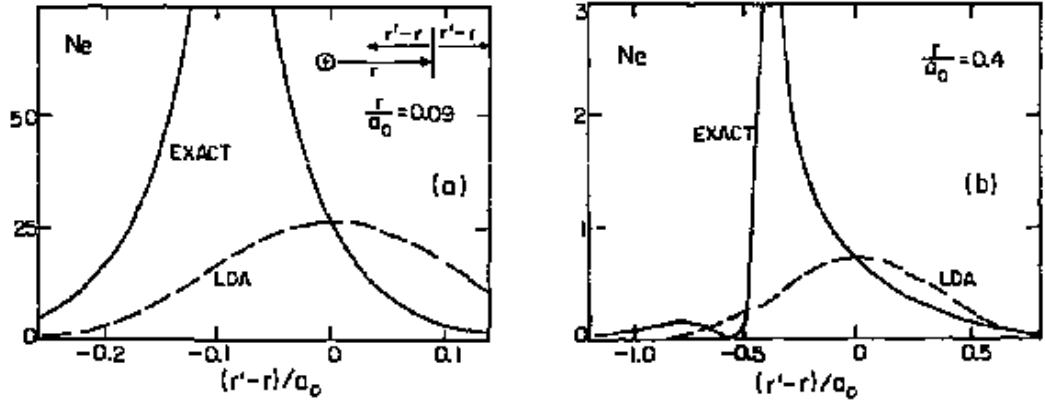}}
\subfigure[Termine di monopolo della buca di scambio e correlazione. Media sferica.]
{\includegraphics[width=14 cm]{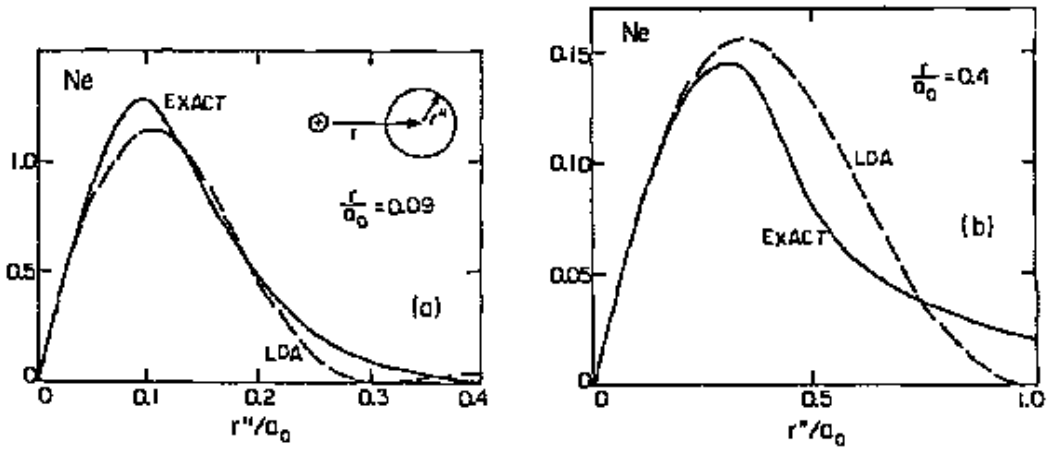}}
\caption{Confronto tra l'andamento della densit\`{a} esatta di scambio e correlazione
         e la sua media sferica.}
\label{fig:LDA_successo}
\rule{\linewidth}{0.2 mm} 
\end{center}
\end{figure}

Il successo dell'approssimazione LDA nasce proprio dal fatto che anche per quest'ultima
$\varrho_{xc}$ ha un andamento di tipo sferico e soddisfa alla regola di somma valida per
il caso esatto. In altre parole pur essendo $\varrho_{xc}$ molto differente da
$\varrho^{LDA}_{xc}$ per sistemi in cui la densit\`{a} $\rho$ varia rapidamente,
risultano molto simili (\ref{fig:LDA_successo}) le due grandezze
$\varrho^{LDA}_{xc}$ e $\varrho^{00}_{xc}$, dove con $00$ abbiamo preso
il termine di monopolo della buca di scambio e correlazione.
%*******************************************************************************************************
%*******************************************************************************************************
%*******************************************************************************************************
\section{Oltre l'approssimazione locale}
\subsection{L'equazione di Sham-Schluter}
L'equazione di Sham-Schluter costituisce un possibile punto di partenza per la costruzione di
un potenziale di scambio e correlazione nella teoria DFT a partire dalla MBPT.
Essa stabilisce infatti una relazione esatta tra il potenziale di scambio e correlazione e
la self-energia del sistema.
Il punto di partenza per giungere a tale equazione \`{e} dato dall'osservazione che la parte diagonale
della funzione di Green coincide con la densit\`{a} del sistema nello stato fondamentale che, come 
sappiamo, pu\`{o} essere calcolata nel formalismo del funzionale densit\`{a}.
\begin{equation}
-iG(1,1^+)=\rho(1)
\hspace{0.5 cm} \text{.}
\end{equation}

Partendo da questa osservazione l'idea \`{e} quella di definire la funzione di Green soluzione 
dell'equazione del moto di KS come $G_{KS}(1,2)$. Se ora cerchiamo di stabilire un legame tra 
la funzione di Green esatta e quella appena definita osservando le due equazioni del moto a cui
queste soddisfano otteniamo
\begin{equation}
G(1,2)=G_{KS}(1,2)+G_{KS}(1,3)\bigg(\Sigma(3,4)-\delta(3,4)v_{xc}(4)\bigg)G(4,2)
\hspace{0.5 cm} \text{.}
\end{equation}
Ricordiamo che della funzione di Green $G_{KS}(1,2)$ solo la parte diagonale ha significato
fisico. Utilizzando ora proprio il fatto che il termine diagonale di entrambe coincide con la 
densit\`{a} del sistema otteniamo
\begin{equation} \label{Sham-Schluter}
G_{KS}(1,3)\bigg(\Sigma^{\star}(3,4)-\delta(3,4)v_{xc}(4)\bigg)G(4,1)=0
\hspace{0.5 cm} \text{.}
\end{equation}

Quella scritta \`{e} l'equazione di Sham-Schluter. Come sempre quando abbiamo a che fare con grandezze
a molti corpi essa non pu\`{o} essere risolta in modo esatto, \`{e} necessario quindi trovarne delle
soluzioni approssimate.
Una possibilit\`{a} \`{e} quella di trovare una soluzione linearizzata di tale equazione
(il che significa approssimare nella (\ref{Sham-Schluter}) $G(4,1)\simeq G_{KS}(4,1)$);
soluzione che pu\`{o} essere
utilizzata ad esempio per ottenere un potenziale di scambio e correlazione con il corretto andamento
asintotico in modo da correggere l'errore prodotto dall'approssimazione LDA \cite{Tesi_dot_Fabien}.

%% file: excitations.tex
\section{Introduzione}
Fino ad ora ci siamo occupati di delineare le due teorie per lo studio dello stato fondamentale
del nostro sistema ed abbiamo mostrato come da entrambe possiamo ottenere l'energia totale.
Abbiamo inoltre gi\`{a} parlato delle energie di eccitazione del sistema per stati che presentano
una particella in pi\`{u} o in meno dello stato fondamentale del sistemi. Non abbiamo invece
detto nulla sullo spettro di eccitazione che riguarda stati che presentino lo stesso numero di
particelle dello stato fondamentale poich\`{e} lo studio di tali energie comporta la necessit\`{a}
di tener conto di un importante effetto fisico, l'accoppiamento particella
buca, che le teorie si qui formulate non sono in grado di descrivere in modo esauriente. Avremo
dunque la necessit\`{a} di svilupparle ulteriormente entrambe; andremo dunque verso quelle che sono
le attuali frontiere delle ricerca, facendo in particolare riferimento, per quanto concerne lo
stato dell'arte alle referenze \cite{Onida,Testo_TDDFT}.

Per quel che riguarda la teoria many body lo spettro di eccitazione pu\`{o} in realt\`{a} gi\`{a}
essere ottenuto a partire dalla rappresentazione di Lehmann della funzione risposta (o polarizzazione).
Studiando per\`{o} tale grandezza vedremo che compariranno delle funzioni ``a quattro punti'', che sono
legate quindi alla funzione di Green a due particelle. Per questo motivo, se vogliamo scrivere
un'equazione del tipo ``equazione di Dyson'' per questa grandezza avremo bisogno di definire una
funzione di risposta a quattro punti. Il risultato di questo tipo di approccio sar\`{a} l'equazione
di Bethe-Salpeter.

Per quel che riguarda la teoria del funzionale densit\`{a}, sebbene in principio a partire
dal teorema di Hohenberg-Kohn potremmo accedere allo spettro di eccitazione del sistema, di
fatto abbiamo bisogno di costruire un metodo pratico per accedere a tale spettro. Come gi\`{a}
anticipato tale metodo sar\`{a} costituito dalla TDDFT o meglio dalla Time-Dependent
Density-Functional Response Theory (TD-DFRT\footnote{Per ragioni pratiche d'ora in avanti ci
limiteremo a parlare di TDDFT}.). Dovremo dunque dimostrare nuovi
teoremi che possano estendere la teoria al caso dipendente dal tempo.

Prima di procedere per\`{o} allo sviluppo delle due teorie vogliamo introdurre la teoria della risposta
lineare e mostrare come, a partire da questa, sar\`{a} possibile ricavare lo spettro di eccitazione
del sistema attraverso la sua rappresentazione di Lehmann.

%*****************************************************************************************************
%*****************************************************************************************************
%*****************************************************************************************************
\section{La teoria della risposta lineare}
Consideriamo un sistema descritto da un'Hamiltoniana $\hat{H}$ indipendente dal tempo.
Immaginiamo quindi di perturbare il sistema accendendo, all'istante $t=t_0>0$, una
perturbazione esterna dipendente dal
tempo $\delta\hat{V}^{ext}(t)$ che possa essere considerata piccola. Scopo della teoria della risposta
lineare \`{e} studiare come varia una grandezza fisica, all'ordine lineare, dopo l'accensione della
perturbazione esterna.
L'evoluzione di uno stato in descrizione di Schro\"{e}dinger
sar\`{a} data dalla solita funzione
esponenziale per la parte di Hamiltoniana indipendente dal tempo e dall'operatore descritto nella
(\ref{ev-interazione}) per la perturbazione che \`{e} dipendente dal tempo.
Risulter\`{a} dunque:
\begin{equation}
\begin{split}
|\Psi_S(t)\rangle =&\ e^{-i\hat{H}t/\hbar}\hat{U}(t,t_0) |\Psi(t_0)\rangle   \\
                  =&\ e^{-i\hat{H}t/\hbar}|\Psi(t_0)\rangle +
                      e^{-i\hat{H}t/\hbar}\int_{t_0}^t dt'\ \delta\hat{V}_{\hat{H}}^{ext}(t')
                        |\Psi(t_0)\rangle + ...
\hspace{0.5 cm} \text{.}
\end{split}
\end{equation}
dove abbiamo sviluppato all'ordine lineare rispetto alla perturbazione esterna l'operatore di
evoluzione temporale $\hat{U}(t,0)$. $t_0$ \`{e} l'istante a cui scegliamo di accendere la perturbazione
esterna, mentre lo stato $\Psi(t_0)$ \`{e} autostato dell'Hamiltoniana $\hat{H}$ ed \`{e}
indipendente dal tempo: $\Psi(t_0)=\Psi(0)$.
Scriviamo ora il valor medio di una grandezza qualsiasi che venga rappresentata da un operatore
$\hat{O}(t)$ rispetto allo stato perturbato
\begin{equation}
\langle \hat{O}(t) \rangle = \langle \Psi(0) | \hat{O}_{\hat{H}}(t) | \Psi(0) \rangle +
                             \int_{t_0}^t dt' \langle \Psi(0) |[\hat{O}_{\hat{H}}(t),
                             \delta\hat{V}_{\hat{H}}^{ext}(t') | \Psi(0) \rangle + ...
\hspace{0.5 cm} \text{,}
\end{equation}
dove abbiamo spostato l'operatore di evoluzione temporale sull'operatore $\hat{O}(t)$ in modo che ora
l'operatore sia scritto in descrizione di Heisenberg rispetto all'Hamiltoniana totale $\hat{H}$.
In questo modo oltre alla possibile dipendenza dal tempo del'operatore, l'operatore dipender\`{a}
dal tempo perch\'{e} scritto in tale rappresentazione ($\hat{O}_{\hat{H}}(t)$)

Di particolare interesse \`{e} la risposta dell'operatore densit\`{a} ad un potenziale esterno che
possa essere scritto nella forma:
\begin{equation}
\delta \hat{V}^{ext}(t)= \sum_{\sigma}\int d^3\mathbf{x}\
\delta V_{\sigma}^{ext}(\mathbf{x},t) \hat{\rho}_{\sigma}(\mathbf{x})
\hspace{0.5 cm} \text{.}
\end{equation}
In questo modo la formula della risposta lineare pu\`{o} essere scritta nel seguente modo
\begin{multline}
\rho_{\sigma}(\mathbf{x},t)=\rho_{\sigma}^0(\mathbf{x},t)+ \sum_{\sigma'}\int_{-\infty}^{+\infty} dt'
                         \int d^3 \mathbf{x}\  \Theta(t-t') \delta V_{\sigma'}^{ext}(\mathbf{x}'t') \\
\langle \Psi(0) |\left[\hat{\rho}_{\sigma}(\mathbf{x},t),
                         \hat{\rho}_{\sigma'}(\mathbf{x}',t')\right]|\Psi(0) \rangle
\hspace{0.5 cm} \text{,}
\end{multline}
dove l'integrazione sul tempo \`{e} stata estesa a tutto l'asse reale sfruttando il fatto che la
perturbazione \`{e} nulla per $t'<t_0$ e inserendo la funzione $\Theta(t-t')$ per i tempi
$t'>t$.
La funzione di risposta pu\`{o} essere definita come (torniamo qui ad utilizzare la notazione
$1=\mathbf{x}_1,t_1,\sigma_1$):
\begin{equation}
\chi^R(1,2)=\langle \Psi(0) |\left[\hat{\rho}(1),
             \hat{\rho}(2)\right] |\Psi(0) \rangle \Theta(t_1-t_2)
\hspace{0.5 cm} \text{.}
\end{equation}
In questo modo risulta infatti
\begin{equation} \label{pol-rit-definizione}
\delta \rho(1)=\chi^R(1,2) V^{ext}(2)
\hspace{0.5 cm} \text{,}
\end{equation}
dove stiamo integrando sulle variabili ripetute. Troviamo cos\`{i} la definizione classica di
funzione risposta come la variazione della densit\`{a} rispetto ad una perturbazione esterna
applicata.

Quella che abbiamo qui definito \`{e} in realt\`{a} la funzione di risposta ``ritardata''. Fino ad ora
avevamo invece lavorato con grandezze ``T-ordinate'', poich\'{e} queste sono le uniche che possiamo 
sviluppare utilizzando in teorema di Wick e quindi i diagrammi di Feynman. La polarizzazione o
funzione di risposta T-ordinata infatti, se calcolata esplicitamente utilizzando la definizione
(\ref{polarizzazione-def}) risulta pari a
\begin{equation}
\chi(1,2)=\langle
\Psi(0) |T\left[\delta\hat{\rho}(1)
                \delta\hat{\rho}(2)\right]|\Psi(0) \rangle
\hspace{0.5 cm} \text{,}
\end{equation}
dove $\delta\hat{\rho}=\hat{\rho}-\rho^0$ \`{e} l'operatore fluttuazione rispetto alla
densit\`{a} media $\rho^0=\langle\Psi(0)|\hat{\rho}|\Psi(0)\rangle$.

I due tipi di grandezze
possono in realt\`{a} essere ricollegate (insieme con le grandezza ``avanzate'') utilizzando la
rappresentazione di Lehmann delle stesse. Vediamo come nel prossimo paragrafo.

%*****************************************************************************************************
\subsection{La rappresentazione di Lehmann della funzione risposta}
Riprendiamo la definizione di rappresentazione di Lehmann (\ref{Lehmann_Green}) della funzione di Green
e svolgendo passaggi del tutto analoghi a quelli gi\`{a} svolti per tale grandezza otteniamo
la rappresentazione di Lehmann delle due funzioni risposta. La sola differenza sar\`{a} questa volta
data dal fatto che il termine $I=0$ non compare poich\'{e} viene cancellato in modo esatto sia
nella funzione T-ordinata che in quella ritardata. Scriviamo dunque i due risultati
\begin{equation} \label{Lehmann-polarizzazioneR}
\chi^R(\mathbf{x},\mathbf{x}',\omega) = \sum_{I\neq0} \frac{\langle\Psi_0|\hat{n}(\mathbf{x})|\Psi_I\rangle
                                                  \langle\Psi_I|\hat{n}(\mathbf{x}')|\Psi_0\rangle}
                                                 {\omega-\omega_I+i\eta}-
                                              \frac{\langle\Psi_0|\hat{n}(\mathbf{x}')|\Psi_I\rangle
                                                  \langle\Psi_I|\hat{n}(\mathbf{x})|\Psi_0\rangle}
                                                 {\omega+\omega_I+i\eta}
\hspace{0.2 cm} \text{,}
\end{equation} 
\begin{equation} \label{Lehmann-polarizzazione}
\chi(\mathbf{x},\mathbf{x}',\omega) = \sum_{I\neq0} \frac{\langle\Psi_0|\hat{n}(\mathbf{x})|\Psi_I\rangle
                                                  \langle\Psi_I|\hat{n}(\mathbf{x}')|\Psi_0\rangle}
                                                 {\omega-\omega_I+i\eta}-
                                              \frac{\langle\Psi_0|\hat{n}(\mathbf{x}')|\Psi_I\rangle
                                                  \langle\Psi_I|\hat{n}(\mathbf{x})|\Psi_0\rangle}
                                                 {\omega+\omega_I-i\eta}
\hspace{0.5 cm} \text{.}
\end{equation}
Vediamo come la funzione ritardata e la seconda funzione T-ordinata differiscano solo per il segno
della parte immaginaria a denominatore.

Lo spettro di eccitazione $\omega_I$ scritto a denominatore della (\ref{Lehmann-polarizzazioneR})
e della (\ref{Lehmann-polarizzazione}) \`{e} costituito dagli autovalori
di tutti quegli stati che hanno lo stesso numero di particelle del ground-state (poich\'{e} l'operatore
densit\`{a} commuta con l'operatore numero di particelle) relativo all'energia dello stato fondamentale:
$\omega_I=E_I-E_O$
Per poter ottenere dunque tali valori non dobbiamo far altro che calcolare i poli di tale grandezza.
Vediamo infine come la funzione ritardata e la funzione T-ordinata differiscano solo per il segno
della parte immaginaria a denominatore.

%*****************************************************************************************************
%*****************************************************************************************************
%*****************************************************************************************************
\section{L'approccio many-body}
\subsection{Perch\'{e} la funzione di Green a particella singola non \`{e} sufficiente}
L'idea per poter studiare la funzione risposta \`{e} quella di scrivere un'equazione di tipo
Dyson per tale grandezza; nell'ambito dell'approccio many-body delle due grandezze studieremo
la funzione risposta T-ordinata.
A questo scopo riscriviamo l'equazione di Hedin (\ref{Hedin-polarizzazione})
per la polarizzazione inserendo al posto del
vertice $\Gamma$ la sua equazione di Hedin (\ref{vertice}) nella versione completa
\begin{multline} \label{vertice-completo}
\Gamma(1;2,3)=\delta(1,2)\delta(1,3)+
              \frac{1}{\hbar} \Gamma(1;4,5)G(6,4)G(7,5)  \\
                \frac{\delta}{\delta G(6,7)}
                \big[\delta(2,3)v_H(3)+\Sigma^{\star}(2,3)\big]
\hspace{0.5 cm} \text{.}
\end{multline}
In questo modo otteniamo
\begin{multline} \label{Polarizzazione-Dyson}
\Pi(1,2)= G(1,2)G(2,1) +
          G(1,3)G(1,4) \\ \frac{\delta}{\delta G(5,6)}
          \big[\delta(2,3)v_H(3)+\Sigma^{\star}(3,4)\big]
          G(5,7)G(6,8) \Gamma(7,8;2)
\hspace{0.5 cm} \text{.}
\end{multline}
Se ora cerchiamo di ricavare un'equazione di tipo Dyson per la polarizzazione, definendo
la grandezza $\Pi_0(1,2)=G(1,2)G(2,1)$ come
ordine zero del nostro sviluppo, ci accorgiamo immediatamente
che non \`{e} possibile ottenere il nostro obiettivo poich\'{e} le due funzioni di Green che 
compaiono a sinistra della derivata funzionale non sono chiuse, cos\`{i} come non lo sono
le due funzioni di Green che compaiono a sinistra della stessa derivata funzionale. 
Se cos\`{i} fosse potremmo infatti riscrivere l'equazione appena ricavata utilizzando la
(\ref{Hedin-polarizzazione}) e la definizione appena introdotta di $\Pi_0:$
\begin{equation}
\Pi= \Pi_0 + \Pi_0 K \Pi
\hspace{0.5 cm} \text{,}
\end{equation}
con la grandezza $K$ (il kernel dell'equazione) definita da
\begin{equation}
K(1,2;3,4)=\frac{\delta}{\delta G(3,4)}
           \big[\delta(1,2)v_H(2)+\Sigma^{\star}(1,2)\big]
\hspace{0.5 cm} \text{.}
\end{equation}
Il problema \`{e} che compare una grandezza sconosciuta, un kernel a quattro punti;
fatto che in qualche modo \`{e} legato alla forma dell'operatore di interazione, che
\`{e} un operatore a due particelle e che porta alla comparsa di grandezze a quattro punti
come avevamo gi\`{a} visto nell'equazione del moto per la funzione di Green 
(\ref{moto1}) in cui compare la funzione di Green a due particelle. Nello
sviluppare la teoria per il ground-state siamo riusciti a superare questa difficolt\`{a} introducendo
la self-energia del sistema e quindi le equazioni di Hedin. Sebbene anche in questa formulazione
compaia lo stesso termine a quattro punti nell'equazione per il vertice, termine che costituisce
la fonte di tutte le difficolt\`{a} che si incontrano del tentativo di risolvere tale set di equazioni
in modo esatto, per lo studio del ground state \`{e} possibile fare approssimazioni (l'approssimazione
GW) che ci permettano di lavorare semplicemente con grandezze a due punti.

Cercando di calcolare lo spettro di eccitazione del sistema non \`{e} invece pi\`{u} conveniente
continuare a formulare la teoria in termini di grandezze a due punti.
La ragione fisica di questa necessit\`{a}
pu\`{o} essere individuata nel fatto che quando eccitiamo un sistema dal suo stato fondamentale
oltre a promuovere un elettrone in uno stato eccitato, creiamo una buca (hole) nel sistema e che
l'interazione tra i due oggetti diventa decisiva nello studio del sistema. Non \`{e} pi\`{u} dunque
possibile relegare il ruolo dell'interazione coulombiana schermata particella-buca in secondo ordine,
all'interno della
self energia. Il problema in altre parole dipende dal fatto che, poich\'{e} non siamo in grado di
risolvere la teoria in modo esatto dobbiamo cercare di ``mettere in evidenza'' le grandezze che hanno
un ruolo dominante dal punto di vista fisico.

%************************************************************************************************
\subsection{L'equazione di Bethe Salpeter}
L'idea per costruire l'equazione di Bethe-Salpeter \`{e} quella di partire dalla
funzione di Green a due particelle. A questo scopo introduciamo una nuova grandezza che definiamo
come la differenza tra la funzione di Green a due particelle, che descrive appunto la propagazione
delle due mentre interagiscono tra loro (oltre che con il resto del sistema), e il prodotto di due
funzioni di Green a particella singola, che descrivono la propagazione delle due particelle senza
tener conto della reciproca interazione.
\begin{equation}
iL(1,2;3,4)=-G_2(1,2;3,4) + G(1,3)G(2,4)
\hspace{0.5 cm} \text{.}
\end{equation}
Questa grandezza descrive evidentemente il moto accoppiato delle due particelle, possiamo immaginarla
come la correzione alla semplice propagazione indipendente delle due.
I passaggi che dobbiamo ora svolgere per la costruzione di un'equazione di tipo Dyson per questa
grandezza sono del tutto simili a quelli che abbiamo utilizzato per costruire l'equazione di Hedin
per la funzione risposta (o polarizzazione) e quindi l'equazione scritta nel paragrafo precedente
con la quale non siamo riusciti a raggiungere il nostro scopo semplicemente perch\'{e} stavamo
lavorando con grandezze a 2 punti.
Partiamo con l'osservare che la funzione $L(1,2;3,4)$ coincide con la definizione del
termine di destra dell'equazione (\ref{two-particels}) se invece di considerare un potenziale
esterno a particella singola ne consideriamo uno a due particelle:
\begin{equation}
iL(1,2;3,4)=\frac{\delta G(1,3)}{\delta \varphi(4,2)}
\hspace{0.5 cm} \text{.}
\end{equation}
Procediamo quindi come in precedenza utilizzando la relazione (\ref{rel-Green})
\begin{equation}
iL(1,2;3,4)= -\ G(1,5)G(6,3)\frac{\delta G^{-1}(5,6)}{\delta \varphi(4,2)}
\end{equation}
e definendo, portando avanti l'analogia con quanto fatto in precedenza, una
funzione di vertice a 4 punti
\begin{equation}
\Gamma_4(1,2;3,4)=\frac{G^{-1}(1,3)}{\delta \varphi(4,2)}
\hspace{0.5 cm} \text{.}
\end{equation}
Possiamo ora scrivere l'equivalente dell'equazione (\ref{vertice}) nel caso che stiamo considerando,
cio\`{e} con un vertice a quattro punti completo
\begin{equation}
\Gamma_4(1,2;3,4)=\delta(1,3)\delta(2,4)
+\frac{\delta\left(\delta(1,3)v_H(1)+\Sigma^{\star}(1,3)\right)}{\delta \varphi(4,2)}
\hspace{0.5 cm} \text{.}
\end{equation}
Utilizziamo infine le regole di derivazione a catena e definiamo (ora si \`{e} possibile)
l'ordine zero della nostra funzione ($L_0(1,2;3,4)=G(1,4)G(2,3)$) insieme con il kernel
\begin{equation}
K(1,2;3,4)=\frac{\delta (\delta(1,3)v_H(1)+\Sigma^{\star}(1,3))}{\delta G(4,2)}
\hspace{0.5 cm} \text{,}
\end{equation}
ottenendo la seguente equazione:
\begin{equation} \label{Bethe-Salpeter}
\begin{split}
L(1,2;3,4)&=G(1,4)G(2,3)+ \\
&\phantom{G(1,4)}G(1,6)G(2,5)
\frac{\delta \big(\delta(5,7)v_H(5)+\Sigma^{\star}(5,7)\big)}{\delta G(8,6)}
\frac{\delta G(7,3)}{\delta \varphi(4,8)}  \\
&=L_0(1,2;3,4)+  \\
&\phantom{L_0(1,2)}L_0(1,2;5,6)K(5,6;7,8)L(7,8;3,4)
\hspace{0.5 cm} \text{.}
\end{split}
\end{equation}
Quella appena scritta \`{e} l'equazione di Bethe-Salpeter ed \`{e} l'equazione utilizzata
nell'approccio many-body per studiare le eccitazioni del sistema. Confrontando infatti le
equazioni (\ref{Polarizzazione-Dyson}) e (\ref{Bethe-Salpeter}) si vede come la contrazione
della funzione $L(1,2;3,4)$ dia come risultato la polarizzazione del sistema
\begin{equation}
L(1,2;1,2)=\Pi(1,2)
\end{equation}
e dunque lo spettro di eccitazione del sistema.

%******************************************************************************************************
%******************************************************************************************************
%******************************************************************************************************
\section{La Time-Dependent DFT}
Per poter studiare il problema delle eccitazioni nell'ambito della teoria del funzionale
densit\`{a} il primo passo da fare \`{e} quello di estendere la validit\`{a} del teorema
di Hohenberg-Kohn al caso in cui l'Hamiltoniana sia dipendente dal tempo.
Se consideriamo infatti un'Hamiltoniana in cui \`{e} presente un potenziale che dipende dal
tempo il nostro sistema non presenta pi\`{u} uno stato fondamentale e dunque non \`{e} pi\`{u}
possibile utilizzare il teorema di HK.
Tale estensione \`{e} stata elaborata da Runge e Gross nel 1984 e definisce una corrispondenza
biunivoca (a meno di una costante additiva dipendente dal tempo per il potenziale e di una fase
dipendente dal tempo per la funzione d'onda) tra la densit\`{a}, la funzione d'onda e il potenziale
esterno. A questo scopo viene richiesto che sia gi\`{a} definito uno stato iniziale al tempo
$t_0$ e che il potenziale esterno sia sviluppabile in serie di Taylor nell'intorno dell'istante
$t_0$ stesso. Non siamo per\`{o} ora interessati ai dettagli della dimostrazione, per la quale,
cos\`{i} come per gli altri dettagli della teoria del funzionale densit\`{a}, rimandiamo chi
fosse interessato alle citazioni della bibliografia. Ci\`{o} a cui siamo ora interessati \`{e}
mostrare come possa essere utilizzata l'estensione al caso dipendente dal tempo della teoria
per ottenere lo spettro di eccitazione del nostro sistema. In particolare in questa sezione
lavoreremo nell'ambito della teoria della risposta lineare, per la quale, oltre che nel risultato
di Runge e Gross, la teoria trova giustificazione formale in un altro lavoro di Gross e Kohn.

%************************************************************************************************
\subsection{Utilizziamo la teoria della risposta lineare nella TDDFT}
L'idea per ottenere lo spettro di eccitazione di un sistema a partire dalla TDDFT \`{e}
semplicemente quella si studiare la relazione tra una perturbazione esterna dipendente dal
tempo introdotta nel nostro sistema e la variazione della densit\`{a}. Entrambe queste grandezze
infatti devono corrispondere a quelle reali nell'ambito di questa teoria.

Consideriamo dunque nuovamente il nostro sistema descritto dall'Hamiltoniana
\begin{equation}
\hat{H}=\hat{T}+\hat{W}+\hat{V}^{ext}
\hspace{0.5 cm} \text{,}
\end{equation}
in cui il potenziale esterno non dipende dal tempo e la corrispondente Hamiltoniana
efficace della DFT
\begin{equation}
\hat{H}_{DFT}=\hat{T}+\hat{V}^{ext}+\hat{v}_H[\rho]+\hat{v}_{xc}[\rho]
\hspace{0.5 cm} \text{.}
\end{equation}

Inseriamo quindi una perturbazione $\delta V^{ext}(t)$ dipendente dal tempo e vediamo
come varia l'Hamiltoniana della DFT
\begin{equation}
\delta \hat{v}_s(t)= \delta \hat{V}^{ext}(t) + \delta \hat{v}_H
                      +\delta \hat{v}_{xc}
\hspace{0.5 cm} \text{,}
\end{equation}
dove $\delta \hat{v}_H = \hat{v}_H[\rho'] - \hat{v}_H[\rho]$ e $\delta \hat{v}_{xc} =
\hat{v}_{xc}[\rho'] - \hat{v}_{xc}[\rho]$. La variazione della densit\`{a} potr\`{a}
essere scritta, all'ordine lineare, come
\begin{equation} \label{risp-KS}
\delta \rho(1) = \ \chi^{KS}_0(1,2) \delta v_s(2)
\hspace{0.5 cm} \text{,}
\end{equation}
dove la funzione di risposta $\chi^{KS}$ \`{e} stata indicata con l'apice $KS$ in
quanto descrive la risposta delle particelle del sistema di KS, che ricordiamo
sono particelle non interagenti. Notiamo come rispetto alla teoria many-body qui utilizziamo
la funzione di risposta ritardata.

Poich\'{e} siamo interessati alla variazione della densit\`{a}
rispetto al potenziale esterno trasformiamo l'equazione precedente scrivendo esplicitamente
la forma del potenziale efficace. In particolare, per consistenza con l'approssimazione in cui
stiamo lavorando, linearizziamo i potenziali di scambio e correlazione e il potenziale di
Hartree rispetto alla variazione della densit\`{a}.
In questo modo risulta
\begin{eqnarray}
v_H[\rho'] \simeq v_H[\rho] + \frac{\delta v_H[\rho]}
                                   {\delta \rho} \delta \rho   \nonumber   \\
v_{xc}[\rho'] \simeq v_{xc}[\rho] + \frac{\delta v_{xc}[\rho]}
                                         {\delta \rho} \delta \rho
\hspace{0.5 cm} \text{.}
\end{eqnarray}
Definiamo ora la grandezza
\begin{equation}
f_{Hxc}[\rho](1,2) = \frac{\delta v_H[\rho](1)}{\delta \rho(2)} +
          \frac{\delta v_{xc}[\rho](1)}{\delta \rho(2)} 
\hspace{0.5 cm} \text{,}
\end{equation}
la sostiamo nell'equazione (\ref{risp-KS}) e, utilizzando la (\ref{pol-rit-definizione}), otteniamo
\begin{multline} \label{chi-tddft}
\delta \rho(1) =\ (\chi^{KS}(1,2)\delta(2,3)\delta(3,4)  \\
                +\chi^{KS}(1,2)f_{Hxc}(2,3)\chi^R(3,4))\delta V^{ext}(4)
\hspace{0.5 cm} \text{.}
\end{multline}
Abbiamo cos\`{i} scritto al relazione tra il potenziale esterno e la densit\`{a} utilizzando
gli elementi della formulazione many-body della TDDFT. Poich\'{e} in particolare sappiamo
che i poli della funzione $\chi_0^{RS}$ non sono altro che le differenze tra gli autovalori
di KS possiamo gi\`{a} vedere come all'ordine zero della teoria (scegliendo cio\`{e} $f_{Hxc}=0$)
sia in effetti giustificato considerare gli autovalori di KS
per valutare le eccitazioni del sistema. Tale approssimazione \`{e} d'altra parte spesso
insufficiente a dare una descrizione soddisfacente delle propriet\`{a} fisiche di un 
sistema qualsiasi. Tutto questo sar\`{a} poi meglio chiarito nella formulazione
di Casida della TDDFT (Capitolo 6). 

%********************************************************************************************
\subsection{Un'equazione di Dyson per la TDDFT}
Il risultato appena ottenuto pu\`{o} essere facilmente riscritto sotto forma di equazione di
Dyson per la funzione di risposta. Un semplice confronto dell'equazione (\ref{chi-tddft})
con l'equazione (\ref{pol-rit-definizione}) ci porta
\begin{equation} \label{Dyson-tddft}
\chi^R(1,2)= \chi^{KS}(1,2) + \ \chi^{KS}(1,3) f_{Hxc}(3,4) \chi^{R}(4,2)
\hspace{0.5 cm} \text{.}
\end{equation}
Sottolineiamo ancora una volta la differenza tra la teoria many-body e la TDDFT. In quest'ultima
infatti siamo riusciti a scrivere un'equazione di Dyson per la funzione risposta utilizzando
grandezze a due punti poich\'{e} a differenza della teoria many-body stiamo utilizzando due
grandezze diagonali nello spazio $\mathbf{r}$. Il fatto che la funzione risposta sia diagonale in
spazio diretto non deve per\`{o} far pensare che stiamo utilizzando grandezze differenti.
Se scriviamo infatti la funzione risposta in seconda quantizzazione vediamo che essa \`{e} in
effetti il valor medio di quattro operatori, sia in TDDFT, sia nelle equazioni di Hedin,
sia nell'equazione di Bethe-Salpeter. La sola differenza \`{e} che nello sviluppare la
teoria nell'ambito del formalismo many-body dobbiamo considerare anche i termini ``non diagonali''
mentre per la teoria del funzionale densit\`{a} questo non \`{e} necessario.
Ancora una volta vedremo come nella formulazione di Casida, quando proietteremo le grandezze nello
spazio degli stati anche l'equazione per la TDDFT diventer\`{a} una grandezza a quattro punti.

%*******************************************************************************************
\section{Approssimazioni}
\subsection{L'approssimazione adiabatica LDA}
Come per il computo dello stato fondamentale del sistema, anche per il computo delle eccitazioni
la teoria esatta non pu\`{o} essere risolta e dobbiamo utilizzare delle approssimazioni.
Poich\'{e} nell'ambito di questa tesi calcoleremo spettri di eccitazioni con la TDDFT indicheremo in
questa sezione quella che \`{e} l'approssimazione utilizzata nell'ambito di tale teoria:
l'approssimazione adiabatica.

Tale approssimazione \`{e} in effetti l'equivalente dell'approssimazione locale, presentata per la
DFT, per la variabile temporale. In termini matematici consiste nell'utilizzare la seguente relazione
\begin{equation}
v_{xc}[\rho](\mathbf{x},t)=\frac{\delta A_{xc}[\rho]}{\delta \rho(\mathbf{x},t)} \simeq
\frac{\delta E_{xc}[\rho_t]}{\delta \rho_t(\mathbf{x})}
\hspace{0.5 cm} \text{,}
\end{equation}
dove nell'ultimo termine abbiamo indicato la variabile temporale a pedice della densit\`{a} per
indicare che non stiamo facendo la derivata funzionale rispetto a tale variabile ma solo rispetto
alla variabile spaziale. L'energia di scambio e correlazione $E_{xc}$ dipende cio\`{e} solo dalla
densit\`{a} al tempo $t$ e nell'approssimazione vengono dunque persi tutti gli ``effetti di memoria''.

Questo \`{e} particolarmente chiaro nella scrittura del kernel della TDDFT nell'ambito di tale
approssimazione che risulta
\begin{equation}
f_{xc}[\rho](\mathbf{x}_1t_1,\mathbf{x}_2t_2) =
\frac{\delta v_{xc}[\rho](\mathbf{x}_1,t_1)}{\delta \rho(\mathbf{x}_2,t_2)}
\simeq \delta(t_1-t_2)\frac{\delta v_{xc}^A[\rho_{t_1}](\mathbf{x}_1)}{\delta \rho_{t_1}(\mathbf{x}_2)}
\hspace{0.5 cm} \text{.}
\end{equation}
Nello scrivere il kernel abbiamo qui trascurato il termine di Hartree, poich\'{e} infatti il
campo di Hartree \`{e} statico l'approssimazione adiabatica non avr\`{a} alcun effetto su
tale campo e quindi sul suo kernel che \`{e} d'altra parte conosciuto in modo esatto
\begin{equation}
f_H[\rho](\mathbf{x}_1,\mathbf{x}_2)=\frac{1}{|\mathbf{x}_1-\mathbf{x}_2|}
\hspace{0.5 cm} \text{.}
\end{equation}

Se mettiamo insieme l'approssimazione adiabatica con l'approssimazione LDA utilizzata per
l'energia di scambio e correlazione otteniamo la seguente espressione per il kernel
\begin{equation} \label{K_alda}
f^{ALDA}_{xc}(1,2)= \delta(1,2)
\left(2\frac{d \epsilon^{hom}_{xc}(\rho)}
{d \rho}
+\rho\frac{d^2 \epsilon^{hom}_{xc}(\rho)}
{d^2 \rho}\right)
\hspace{0.5 cm} \text{.}
\end{equation}

Possiamo infine inserire una parametrizzazione per l'energia di stato omogeneo in funzione della
densit\`{a} scrivendo cos\`{i} in modo esplicito il kernel in funzione della densit\`{a}. A
questo scopo scriviamo quella che \`{e} l'approssimazione locale dell'energia di scambio, trascurando
l'energia di correlazione. Riprendendo lo sviluppo fatto nel capitolo 4 e
in particolare la (\ref{xLDA1}) otteniamo
\begin{eqnarray}
&&\epsilon_x^{hom}(\rho_t(\mathbf{x}))=-d_0 \rho_t^{1/3}(\mathbf{x}) \nonumber \\
&&f^{ALDA}_{x}(1,2)=-\delta(1,2)
\left(\frac{4d_0}{9} \rho_{t_1}^{-2/3}(\mathbf{x}_1)\right)  
\hspace{0.5 cm} \text{,}
\end{eqnarray}
da cui possiamo osservare l'andamento divergente del kernel nelle regioni in cui la densit\'{a}
tende a zero. Inserendo anche il termine di correlazione (\ref{cLDA}) \`{e} possibile vedere
che anche questo diverge in zone in cui la densit\`{a} tende a zero.

\subsection{L'approssimazione adiabatica LSDA}
Per il nostro lavoro dobbiamo cosiderare un kernel che dipenda dallo spin e dunque
anzich\`{e} utilizzare, nell'ambito dell'approssimazione adiabatica, il funzionale LDA utilizzeremo
il funzionale LSDA. Avremo dunque un kernel definito come
\begin{equation}
f_{Hxc}(1,2)=
\frac{\delta v_{xc}(1)}{\delta \rho(2)}
\hspace{0.5 cm} \text{,}
\end{equation}
dove le variabili di spin sono come sempre incluse nella notazione $1=(\mathbf{x}_1,t_1,\sigma_1)$.
In approssimazione A-LSDA otteniamo
\begin{multline} \label{K_alsda}
f^{ALDA}_{xc}(1,2)= \delta(t_1-t_2)\delta(\mathbf{x_1}-\mathbf{x}_2)
\bigg(\frac{d \epsilon^{LSDA}_{xc}(\rho,\zeta)}
{d \rho_{\sigma_1}}+ \\
\frac{d \epsilon^{LSDA}_{xc}(\rho,\zeta)}{d \rho_{\sigma_2}}
+\rho\frac{d^2 \epsilon^{LSDA}_{xc}(\rho,\zeta)}
{d\rho_{\sigma_1}d\rho_{\sigma_2}} \bigg)
\hspace{0.5 cm} \text{,}
\end{multline}
dove la derivazione rispetto alla componente $\sigma$ della densit\`{a} \`{e} intesa
come (utilizzando la regola di derivazione per le funzioni composte)
\begin{equation}
\frac{d \epsilon^{LSDA}_{xc}(\rho,\zeta)}{d \rho_{\sigma}}=
\frac{\partial \epsilon^{LSDA}_{xc}(\rho,\zeta)}{\partial \rho}\pm
\frac{\delta \epsilon^{LSDA}_{xc}(\rho,\zeta)}{\partial \zeta}
\hspace{0.5 cm} \text{;}
\end{equation}
$\pm$ dipende dalla componente della densit\`{a} che stiamo utilizzando: 
$+$ per $\rho_{\uparrow}$; $-$ per $\rho_{\downarrow}$.

%% file: casida.tex
\section{Introduzione}
L'idea di Casida \`{e} essenzialmente quella di proiettare la formulazione
della TDDFT nello spazio degli stati o spazio delle configurazioni. In altre parole
invece di rappresentare le grandezze della teoria in spazio diretto
($\mathbf{x},\sigma$) o in spazio reciproco ($\mathbf{k},\sigma$) sceglie
di rappresentarle rispetto ad una base di funzioni
d'onda di particella singola. In particolare la base scelta \`{e} quella degli
autostati dell'Hamiltoniana di KS della DFT, in modo che, come vedremo, la funzione di
risposta libera sia diagonale in tale spazio. La scelta fatta \`{e} soprattutto
mirata allo studio di sistemi isolati, per i quali lo spazio delle configurazioni
\`{e} sicuramente la descrizione migliore poich\'{e} la mancanza di periodicit\`{a}
del sistema elimina i vantaggi dell'utilizzo dello spazio reciproco.

In questa nuova rappresentazione l'idea di Casida \`{e} in particolare quella di
studiare come grandezza di riferimento non la funzione di risposta ma la polarizzazione
del sistema, definita come:
\begin{equation}
\alpha = \left[\frac{\delta \mathbf{\mu}(\mathbf{E},\mathbf{B})}
              {\delta \mathbf{E}}\right]_
              {\mathbf{E}=\mathbf{B}=0}
\hspace{0.5 cm} \text{,}
\end{equation}
dove $\mathbf{\mu}$ \`{e} il momento di dipolo elettrico del sistema, mentre $\mathbf{E}$ e
$\mathbf{B}$ sono il campo elettrico e il campo magnetico che vengono applicati al sistema.
\begin{comment}
La scelta viene fatta anche perch\'{e} studiando tale grandezza possiamo ricavare oltre che
lo spettro delle eccitazioni anche la forze di oscillatore associate ad ogni eccitazione. 
\end{comment}

%*********************************************************************************************
%*********************************************************************************************
%*********************************************************************************************
\section{La polarizzabilit\`{a} del sistema}
Vediamo innanzi tutto come la polarizzabilit\`{a} \`{e} legata alla funzione di risposta e
quindi alle grandezze fisiche del sistema.
A questo scopo consideriamo ora il caso in cui il potenziale applicato sia del tipo
\begin{equation}
\delta V^{ext}(t) = z E_{z}(t)
\hspace{0.5 cm} \text{.}
\end{equation}
Consideriamo quindi il momento di dipolo in direzione $x$, esso sar\`{a} definito come 
\begin{equation}
\delta \mu_x = - q \delta x
\hspace{0.5 cm} \text{,}
\end{equation}
Qui $q$ \`{e} la carica elettrica che d'ora in avanti supporremo pari ad uno,
mentre $\delta x$ sar\`{a} la variazione del valor medio dell'operatore
$\hat{x}=\int d^3\mathbf{x}\ \hat{\rho}(\mathbf{x})x$ (il termine dopo l'uguale \`{e}
la sua rappresentazione in seconda quantizzazione). Ovvero
\begin{equation}
\delta x= \int d^3\mathbf{x}\ \delta \rho(\mathbf{x},t) x
\hspace{0.5 cm} \text{,}
\end{equation}
da cui la componente $xz$ polarizzabilit\`{a} nello spazio delle frequenze risulta:
\begin{equation} \label{polarizzabilita2}
\alpha_{xz}(\omega)= -\int d^3\mathbf{x}\
\frac{\delta \rho(\mathbf{x})x}{E_z(\omega)}
\hspace{0.5 cm} \text{.}
\end{equation}

La relazione con lo spettro di eccitazione del sistema pu\`{o} infine essere scritta 
utilizzando la rappresentazione di Lehmann di quest'ultima insieme con la relazione
\begin{equation}
\langle \Psi_0 | \hat{x} | \Psi_I \rangle \langle \Psi_I | \hat{z} | \Psi_0 \rangle =
\langle \Psi_0 | \hat{z} | \Psi_I \rangle \langle \Psi_I | \hat{x} | \Psi_0 \rangle
\hspace{0.5 cm} \text{,}
\end{equation}
ottenendo:
\begin{equation} \label{polarizzabilita}
\alpha_{xz}=\sum_I \frac{2(E_I-E_0)\langle \Psi_0 | \hat{x} | \Psi_I \rangle
                         \langle \Psi_I | \hat{z} | \Psi_0 \rangle}
                        {(E_I-E_0)^2 - \omega^2}
\hspace{0.5 cm} \text{.}
\end{equation}

Nelle sezioni seguenti ci occuperemo dunque di riscrivere la TDDFT nella nuova base e 
quindi di calcolare l'espressione della polarizzabilit\`{a} con tale formalismo.
Confrontando infine il nostro risultato con l'espressione appena scritta potremo
ricavare lo spettro di eccitazione e le forze di oscillatore.

%****************************************************************************************
%****************************************************************************************
%****************************************************************************************
\section{La TDDFT nello spazio delle configurazioni}
\subsection{Cambio di base}
Ci\`{o} che dobbiamo fare \`{e} semplicemente scrivere le relazioni tra matrici e vettori
dello spazio di Hilbert esplicitandone gli elementi in funzione della base scelta in modo
da avere alla fine delle relazioni tra elementi di matrice, ovvero dei numeri complessi.
I dettagli dei passaggi eseguiti sono piuttosto noiosi, riportiamo qui semplicemente la regola
necessaria per ottenere questo risultato. Ci\`{o} che in pratica si fa \`{e} passare dalle
formule scritte in spazio $(\mathbf{x},\sigma)$ alle formule scritte nello spazio delle configurazioni
utilizzando le formule per il cambio di base in seconda quantizzazione.
La regola da utilizzare \`{e} dunque la seguente:
\begin{eqnarray} \label{cambio-2}
&&\hat{a}_i = \sum_j \hat{a}_j \langle i| j \rangle   \nonumber  \\
&&\hat{a}^{\dag}_i = \sum_j \hat{a}^{\dag}_j \langle j | i \rangle
\hspace{0.5 cm} \text{,}
\end{eqnarray}
dove abbiamo scritto un operatore di campo di base $i$ in una nuova base $j$; nel caso in cui
la base "di arrivo" sia continua le sommatorie andranno sostituite con degli integrali, ma la
regola del cambio di base risulter\`{a} la stessa.
Se preferite, poich\'{e} stimiamo lavorando con i valori medi e non con gli operatori direttamente
\`{e} possibile utilizzare anche la regola per proiettare una grandezza sugli stati, (facendo
per\`{o} attenzione al fatto che stiamo utilizzando grandezze che sono diagonali in spazio
$\mathbf{x}$):
\begin{multline} \label{cambio-1}
f_{j_1,\ldots,j_n} = \int d^3\mathbf{x}_1 \ldots d^3\mathbf{x}_n\
       \psi_{j_1}(\mathbf{x}_1) \psi^*_{j_2}(\mathbf{x}_2)\ldots  \\
f(\mathbf{x}_1,\ldots,\mathbf{x}_n)
                   \ldots \psi^*_{j_{n-1}}(\mathbf{x}_{n-1}) \psi_{j_n}(\mathbf{x}_n)
\hspace{0.5 cm} \text{.}
\end{multline}
Effettuando il cambio di base per l'operatore densit\`{a} otteniamo:
\begin{equation}
\hat{n}_{\sigma}(\mathbf{x},t) = \sum_{i,j} \psi^*_{i\sigma}(\mathbf{x}) \psi_{j\sigma}(\mathbf{x})
                              \hat{a}^{\dag}_{i\sigma}(t) \hat{a}_{j\sigma}(t)   
\hspace{0.5 cm} \text{.}
\end{equation}
Definiamo ora la grandezza
\begin{equation}
P_{ij\sigma}(t)= \langle \Psi_{KS} | \hat{a}^{\dag}_{i\sigma}(t)
                   \hat{a}_{j\sigma}(t) | \Psi_{KS} \rangle
\hspace{0.5 cm} \text{,}
\end{equation}
che corrisponde ad una densit\`{a} nello spazio delle configurazioni e le equazioni
(\ref{risp-KS}) e (\ref{chi-tddft}) nel nuovo spazio diventano
\begin{eqnarray}
&&\delta P_{ij\sigma}(t_1) = \int dt'\ \chi^{KS}_{ij\sigma,hk\tau}(t_1-t_2) (v_s)_{hk\tau}(t_2)   \\
&&\delta P_{ij\sigma}(t_1) = \int dt'\ \chi_{ij\sigma,hk\tau}(t_1-t_2) V^{ext}_{hk\tau}(t_2)
\hspace{0.5 cm} \text{,}
\end{eqnarray}
dove stiamo sommando sugli indici ripetuti. Scriviamo ora anche il Kernel $K$
nello spazio delle configurazioni
\begin{equation}
K_{ij\sigma,hk\tau}^{Hxc}(t_1-t_2) = \frac{\delta v^H_{ij\sigma}(t_1)}{\delta P_{hk\tau}(t_2)}+
                               \frac{\delta v^{xc}_{ij\sigma}(t_1)}{\delta P_{hk\tau}(t_2)}
\end{equation}
e abbiamo tutti gli elementi per riscrivere l'equazione (\ref{chi-tddft}) nella nuova base.
Prima di scrivere la nuova equazione vogliamo per\`{o} anche cambiare la base
della variabile temporale e passare nello spazio delle frequenze poich\'{e} in questo modo le
relazioni di convoluzione che compaiono diventeranno dei semplici prodotti nel nuovo spazio.
Applichiamo dunque la trasformata di Fourier a tutte le grandezze che compaiono nelle formule
e siamo finalmente pronti a scrivere il nostro risultato
\begin{equation}
\left[\left(\chi^{KS}_{ij\sigma,hk\tau}(\omega)\right)^{-1} -
            K_{ij\sigma,hk\tau}(\omega)  \right]
\delta P_{hk\tau}(\omega) = \delta V^{ext}_{ij\sigma}(\omega)
\hspace{0.5 cm} \text{,}
\end{equation}
dove come sempre sommiamo sugli indici ripetuti. Abbiamo qui direttamente invertito la 
funzione risposta in modo da scrivere l'equazione in modo pi\`{u} compatto, in particolare
nel farlo abbiamo invertito la funzione $\chi^{KS}$ che potrebbe per\`{o} essere non invertibile.
In realt\`{a} sappiamo che tale matrice (lo vedremo nel prossimo paragrafo) \`{e} diagonale 
nella rappresentazione scelta e dunque certamente invertibile. L'unico pericolo risulterebbe
costituito dai termini diagonali nulli; questi per\`{o} possono essere esclusi dalla teoria
poich\'{e} quando $\chi^{KS}$ \`{e} pari a zero il sistema presenta una variazione di
densit\`{a} nulla (cfr. eq (\ref{risp-KS})) e tutte le equazioni scritte non portano alcuna
informazione.

%*********************************************************************************************
\subsection{La funzione risposta libera e il kernel}
Prima di procedere studiamo pi\`{u} in dettaglio la funzione di risposta
di un sistema di particelle non interagenti. Questa \`{e} infatti conosciuta in modo esatto
e pu\`{o} essere facilmente ricavata nello spazio degli stati a partire dalla definizione
di funzione di risposta.
Partiamo dalla rappresentazione di Lehmann della funzione di risposta (\ref{Lehmann-polarizzazione})
scritta per\`{o} nella nuova base scelta:
\begin{multline}
\chi^{KS}_{ij\sigma,hk\tau}(\omega) = \sum_I
\frac{\langle \Psi^{KS}_0 |\hat{a}^{\dag}_{j\sigma}\hat{a}_{i\sigma} | \Psi^{KS}_I \rangle
      \langle \Psi^{KS}_I |\hat{a}^{\dag}_{h\tau}\hat{a}_{k\tau} | \Psi^{KS}_0 \rangle}
     {\omega-(E_I-E_0)+i\eta} - \\
\frac{\langle \Psi^{KS}_0 |\hat{a}^{\dag}_{h\tau}\hat{a}_{k\tau} | \Psi^{KS}_I \rangle
      \langle \Psi^{KS}_I |\hat{a}^{\dag}_{j\sigma}\hat{a}_{i\sigma} | \Psi^{KS}_0 \rangle}
     {\omega+(E_I-E_0)+i\eta}
\hspace{0.5 cm} \text{.}
\end{multline} 
Il fatto che stiamo considerando la funzione di risposta libera (ovvero quella di un sistema
non interagente) significa ora che le funzioni d'onda a molti corpi sono semplicemente il
prodotto di funzioni d'onda a particella singola e che quindi gli stati eccitati non sono
altro che eccitazioni di singola particella. Questo si pu\`{o} vedere chiaramente nella formula
poich\`{e}, essendo gli stati di particella singola ortogonali tra loro, le funzioni d'onda 
$\Psi_I$ non possono differire da $\Psi_0$ che per una particella passata dallo stato $i$ allo
stato $j$.

Come conseguenza di tutto ci\`{o} la funzione di risposta libera risulta diagonale in questa
rappresentazione. Inoltre poich\'{e} gli stati eccitati differiscono dallo stato fondamentale
per un solo autovalore di KS (lo stato $i$ al posto di quello $j$) risulter\`{a}
$E_I-E_0 = \epsilon^{KS}_j - \epsilon^{KS}_i$. Infine valutando i numeratori dell'espressione,
se ammettiamo che gli stati possano avere numeri di occupazione frazionaria, otterremo
dall'applicazione dell'operatore di distruzione a numeratore esattamente i numeri di occupazione.
Possiamo quindi scrivere la funzione di risposta libera (la somma sugli stati eccitati
scompare perch\'{e} sopravvive solo quello stato eccitato con eccitazione 
$i\rightarrow j$ e per lo stesso motivo abbiamo solo uno dei due termini della somma):
\begin{equation}
\chi^{KS}_{ij\sigma,hk\tau}(\omega) =
\delta_{\sigma,\tau}\delta_{j,k}\delta_{i,h}\frac{f_{j\sigma}-f_{i\sigma}}
     {\omega-(\epsilon^{KS}_i-\epsilon^{KS}_j)}
\hspace{0.5 cm} \text{.}
\end{equation}

Per esprimere il kernel in funzione di grandezze note non possiamo invece far altro che scriverlo
in funzione del kernel in spazio $(\mathbf{x},\sigma)$ poich\'{e} solo in questa base siamo in grado di
calcolarlo.
L'espressione \`{e} facilmente scrivibile utilizzando le regole di cambio base della prima 
quantizzazione (\ref{cambio-1}) e ricordando che il nostro kernel \`{e} diagonale nello spazio
e nello spin
\begin{equation} \label{cambio-kernel}
K_{ij\sigma,hk\tau}(\omega)=\int d^3\mathbf{x}_1 d^3\mathbf{x}_2\
          \psi^*_{i\sigma}(\mathbf{x}_1)\psi_{j\sigma}(\mathbf{x}_1)
          K_{\sigma,\tau}[\rho](\mathbf{x}_1,\mathbf{x}_2;\omega)
          \psi_{h\tau}(\mathbf{x}_2)\psi_{k\tau}^*(\mathbf{x}_2)
\hspace{0.5 cm} \text{.}
\end{equation}

%*********************************************************************************************
\subsection{L'equazione di Casida}
Abbiamo finalmente tutti gli elementi per scrivere l'equazione che \`{e} possiamo dire alla
base del lavoro di Casida e che non \`{e} altro che la formulazione dell'equazione della TDDFT
nello spazio delle configurazioni:
\begin{equation} \label{Casida}
\sum_{kl\tau}^{f_{k\tau}-f_{h\tau}\neq 0}
\left[ \delta_{\sigma,\tau}\delta_{j,h}\delta_{i,k}
\frac{\omega-(\epsilon_{h\tau}-\epsilon_{k\tau})}{f_{k\tau}-f_{h\tau}} -
K_{ij\sigma,hk\tau}(\omega)  \right] \delta P_{hk\tau}(\omega) =
\delta V^{ext}_{ij\sigma}(\omega)
\hspace{0.5 cm} \text{.}
\end{equation}

\begin{comment}
Partendo da questa equazione potremmo ora costruire, raccogliendo a destra e a sinistra delle
parentesi quadre la radice della differenza dei numeri di occupazione, la seguente matrice
\begin{eqnarray} \label{Omega_tilde}
\tilde{\Omega}_{ij\sigma,hk\tau}(\omega) = \delta_{i,h} \delta_{j,k} \delta_{\sigma,\tau}
      (\epsilon_{k\sigma}-\epsilon_{h\sigma}) \\ \nonumber
    +\sqrt{(f_{i\sigma}-f_{j\sigma})}
      K_{ij\sigma,hk\tau}(\omega) 
      \sqrt{(f_{h\tau}-f_{k\tau})}
\end{eqnarray}
e scrivere l'espressione come
\begin{equation}
\sum_{kl\tau}^{f_{k\tau}-f_{h\tau}\neq 0}\frac{1}{\sqrt{(f_{i\sigma}-f_{j\sigma})}}
\left[ \delta_{\sigma,\tau}\delta_{j,h}\delta_{i,k}
\omega-\tilde{\Omega}_{ij\sigma,hk\tau} \right]\frac{1}{\sqrt{(f_{h\tau}-f_{k\tau})}}
 \delta P_{hk\tau}(\omega) =
\delta V^{ext}_{ij\sigma}(\omega)
\hspace{0.5 cm} \text{,}
\end{equation}
da cui vediamo che per trovare lo spettro di eccitazione del sistema non dobbiamo far altro che
diagonalizzare la matrice scritta. Poich\'{e} per\`{o} la matrice appena scritta non \`{e} hermetiana,
l'idea di Casida \`{e} quella di riordinare la base e di considerare solo termini del tipo particella-buca.
In questo modo, facendo alcune trasformazioni, riesce ad ottenere un'equazione simile che lega per\`{o} la
parte reale della densit\`{a} alla parte reale del potenziale esterno. Il vantaggio di questa trasformazione
sar\`{a} quello di ottenere una nuova matrice che sar\`{a} hermetiana e di dimensione dimezzata rispetto a
quella scritta in precedenza.
\end{comment}

Poich\'{e} siamo passati ad una base in cui n\'{e} la densit\`{a} n\'{e} il potenziale esterno sono
diagonali avremo a che fare sia con la parte reale che quella immaginaria di tali grandezze, poich\'{e}
come sappiamo le grandezze fisiche vengono rappresentate da matrici hermitiane i cui termini non diagonali
possono dunque essere complessi.

Per mettere in luce la parte immaginaria e quella reale 
l'idea \`{e} quella di riordinare la base del nostro spazio nel seguente modo: 
$i < j \rightarrow f_{i\sigma} \geq f_{j\sigma}$. Poi di dividere l'equazione in due, la prima
per la quale consideriamo i termini in cui $f_{i\sigma} > f_{j\sigma}$ e la seconda per
cui $f_{i\sigma} < f_{j\sigma}$. In ciascuna delle due equazioni inoltre separiamo in due parti
la sommatoria, la prima con $f_{h\tau} > f_{k\tau}$ e la seconda con $f_{h\tau} < f_{k\tau}$.
Ora invertiamo a coppie il nome di tutti quegli indici per cui quello con numero di occupazione
minore viene prima in modo da ottenere ovunque $f_{i\sigma} > f_{j\sigma}$
e $f_{h\tau} > f_{k\tau}$:

\begin{multline}
\sum_{kl\tau}^{f_{h\tau}-f_{k\tau}> 0}
\left[ \delta_{\sigma,\tau}\delta_{i,h}\delta_{j,k}
\frac{\omega-(\epsilon_{h\tau}-\epsilon_{k\tau})}{f_{k\tau}-f_{h\tau}} -
K_{ij\sigma,hk\tau}(\omega)  \right] \delta P_{hk\tau}(\omega)-    \\
\sum_{kl\tau}^{f_{h\tau}-f{k\tau}> 0}
K_{ij\sigma,kh\tau}(\omega) \delta P_{kh\tau}(\omega) =
\delta V^{ext}_{ij\sigma}(\omega)
\end{multline}
\begin{multline}
\sum_{kl\tau}^{f_{h\tau}-f_{k\tau}> 0}
\left[ \delta_{\sigma,\tau}\delta_{i,h}\delta_{j,k}
\frac{\omega-(\epsilon_{k\tau}-\epsilon_{h\tau})}{f_{h\tau}-f_{k\tau}} -
K_{ji\sigma,kh\tau}(\omega)  \right] \delta P_{kh\tau}(\omega)-    \\
\sum_{kl\tau}^{f_{h\tau}-f{k\tau}> 0}
K_{ji\sigma,hk\tau}(\omega) \delta P_{hk\tau}(\omega) =
\delta V^{ext}_{ji\sigma}(\omega)
\hspace{0.5 cm} \text{,}
\end{multline}
Possiamo riscrivere le due equazioni in forma matriciale sfruttando il fatto che il
kernel $K_{Hxc}(\mathbf{x_1}t_1,\mathbf{x_2}t_2)$ \`{e} una matrice hermetiana.
Otteniamo cos\`{i}:
\begin{equation} \label{excitations-tddft}
\left[
\left( \begin{array}{cc}
A(\omega)   & B(\omega) \\
B^*(\omega) & A^*(\omega) \end{array} \right)
- \omega \left( \begin{array}{cc}
C & 0 \\
0 & -C \end{array} \right)
\right]
\left( \begin{array}{c}
\delta P(\omega) \\
\delta P^*(\omega) \end{array} \right)
= \left( \begin{array}{c}
\delta V_{ext}(\omega) \\
\delta V^*_{ext}(\omega) \end{array} \right)
\hspace{0.5 cm} \text{,}
\end{equation}
dove abbiamo definito le grandezze:
\begin{eqnarray}
&&A_{ij\sigma,hk\tau}(\omega)= \delta_{\sigma,\tau}
\delta_{i,h} \delta_{j,k} \frac{\epsilon_{h\tau}-\epsilon_{k\tau}}
{f_{h\tau}-f_{k\tau}}- K_{ij\sigma,hk\tau}(\omega)              \nonumber  \\
&&B_{ij\sigma,hk\tau}(\omega)= -K_{ij\sigma,kh\tau}(\omega)
                                              \nonumber  \\
&&C_{ij\sigma,hk\tau}=\frac{\delta_{\sigma,\tau} \delta_{i,h}
\delta_{j,k}}{f_{h\tau}-f_{k\tau}}
\hspace{0.5 cm} \text{.}
\end{eqnarray}

Se ora facciamo l'ulteriore assunzione che gli orbitali di KS siano reali e che
il kernel in spazio $(\mathbf{x},\sigma)$ sia una quantit\`{a} reale\footnote{Il che coincide,
in accordo a quanto afferma Casida \cite{Casida1}, ad interpretare come infinito il tempo di
vita delle eccitazioni.} otteniamo le relazioni $A=A^*$ e $B=B^*$ e possiamo quindi diagonalizzare 
la prima matrice da sinistra dell'espressione ricavando un'equazione
in cui compaiono la parte reale e quella immaginaria sia del potenziale che della densit\`{a}:
\begin{multline}
\left[
\left( \begin{array}{cc}
A(\omega)+B(\omega) & 0 \\
0 & A(\omega)-B(\omega) \end{array} \right)
- \omega \left( \begin{array}{cc}
0 & -C \\
-C & 0 \end{array} \right)
\right] \nonumber \\ 
\left( \begin{array}{c}
\Re[\delta P](\omega) \\
-i\Im[\delta P](\omega) \end{array} \right)
= \left( \begin{array}{c}
\Re[\delta V_{ext}](\omega) \\
-i\Im[\delta V_{ext}](\omega) \end{array} \right)
\hspace{0.5 cm} \text{.}
\end{multline}
Se ora esplicitiamo l'equazione per la parte reale della densit\`{a} da quella per la
parte immaginaria otteniamo
\begin{multline} \label{Casida-re}
\left[ (A(\omega)+B(\omega))-\omega^2 C(A(\omega)-B(\omega))^{-1}C\right]
\Re[\delta P](\omega)=  \\
\Re[\delta V^{ext}](\omega)-i\omega C(A(\omega)-B(\omega))^{-1}\Im[\delta V^{ext}](\omega)
\end{multline}
\begin{multline} \label{Casida-im}
\left[ (A(\omega)-B(\omega))-\omega^2 C(A(\omega)+B(\omega))^{-1}C\right]
\Im[\delta P](\omega)= \\
\Im[\delta V^{ext}](\omega)-i\omega C(A(\omega)+B(\omega))^{-1}\Re[\delta V^{ext}](\omega)
\hspace{0.5 cm} \text{.}
\end{multline}
Infine nel caso di una perturbazione reale se consideriamo solo la variazione della parte
reale della densit\`{a} possiamo riscrivere nel seguente modo la prima delle due equazioni:
\begin{equation} \label{Casida-re-matr}
\Re[\delta P](\omega) = S^{-1/2} (\omega^2 -\Omega(\omega))^{-1} S^{-1/2}
                                   \delta V_{ext}(\omega)
\hspace{0.5 cm} \text{,}
\end{equation}
dove abbiamo definito le due grandezze
\begin{eqnarray} \label{Omega_matr}
&&S(\omega) = C (A-B)^{-1} C \nonumber \\
&&\Omega(\omega) = S^{-1/2} (A+B) S^{-1/2}
\hspace{0.5 cm} \text{.}
\end{eqnarray}
A partire dall'espressione (\ref{Casida-re-matr}) la polarizzabilit\`{a} del sistema
pu\`{o} essere facilmente scritta come (scriviamo qui la componente $\alpha_{xz}$)
utilizzando la (\ref{polarizzabilita2}) proiettata nello spazio degli stati:
\begin{equation}
\alpha_{xz}(\omega)=2 \vec{x}^{\dag} S^{-1/2} (\Omega(\omega)-\omega^2)^{-1} S^{-1/2} \vec{z}
\hspace{0.5 cm} \text{,}
\end{equation}
dove tutte le grandezze vanno ovviamente pensate nello spazio degli stati.
Il simbolo di vettore sopra le grandezze $x$ e $z$ indica che queste sono vettori
in tale rappresentazione.

Confrontando questa espressione con la (\ref{polarizzabilita}) possiamo
immediatamente osservare che lo spettro di eccitazione del sistema pu\`{o} essere ottenuto
diagonalizzando la matrice $\Omega$.
L'equazione agli autovalori di cui parlavamo all'inizio di questa sezione
sar\`{a} dunque:
\begin{equation}
\Omega(\omega_I) F_I = \omega_I^2 F_I
\hspace{0.5 cm} \text{.}
\end{equation}
Tale equazione costituisce il risultato principale del lavoro di Casida ed \`{e}
di fatto l'equazione che viene utilizzata per calcolare lo spettro di eccitazioni
di sistemi finiti nell'ambito della TDDFT. Segnaliamo che il risultato sin qui
ottenuto \`{e} esatto e del tutto generale, in particolare lo stesso formalismo
potrebbe essere ad esempio applicato partendo, invece che dal sistema di Kohn e Sham
dall'Hamiltoniana di Hartree-Fock o da una qualsiasi Hamiltoniana a particella singola.

%Equazione per $\delta P$ scrivendo gli indici, pu\`{o} sempre tornare utile
%\begin{eqnarray}
%\sum_{kl\tau}^{f_{k\tau}-f{h\tau}> 0}
%\left[ \delta_{\sigma,\tau}\delta_{i,h}\delta_{j,k}
%\frac{(\epsilon_{h\tau}-\epsilon_{k\tau})}{f_{h\tau}-f_{k\tau}} -
%2K_{ij\sigma,hk\tau}(\omega)
%- \omega^2 \frac{\delta_{\sigma,\tau}
%\delta_{i,h}\delta_{j,k}}{(\epsilon_{h\tau}-\epsilon_{k\tau})
%(f_{h\tau}-f_{k\tau})}  \right] Re[\delta P_{hk\tau}(\omega)] =
%\delta V^{ext}_{ij\sigma}(\omega)
%\end{eqnarray}

Nell'ambito della TDDFT tale formulazione offre per\`{o} l'indubbio vantaggio di essere molto pi\`{u}
semplice. Se infatti consideriamo particelle a funzione d'onda reale possiamo facilmente osservare
che il kernel risulta simmetrico rispetto all'inversione delle coppie di
indici ($i,j$) o ($h,k$) e la matrice $\Omega$
pu\`{o} essere scritta come
\begin{multline} \label{Omega}
\Omega_{ij\sigma,hk\tau}(\omega) = \delta_{i,h} \delta_{j,k} \delta_{\sigma,\tau}
      (\epsilon_{k\sigma}-\epsilon_{h\sigma})^2 \\ 
    +2\sqrt{(f_{i\sigma}-f_{j\sigma})(\epsilon_{j\sigma}-\epsilon_{i\sigma})}
      K_{ij\sigma,hk\tau}(\omega) 
      \sqrt{(f_{h\tau}-f_{k\tau})(\epsilon_{k\tau}-\epsilon_{h\tau})}
\hspace{0.5 cm} \text{.}
\end{multline}
Anche quest'ultimo risultato \`{e} esatto. Nessuna ulteriore approssimazione \`{e} stata sin qui
introdotta.

Ci\`{o} che in pratica viene fatto \`{e} diagonalizzare questa equazione scegliendo
una certa approssimazione per l'energia di scambio e correlazione (che rappresenta come sempre
il termine approssimato della TDDFT) e di conseguenza per il kernel, che ricordiamo \`{e} la
derivata seconda dell'energia. L'approssimazione che solitamente viene fatta e che utilizzeremo
anche noi nelle molecole da studiare nell'ambito della tesi, \`{e} l'approssimazione adiabatica,
gi\`{a} introdotta nel capitolo precedente.
Tale approssimazione, nel formalismo introdotto da Casida, corrisponde a considerare un kernel
statico e quindi una matrice $\Omega$ statica, partendo dalla trasformata di Fourier della
(\ref{K_alda}).

%**********************************************************************************************
%**********************************************************************************************
%**********************************************************************************************
\section{Studio di sistemi a shell chiusa}

\begin{figure}[!h] 
\begin{center}
%\rule{\linewidth}{0.2 mm}
\includegraphics[width=10 cm, angle=0]{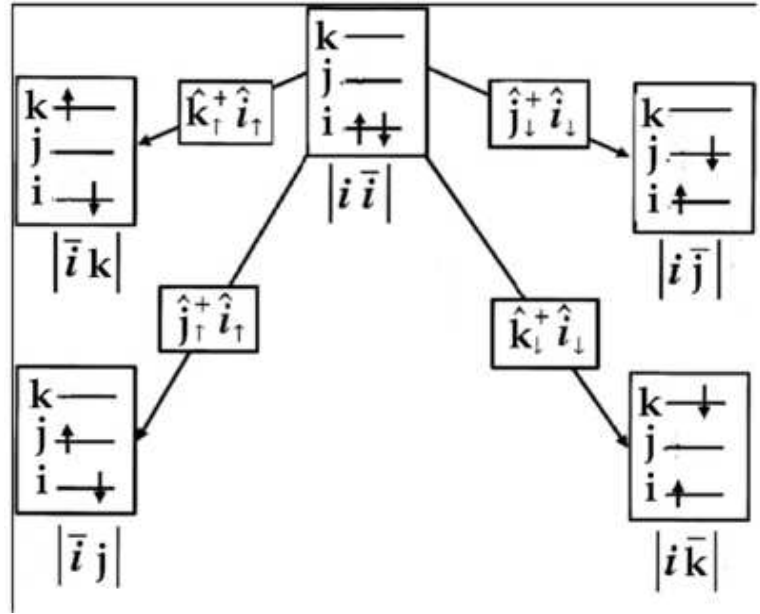}
\caption{Descrizione degli stati eccitati per un modello a $3$ livelli, 2 elettroni.}
\label{fig:modello1} 
\rule{\linewidth}{0.2 mm}
\end{center}
\end{figure}

I primi studi di Casida sulla TDDFT sono stati applicati in particolare per l'analisi di sistemi a
shell chiusa. Tali sistemi hanno la particolarit\`{a} di avere la funzione d'onda a spin up
identica alla funzione d'onda a spin down (spin lungo l'asse $z$) il che permette di trasformare
la matrice $\Omega$ in una matrice a blocchi sfruttando anche le simmetrie risultanti dal
considerare l'approssimazione adiabatica. Per illustrare tali simmetrie scegliamo di utilizzare
un sistema modello (fig. \ref{fig:modello1}):
un sistema con tre stati, che chiameremo $i$, $j$ e $k$ in cui
lo stato $i$ \`{e} occupato mentre gli stati $j$ e $k$ sono liberi.

Scriviamo quindi la matrice $\Omega$ per tale sistema:
\[
\left( \begin{array}{llll}
A_{\uparrow\uparrow}(\omega)       &  B_{\uparrow\downarrow}(\omega)         &
\alpha_{\uparrow\uparrow}(\omega)  &  \beta_{\uparrow\downarrow}(\omega)      \\
B_{\downarrow\uparrow}(-\omega)    &  A_{\downarrow\downarrow}(\omega)       &
\beta_{\downarrow\uparrow}(\omega) &  \alpha_{\downarrow\downarrow}(\omega)   \\
\alpha_{\uparrow\uparrow}(-\omega) &  \beta_{\uparrow\downarrow}(-\omega)    &
C_{\uparrow\uparrow}(-\omega)      &  D_{\uparrow\downarrow}(-\omega)         \\
\beta_{\downarrow\uparrow}(-\omega)&  \alpha_{\downarrow\downarrow}(-\omega) &
D_{\downarrow\uparrow}(-\omega)    &  C_{\downarrow\downarrow}(\omega)        \\
\end{array} \right)
\hspace{0.5 cm} \text{,}
\]
dove gli indici che variano sia sulla righe che sulle colonne sono $1=ij\uparrow$, $2=ij\downarrow$,
$3=ik\uparrow$, $4=ik\downarrow$. Queste sono infatti le uniche eccitazioni possibili del sistema.
$A$, $B$, $C$, $D$, $\alpha$ e $\beta$ indicano qui gli elementi della matrice $\Omega$
e possono essere
costruiti a partire dalla (\ref{Omega})
(e non devono essere confuse con la notazione utilizzata in precedenza per le grandezze $A$, $B$ e $C$);
la scelta di nominarle in questo modo \`{e} stata fatta
per mettere in evidenza le grandezze che risultano poi identiche nell'ambito delle approssimazioni
eseguite.
Ora l'approssimazione adiabatica per il kernel della TDDFT
fa si che gli elementi con argomento $\omega$ siano identici a
a quelli con argomento $-\omega$
(i.e.: $A(\omega)=A(-\omega)=A(0)$), mentre il fatto che stiamo considerando un sistema a shell
chiusa fa si che risulti $A_{\uparrow\uparrow}=A_{\downarrow\downarrow}$,
$C_{\uparrow\uparrow}=C_{\downarrow\downarrow}$, $\alpha_{\uparrow\uparrow}=\alpha_{\downarrow\downarrow}$
e infine $\beta_{\uparrow\downarrow}=\beta_{\downarrow\uparrow}$.%
\footnote{Le identit\`{a} $D_{\uparrow\downarrow}=D_{\downarrow\uparrow}$ e
$B_{\downarrow\uparrow}=B_{\uparrow\downarrow}$ sono sempre vere, anche nella formulazione esatta.}
In questo modo la matrice risulta presentare gli autovettori nella seguente forma:
\[
\left( \begin{array}{c}
a \\
a \\
b \\
b   
\end{array} \right)
\left( \begin{array}{c}
 c \\
 c \\
 d \\
 d   
\end{array} \right)
\left( \begin{array}{c}
 e \\
-e \\
 f \\
-f   
\end{array} \right)
\left( \begin{array}{c}
 g \\
-g \\
 h \\
-h   
\end{array} \right)
\hspace{0.5 cm} \text{.}
\]
La matrice pu\`{o} essere di conseguenza ridotta in due blocchi e di fatto possiamo pensare
di risolvere due sistemi agli autovalori con matrici di dimensione dimezzata:
\[
\left( \begin{array}{cc}
A+B          & \alpha+\beta  \\
\alpha+\beta & C+D           \\  
\end{array} \right)
\qquad
\left( \begin{array}{cc}
A-B          & \alpha-\beta  \\
\alpha-\beta & C-D           \\  
\end{array} \right)
\hspace{0.5 cm} \text{.}
\]

Cercare le eccitazioni di un sistema a shell chiusa utilizzando il formalismo di Casida consiste
nel cercare di diagonalizzare le due matrici appena scritte; troveremo dunque due gruppi di
eccitazioni. Nella prossima sezione cercheremo dunque di interpretare il significato fisico delle
due matrici; a questo scopo sar\`{a} necessario passare allo studio delle funzioni d'onda del
sistema sebbene queste dal punto di vista formale non abbiano alcun significato fisico (ricordiamo
che sono le funzioni d'onda di Kohn e Sham). 

%**********************************************************************************************
\subsection{Interpretazione delle funzioni d'onda degli stati eccitati e regole di selezione}
Questa parte come gi\`{a} detto non trova una giustificazione formale altrettanto rigorosa che
le sezioni precedenti ma, come afferma lo stesso Casida, risulta indispensabile per interpretare
la teoria stessa. Per farlo partiamo da un modello ancora pi\`{u} semplice di quello considerato
in precedenza che \`{e} costituito da un sistema con due livelli, una libera ed una occupata. In
analogia a quanto fa Casida possiamo immaginare che questo modello rappresenti la molecola
dell'Idrogeno.

Consideriamo dunque la molecola di idrogeno tenendo in considerazione solo i primi due stati
elettronici $i$ e $j$. Lo stato $i$ corrisponde all'orbitale sigma legante e lo stato $j$
all'orbitale sigma anti-legante.
Lo stato fondamentale sar\`{a} il singolo determinante di Slater:
\[
\Phi_0^1=|\psi_{i\uparrow} \psi_{i\downarrow}|
\hspace{0.5 cm} \text{;}
\]
gli stati eccitati saranno invece
\[
\begin{split}
\Phi_0^1    &=\ \frac{1}{\sqrt{2}}|\psi_{j\uparrow}\psi_{i\downarrow}|+|\psi_{i\uparrow}\psi_{j\downarrow}|  \\
\Phi_1^3    &=\ |\psi_{i\uparrow} \psi_{j\uparrow}|                                                          \\
\Phi_0^3    &=\ \frac{1}{\sqrt{2}}|\psi_{j\uparrow}\psi_{i\downarrow}|-|\psi_{i\uparrow}\psi_{j\downarrow}|  \\  
\Phi_{-1}^3 &=\ |\psi_{i\downarrow} \psi_{j\downarrow}| \hspace{0.5 cm} \text{.}
\end{split}
\]
Questi stati possono essere costruiti cercando funzioni che siano autostati degli operatori
$\hat{h}_{KS}$, $\hat{S}^2$ e $\hat{S}_z$.

Se ora guardiamo al potenziale esterno applicato al nostro sistema vediamo che esso \`{e} scritto nella forma
\begin{equation}
\begin{split}
\hat{V}^{ext}(t)=&\sum_{\sigma}\int d^3\mathbf{x}\ \hat{\psi}_{\sigma}^{\dag}(\mathbf{x})
                   \hat{\psi}_{\sigma}(\mathbf{x}) V_{\sigma}^{ext}(\mathbf{x},t) \\
                =&\sum_{\sigma}\int d^3\mathbf{x}\ \hat{\rho}_{\sigma}(\mathbf{x})
                   V_{\sigma}^{ext}(\mathbf{x},t)
\hspace{0.5 cm} \text{.}
\end{split}
\end{equation}
Possiamo per\`{o} riscriverlo come
\begin{equation} \label{trasf-spin-real}
\begin{split}
\hat{V}^{ext}(t)= \int d^3\mathbf{x}\ \bigg{[}
              \frac{1}{2}[V_{\uparrow}^{ext}(\mathbf{x},t)+V_{\downarrow}^{ext}(\mathbf{x},t)]
                    [\hat{\rho}_{\uparrow}(\mathbf{x})+\hat{\rho}_{\downarrow}(\mathbf{x})]+  \\
              \frac{1}{2}[V_{\uparrow}^{ext}(\mathbf{x},t)-V_{\downarrow}^{ext}(\mathbf{x},t)]
                    [\hat{\rho}_{\uparrow}(\mathbf{x})-\hat{\rho}_{\downarrow}(\mathbf{x})] \bigg{]}
\end{split}
\end{equation}
e quindi otteniamo:
\begin{equation}
\hat{V}^{ext}(t)=\int d^3\mathbf{x}\ \big[
V(\mathbf{x},t)\hat{\rho}(\mathbf{x},t)+B_z(\mathbf{x},t)\hat{m}_z(\mathbf{x},t) \big]
\hspace{0.5 cm} \text{.}
\end{equation}

Da questo possiamo ricavare le regole di selezione del sistema che corrispondono a
\begin{equation}
\delta S = 0,1  \hspace{10 mm} \delta S_z = 0
\end{equation}
e vediamo quindi che le uniche eccitazioni possibili per il sistema considerato sono quelle verso gli stati
\begin{equation}
\Phi_0^1 \hspace{10 mm} \Phi_0^3  
\hspace{0.5 cm} \text{,}
\end{equation}
dunque un'eccitazione di singoletto ed un'eccitazione di tripletto.

%************************************************************************************************************
\subsection{Interpretazione delle matrici di Casida}
Se analizziamo gli stati verso i quali le eccitazioni sono rese possibili dalle regole di selezione
appena trovate notiamo come essi sono formati dalla somma e dalla
differenza delle singole eccitazioni tra stati di KS.
In altre parole la trasformazione che ci farebbe passare dalla base degli stati di KS alla base degli
stati eccitati \`{e} caratterizzata dai vettori
\[
\left( \begin{array}{c}
1 \\
1
\end{array}  \right)
\left( \begin{array}{c}
 1 \\
-1
\end{array}  \right)
\hspace{0.5 cm} \text{.}
\]

Se confrontiamo questi con gli autovettori che dividono in due blocchi la matrice di Casida capiamo come
i due gruppi di eccitazioni che otteniamo sono
costituiti dal gruppo delle eccitazioni di tripletto e dal gruppo delle eccitazioni di singoletto.
A questo punto pu\`{o} essere utile scrivere in modo esplicito come sono fatti gli
elementi delle due matrici rispetto agli indici di spin. Utilizzando nuovamente le simmetrie dei
sistemi a shell chiusa otteniamo
\[
\left( \begin{array}{cc}
A_{\uparrow\uparrow}+B_{\uparrow\downarrow}          & \alpha_{\uparrow\uparrow}+\beta_{\uparrow\downarrow} \\
\alpha_{\uparrow\uparrow}+\beta_{\uparrow\downarrow} & C_{\uparrow\uparrow}+D_{\uparrow\downarrow}          \\  
\end{array} \right)
\qquad
\left( \begin{array}{cc}
A_{\uparrow\uparrow}-B_{\uparrow\downarrow}          & \alpha_{\uparrow\uparrow}-\beta_{\uparrow\downarrow} \\
\alpha_{\uparrow\uparrow}-\beta_{\uparrow\downarrow} & C_{\uparrow\uparrow}-D_{\uparrow\downarrow}          \\  
\end{array} \right)
\hspace{0.5 cm} \text{,}
\]
da cui abbiamo evidenziato come lo spin venga trattato nella divisione in due blocchi della matrice.
In altre parole la matrice viene divisa in due blocchi:
\[
\Omega_S=\Omega_{\uparrow\uparrow}+\Omega_{\uparrow\downarrow}
\hspace{8 mm}
\Omega_T=\Omega_{\uparrow\uparrow}-\Omega_{\uparrow\downarrow}
\hspace{0.5 cm} \text{,}
\]
un blocco per le eccitazioni di singoletto ed un blocco per le eccitazioni di tripletto.
%**********************************************************************************************
%**********************************************************************************************
%**********************************************************************************************
\section{Studio di sistemi a shell aperta}

\begin{figure}[t] 
\begin{center}
%\rule{\linewidth}{0.2 mm}
\includegraphics[width=10 cm, angle=0]{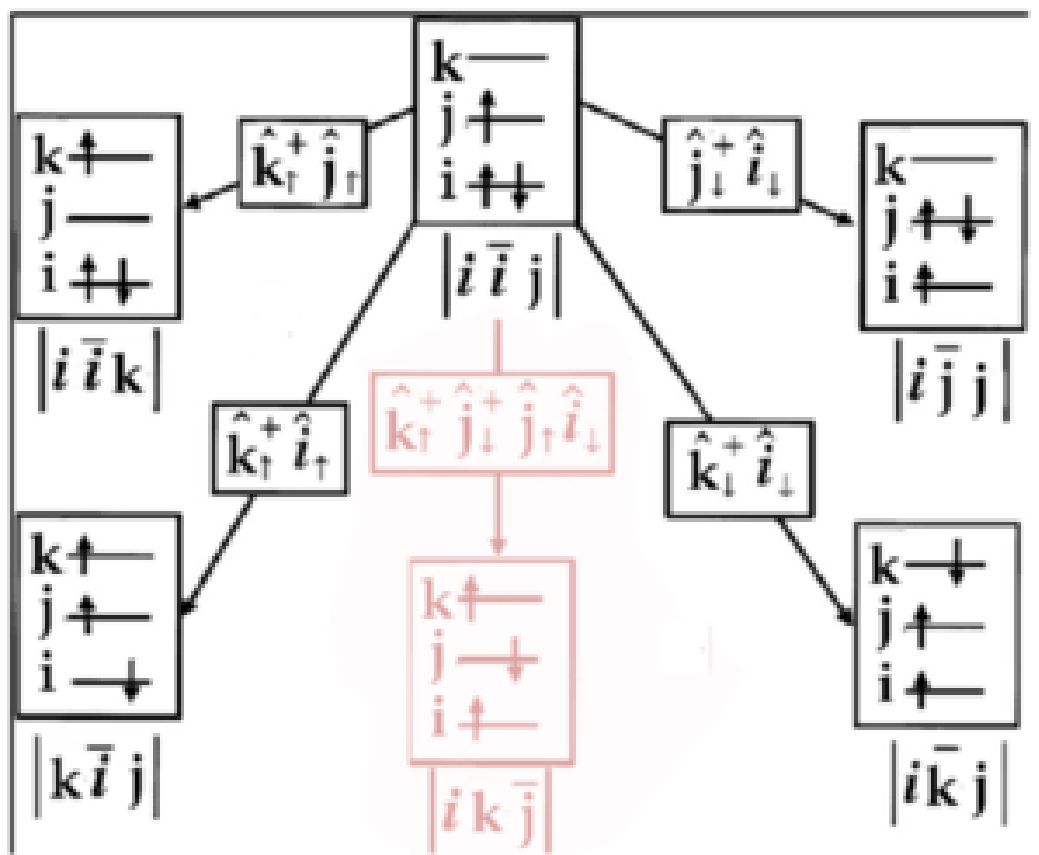}
\caption{Descrizione degli stati eccitati per un modello a $3$ livelli, $3$ elettroni.}
\label{fig:modello2} 
\rule{\linewidth}{0.2 mm}
\end{center}
\end{figure}

La teoria sin qui scritta \`{e} di fatto valida ed esatta, (se considerata nella sua forma esatta)
anche per sistemi a shell-aperta. La matrice $\Omega$ da diagonalizzare sar\`{a} sempre esprimibile
nella forma (\ref{Omega}) dato che di fatto fino a quel punto nessuna ipotesi \`{e} stata fatta da questo
punto di vista. Poich\'{e} d'altra parte abbiamo visto come lo studio delle simmetrie rispetto allo
spin possano semplificare lo studio della matrice per lo studio di sistemi con stato fondamentale
non polarizzato cerchiamo di capire anche nel caso di sistemi con stato fondamentale spin polarizzato
(polarizzazione collineare) se la matrice di Casida pu\`{o} essere separata a blocchi.
Come in precedenza lavoriamo in approssimazione adiabatica per il kernel della TDDFT.

Partiamo nuovamente da un sistema modello con tre bande (fig. \ref{fig:modello2}) 
$i$, $j$ e $k$ di cui il primo
occupato da due elettroni, il secondo solo da un elettrone ed il terzo vuoto. Nella matrice $\Omega$
compariranno gli elementi che descrivono le transizioni
$1=jk\uparrow$, $2=ij\downarrow$, $3=ik\uparrow$, $4=ik\downarrow$,
abbiamo dunque una matrice di dimensione  $4\times4$. In questo caso per\`{o} notiamo
subito che la matrice non presenta pi\`{u} alcuna simmetria e dunque non pu\`{o} pi\`{u} essere 
separata in due blocchi.
La simmetria della matrice rispetto allo spin sembra dunque scomparsa.
D'altra parte sappiamo che anche per sistemi a shell aperta valgono le stesse regole di selezione
individuate in precedenza, e dunque partendo da un sistema come il nostro in cui
$S=1/2$ e $S_z=1/2$ dovremmo avere due possibili blocchi di eccitazioni, quelle che conservano
lo spin, eccitazioni di doppietto, e quelle per cui $\Delta S=1$, eccitazioni di quadrupletto.
Cerchiamo di capire perch\'{e} non ritroviamo tali simmetrie nella matrice costruita.

%************************************************************************************************
\subsection{Il problema delle eccitazioni multiple}
Partiamo sempre dal sistema modello con tre elettroni.
Imponendo come in precedenza che lo stato fondamentale e gli stati eccitati siano contemporaneamente
autostati di $\hat{h}_{KS}$ $\hat{S}^2$ e $\hat{S}_z$ otteniamo le seguenti funzioni d'onda:
lo stato fondamentale con un doppietto
\[
\Phi_{1/2}^2=|\psi_{i\uparrow} \psi_{i\downarrow} \psi_{j\uparrow}|=|i\bar{i}j \rangle
\]
e i suoi stati eccitati con quattro doppietti
\begin{eqnarray*}
\Phi_{1/2}^2 &=& |i\bar{i}k \rangle  \\
\Phi_{1/2}^2 &=& |i\bar{j}j \rangle  \\
\Phi_{1/2}^2 &=& \frac{1}{\sqrt{2}}(|\bar{i}jk\rangle - |ij\bar{k}\rangle) \\
\Phi_{1/2}^2 &=& \frac{1}{\sqrt{6}}(|\bar{i}jk + |ij\bar{k}\rangle - 2|i\bar{j}k\rangle)
\end{eqnarray*}
e uno stato di quandrupletto
\[
\Phi_{1/2}^4=\frac{1}{\sqrt{3}}(|\bar{i}jk\rangle + |ij\bar{k}\rangle + |i\bar{j}k\rangle)
\hspace{0.5 cm} \text{.}
\]
Abbiamo qui gi\`{a} tenuto in considerazione le regole di selezione per lo spin,
la barra sopra uno stato indica che questo \`{e} occupato da un elettrone a spin down.
Notiamo immediatamente che l'aver imposto che le funzioni d'onda siano autostati di $\hat{S}^2$
fa si che gli stati eccitati di KS debbano mischiarsi con uno stato
che dal punto di vista delle funzioni d'onda di Kohn e Sham \`{e} uno stato a eccitazione doppia
($|i\bar{j}k\rangle$).
Questo poich\'{e} in sistemi a shell aperta compaiono stati con tre bande semi occupate e come
conseguenza l'operatore $\hat{S}^2$ di fatto mischia tali stati con altri in cui lo spin di due
particelle del sistema \`{e} stato invertito (questo accadeva anche per sistemi a shell chiusa
ma in questo caso venivano tra loro mischiati solo stati a eccitazione singola, poich\'{e} erano
presenti solo stati eccitati con due bande semi-occupate).
Mancando di fatto tale stato nella formulazione da noi fatta
la matrice non pu\`{o} presentare la simmetria richiesta
e dunque essere separata a blocchi.

Se cercassimo invece di costruire autostati di $\hat{h}_{KS}$, $\hat{S}_z$ e $\hat{S}^2$ con
le sole eccitazioni a particella singola otterremmo tre doppietti
\begin{eqnarray*}
\Phi_{1/2}^2 &=& |i\bar{i}k \rangle  \\
\Phi_{1/2}^2 &=& |i\bar{j}j \rangle  \\
\Phi_{1/2}^2 &=& \frac{1}{\sqrt{2}}(|\bar{i}jk\rangle - |ij\bar{k}\rangle) 
\end{eqnarray*}
e uno stato di tripletto \cite{Casida2}
\[
\Phi_{1/2}^3=\frac{1}{\sqrt{2}}(|\bar{i}jk\rangle + |ij\bar{k}\rangle)
\hspace{0.5 cm} \text{.}
\]
Vediamo dunque come manchi uno dei doppietti, mentre abbiamo un tripletto invece di un quandrupletto.
Tra le energie di eccitazione che otterremo diagonalizzando la matrice ci aspettiamo dunque che 
potranno essere in qualche modo considerate solo quelle eccitazioni che conservano $\hat{S}^2$.

Poich\'{e} d'altra parte la teoria sopra formulata \`{e} in principio
esatta dobbiamo immaginare
che tale deficienza vada individuata nell'approssimazione eseguita e dunque nell'approssimazione
adiabatica. In effetti mostreremo nell'ultimo capitolo di questa tesi il modello che Casida
propone per correggere tale problema e come nell'analizzare il problema lo stesso Casida
individui proprio nell'approssimazione adiabatica per il kernel della TDDFT
 il motivo per cui in tale formalismo non
siano presenti tutti gli stati ad eccitazione doppia.

La correzione di questo problema, che come vedremo si tradurr\`{a} nella necessit\`{a} di scartare
alcune delle 
eccitazioni calcolabili diagonalizzando la matrice di Casida, esula per\`{o} dagli
scopi di questa tesi. Per i calcoli che effettueremo ci accontenteremo dunque di utilizzare la 
teoria sin qui costruita e di diagonalizzare la matrice completa in approssimazione adiabatica.
Abbiamo in altre parole terminato di illustrare
il formalismo che abbiamo implementato e che utilizzeremo nei capitoli successivi per
lo studio degli stati eccitati di molecole con stato fondamentale spin polarizzato.

L'ultima sezione di questo capitolo costituisce invece un tentativo di generalizzare il formalismo
sin qui presentato. Nei calcoli che faremo le generalizzazioni proposte non verranno considerate.

%**********************************************************************************************
%**********************************************************************************************
%**********************************************************************************************
\section{Proposte per estendere la teoria}
In questa sezione infine cercheremo di mostrare come sia possibile estendere la teoria
di Casida sin qui descritta a due situazioni che non sono state considerate.

La prima situazione \`{e} quella
in cui venga considerata una perturbazione esterna di questo tipo
\begin{equation}
\hat{V}^{ext}(t) = \sum_{\alpha\beta} \int d^3\mathbf{x}\ V_{\alpha\beta}^{ext}(\mathbf{x},t)
                   \hat{\psi}_{\alpha}^{\dag}(\mathbf{x}) \hat{\psi}_{\beta}(\mathbf{x}) 
\hspace{0.5 cm} \text{,}
\end{equation}
in cui cio\`{e} compaiano anche termini non diagonali nello spin.

La seconda situazione \`{e} quella in cui
\begin{equation}
[\hat{H},\hat{S}_z] \neq 0
\hspace{0.5 cm} \text{,}
\end{equation}
in cui quindi l'Hamiltoniana e l'operatore $\hat{S}_z$ non hanno autostati in comune e per cui
\`{e} dunque necessario considerare degli spinori.

Per entrambe le estensioni non dovremo far altro che scegliere in modo adeguato lo spazio verso cui
proiettare la TDDFT in modo da ottenere la corretta formulazione. Altra necessit\`{a} che comparir\`{a}
in entrambe le teorie sar\`{a} quella di lavorare con un kernel non diagonale rispetto allo spin e
dunque che dovr\`{a} partire dalla forma completa dell'energia di scambio e correlazione come funzionale
della matrice densit\`{a}
\begin{equation}
K_{\sigma\tau,\delta\gamma}=\frac{\delta E_{xc}[\rho_{\sigma\tau}]}{\delta \rho_{\delta\gamma}}
\hspace{0.5 cm} \text{.}
\end{equation}

%******************************************************************************************************
\subsection{Includere la possibilit\`{a} di spin flip}
Considerare una perturbazione in cui siano presenti anche le componenti non diagonali rispetto
allo spin significa di fatto considerare la possibilit\`{a} di una perturbazione in cui siano
presenti oltre ai termini gi\`{a} individuati per il caso diagonale rispetto allo spin anche le
componenti lungo le direzioni $x$ e $y$ del campo magnetico.
La perturbazione pu\`{o} in altre parole essere scritta come
\begin{eqnarray}
\hat{V}^{ext}(t)=\int d^3\mathbf{x} [
V(\mathbf{x},t)\hat{\rho}(\mathbf{x},t)+B_z(\mathbf{x},t)\hat{m}_z(\mathbf{x},t) \nonumber \\
+B_y(\mathbf{x},t)\hat{m}_y(\mathbf{x},t)+B_x(\mathbf{x},t)\hat{m}_x(\mathbf{x},t) ]
\hspace{0.5 cm} \text{.}
\end{eqnarray}

La principale conseguenza di questa scelta \`{e}, come gi\`{a} detto, che ora anche le componenti
non diagonali della
densit\`{a} rispetto allo spin avranno una variazione differente da zero, questo perch\'{e} anche
nella perturbazione compaiono tali componenti della densit\`{a}.
Scriviamo dunque l'espressione della variazione di densit\`{a} considerando tali componenti:
\begin{equation}
\delta n_{\alpha\beta}(\mathbf{x}t)= \int d^3\mathbf{x'} \int dt' \chi_{\alpha\beta,\gamma\delta}
         (\mathbf{x}t,\mathbf{x}'t') V_{\gamma\delta}^{ext}(\mathbf{x}'t')
\hspace{0.5 cm} \text{.}
\end{equation}
Questa in realt\`{a} \`{e} una generalizzazione della (\ref{Dyson-tddft}) nel caso in cui
si considerino componenti della densit\`{a} non diagonali rispetto allo spin. La generalizzazione
appena fatta ci appare qui del tutto intuitiva e dunque non ne riportiamo una dimostrazione.
Scriviamo quindi la funzione di risposta ed il kernel nello spazio delle configurazioni.
\begin{equation}
\begin{split}
&\chi_{i\alpha j\beta,h\gamma k\delta} = \int \int d^3\mathbf{x} d^3\mathbf{x}'
    \psi_{i\alpha}(\mathbf{x}) \psi_{j\beta}^*(\mathbf{x})
    \chi_{\alpha\beta,\gamma\delta}(\mathbf{x}t,\mathbf{x}'t')         
    \psi_{h\gamma}^*(\mathbf{x}') \psi_{k\delta}(\mathbf{x}')                   \\
&K_{i\alpha j\beta,h\gamma k\delta} = \int \int d^3\mathbf{x} d^3\mathbf{x}'
        \psi_{i\alpha}(\mathbf{x}) \psi_{j\beta}^*(\mathbf{x}')
        K_{\alpha\beta,\gamma\delta}(\mathbf{x}t,\mathbf{x}'t')
        \psi_{h\gamma}^*(\mathbf{x}') \psi_{k\delta}(\mathbf{x}')
\hspace{0.5 cm} \text{.}
\end{split}
\end{equation}

Vediamo dunque che cos\`{i} come per la densit\`{a} anche per la funzione di risposta e per il
kernel compaiono dei termini non diagonali nello spin. 
La prima differenza evidente che compare in questa situazione \`{e} dunque la presenza di un ulteriore
indice di spin in tutte le formule.
Prima di procedere con il tentativo di scrivere l'equazione di Casida dobbiamo per\`{o} fare un'ultima
considerazione sul potenziale scelto. Come sappiamo infatti l'idea di Casida parte dal fatto che, riordinando
la base scelta possiamo scrivere una relazione tra la variazione della parte reale della densit\`{a}
rispetto alla parte reale del potenziale esterno in modo da ottenere una matrice che \`{e} di dimensione 
dimezzata e di tipo hermetiano. La condizione di realt\`{a} per il potenziale ha per\`{o} senso in spazio
reale e non nello spazio dello spin. Per questo per verificare la condizione di realt\`{a} del potenziale in
quest'ultimo dobbiamo scrivere le trasformazioni dal primo.
Per la situazione studiata da Casida queste coincidono con la (\ref{trasf-spin-real}) in cui compaiono solo
coefficienti reali e dunque la condizione di realt\`{a} risulta equivalente nelle due formulazioni. Se
consideriamo invece un potenziale non diagonale rispetto allo spin le trasformazioni risultano essere
\begin{eqnarray}
&&V(\mathbf{x})=1/2(V_{\uparrow\uparrow}+V{\downarrow\downarrow})   \nonumber \\
&&B_z(\mathbf{x})=1/2(V_{\uparrow\uparrow}-V_{\downarrow\downarrow}) \nonumber \\
&&B_x(\mathbf{x})=1/2(V_{\uparrow\downarrow}+V_{\downarrow\uparrow})  \nonumber \\
&&B_y(\mathbf{x})=-i/2(V_{\uparrow\downarrow}-V_{\downarrow\uparrow})
\hspace{0.5 cm} \text{.}
\end{eqnarray}
Compaiono dunque dei coefficienti immaginari e la relazione tra la realt\`{a} del potenziale esterno in spazio
reale e in spazio di spin non \`{e} pi\`{u} quella banale.

In questo caso, in altre parole, ci\`{o} che possiamo fare \`{e} partire dall'equazione 
(\ref{excitations-tddft}) e cercare di estrarre le eccitazioni da questa. Una possibilit\`{a}
\`{e} senz'altro costituita dalla scelta di lavorare in approssimazione Tamm-Dancoff (TDA), il
che consiste nell'ignorare la seconda riga di tale equazione ottenendo
\begin{equation}
(A-\omega C) \delta P(\omega)=\delta V^{ext}(\omega)
\hspace{0.5 cm} \text{.}
\end{equation}
Non sar\`{a} invece pi\`{u} possibile costruire la matrice $\Omega(\omega)$ di Casida.

Concludiamo il paragrafo segnalando come di fatto l'introduzione di questo tipo di perturbazione
modifichi evidentemente le regole di selezione per lo spin. In questo caso avremo:
\[
\Delta S=0,1 \hspace{10 mm} \text{;} \Delta S_z=0,\pm 1
\hspace{0.5 cm} \text{.}
\]
%\begin{eqnarray}
%\left[\hat{S}_z, \hat{n}_{\uparrow\downarrow}\right] &=&  \hat{n}_{\uparrow\downarrow}  \nonumber \\
%\left[\hat{S}_z, \hat{n}_{\downarrow\uparrow}\right] &=& -\hat{n}_{\downarrow\uparrow}
%\end{eqnarray}

%**********************************************************************************************
\subsection{Sistemi non collineari}
La teoria di Casida fin qui esposta prevede che l'operatore $\hat{S_z}$ commuti con l'Hamiltoniana
totale del sistema e che quindi gli autostati dell'Hamiltoniana siano funzioni d'onda con $S_z$
fissato. Cerchiamo ora di studiare come potrebbe essere riformulata la teoria nel caso in cui
questo non accada.
La prima conseguenza sarebbe quella che le auto-funzioni a particella singola della DFT non sarebbero
pi\`{u} degli spinori con una singola componenete differente da zero, il che ci permetteva di lavorare
con semplici funzioni d'onda, ma spinori con entrambe le componenti differenti da zero.
Quello che dobbiamo quindi
fare \`{e} molto semplicemente utilizzare le regole del cambio di base gi\`{a} viste per passare 
dallo spazio $\mathbf{r},\sigma$ allo spazio degli spinori $n$.
Vediamo quindi come risulterebbero le equazioni in questo nuovo spazio.
Iniziamo con lo scrivere la matrice densit\`{a}
\begin{equation}
\hat{n}_{\alpha\beta}(\mathbf{x},t)= \sum_{n,m} \phi^*_n(\mathbf{x}\alpha) \phi_m(\mathbf{x}\beta)
                                  \hat{b}^{\dag}_n(t) \hat{b}_m(t)
\hspace{0.5 cm} \text{,}
\end{equation}
dove con la notazione $\phi_n$ intendiamo uno spinore a due componenti
$\psi_{n\uparrow}$ e $\psi_{n\downarrow}$.
Sottolineiamo come in generale le nuove funzioni $\psi_{n\sigma}$ non siano per\`{o} un
s.o.n.c., ma semplicemente la componente dello spinore a spin up e down. Le relazioni di
ortonormalit\`{a} saranno invece date da
\begin{eqnarray}
&&\int d^3\mathbf{r} \phi^*_n(\mathbf{r}) \phi_m(\mathbf{r}) = \delta_{n,m}  \nonumber \\
&&\sum_{\sigma} \int d^3\mathbf{r} \psi^*_{n\sigma}(\mathbf{r}) \psi_{m,\sigma}(\mathbf{r})
                                                                = \delta_{n,m}
\hspace{0.5 cm} \text{.}
\end{eqnarray}

Definita quindi la nuova grandezza
\begin{equation}
Q_{nm}(t) = \langle \Psi_{KS} | \hat{b}^{\dag}_n \hat{b}_m | \Psi_{KS} \rangle
\hspace{0.5 cm} \text{,}
\end{equation}
otterremo le due equazioni
\begin{eqnarray}
&&\delta Q_{nm}(t) = \chi^{KS}_{nm,lt}(t-t') V^{eff}_{lt}(t')   \\
&&\delta Q_{nm}(t) = \chi_{nm,lt}(t-t') V^{ext}_{lt}(t')
\hspace{0.5 cm} \text{.}
\end{eqnarray}
Proseguendo in modo identico a quanto fatto in precedenza possiamo scrivere il kernel nella
nuova rappresentazione, esprimere poi in modo esplicito la funzione di risposta libera e infine
scrivere la nostra equazione per la funzione di risposta totale. Non ripetiamo quindi le
considerazioni gi\`{a} fatte ma ci limitiamo a scrivere i risultati per il kernel,
\begin{equation}
K_{nm,lt}(t-t') = \frac{\delta V^H_{nm}(t)}{\delta Q_{lt}(t')}+
                               \frac{\delta V^{xc}_{nm}(t)}{\delta Q_{lt}(t')}
\hspace{0.5 cm} \text{,}
\end{equation}
e per la funzione di risposta libera
\begin{equation}
\chi^{KS}_{nm,lt}(\omega) =
\delta_{n,t}\delta_{m,l}\frac{f_{n}-f_{m}}
     {\omega-(\epsilon^{KS}_m-\epsilon^{KS}_n)}
\hspace{0.5 cm} \text{.}
\end{equation}

Infine, come in precedenza esprimiamo il kernel nella nuova base rispetto alla base $\mathbf{r}\sigma$:
\begin{equation}
K_{nm,lt}(t-t')=\sum_{\alpha\beta,\delta\gamma}\int d^3\mathbf{r} d^3\mathbf{r}'
   \phi_{n}^*(\mathbf{r},\alpha)\phi_{m}(\mathbf{r},\beta)
   K_{\alpha\beta,\delta\gamma}[\rho](\mathbf{r}t,\mathbf{r}'t')
   \phi_{l}^*(\mathbf{r}',\delta)\phi_{t}(\mathbf{r}',\gamma)
\hspace{0.5 cm} \text{.}
\end{equation}
In questo modo abbiamo riscritto il formalismo di Casida per un sistema in cui l'Hamiltoniana non 
commuta con lo l'operatore $\hat{S_z}$

Rispetto al caso studiato da Casida dovremo ora tenere in considerazione un'importanti differenza
dovuta al fatto che avremo a che fare ma con degli spinori.
Poich\`{e} infatti gli spinori sono grandezze intrinsecamente complesse
(per vedere questo \`{e} sufficiente provare a scrivere gli autostati delle matrici di Pauli) non
sar\`{a} in alcun modo valida la relazione
\begin{equation}
\phi_n^*(\mathbf{r},\gamma)\phi_m(\mathbf{r},\delta)=\phi_n(\mathbf{r},\gamma)\phi_m^*(\mathbf{r},\delta)
\end{equation}
e come conseguenza
la matrice del kernel nello spazio degli stati non avr\`{a} pi\`{u} quelle simmetrie che
abbiamo utilizzato per giungere alla formulazione della matrice $\Omega$ (\ref{Omega_matr}).
Come nel caso precedente dunque lo sviluppo della teoria dovr\`{a} partire dall'equazione
(\ref{excitations-tddft}) e dunque una possibile soluzione \`{e} nuovamente quella di lavorare
in TDA (Tamm-Dancoff Approximation).

Un'ultima breve considerazione sulle regole di selezione per il caso non collineare. Poich\'{e} a determinare
le regole di selezione \`{e} la perturbazione esterna, le regole di selezione saranno identiche a
quelle sin qui trovate per i sistemi collineari. La non collinearit\`{a} non porta da questo punto di vista
niente di nuovo se non forse che perde in effetti di senso considerare la conservazione della componente $z$
dello spin.

%% file: sistemi.tex
\section{La molecola di BeH}

%*******************************************************************************************
\subsection{Calcoli eseguiti}
Come gi\`{a} detto nell'introduzione, il lavoro di questa tesi \`{e} stato focalizzato su una
modifica dell'implementazione del formalismo di Casida per la TDDFT all'interno del software
ABINIT \cite{Abinit} che permettesse di studiare anche sistemi a shell aperta
e sbilanciati in spin.

Prima di utilizzare il programma per lo studio di un sistema con stato fondamentale
a spin polarizzato abbiamo eseguito alcuni test su di un sistema che non presenta effetti
di polarizzazione rispetto allo spin. A questo scopo abbiamo calcolato con il nostro programma
\cite{Abinit} lo spettro di eccitazione della molecola $N_2$ utilizzando prima l'opzione
$nsppol=1$, in modo che il software utilizzasse l'implementazione gi\`{a} presente ed inseguito
con l'opzione $nsppol=2$ in modo che il software utilizzasse la nuova implementazione includendo
in modo separato gli stati a spin up e quelli a spin down.

Tutti i test hanno dato gli stessi risultati per lo spettro di eccitazione del sistema.
Abbiamo dunque deciso di proseguire il nostro lavoro cercando di studiare un sistema che presenti
stato fondamentale spin-polarizzato.

%*************************************************************************************************
\subsection{Presentazione della molecola}

\begin{figure}[t]
\begin{center}
\includegraphics[width=12 cm]{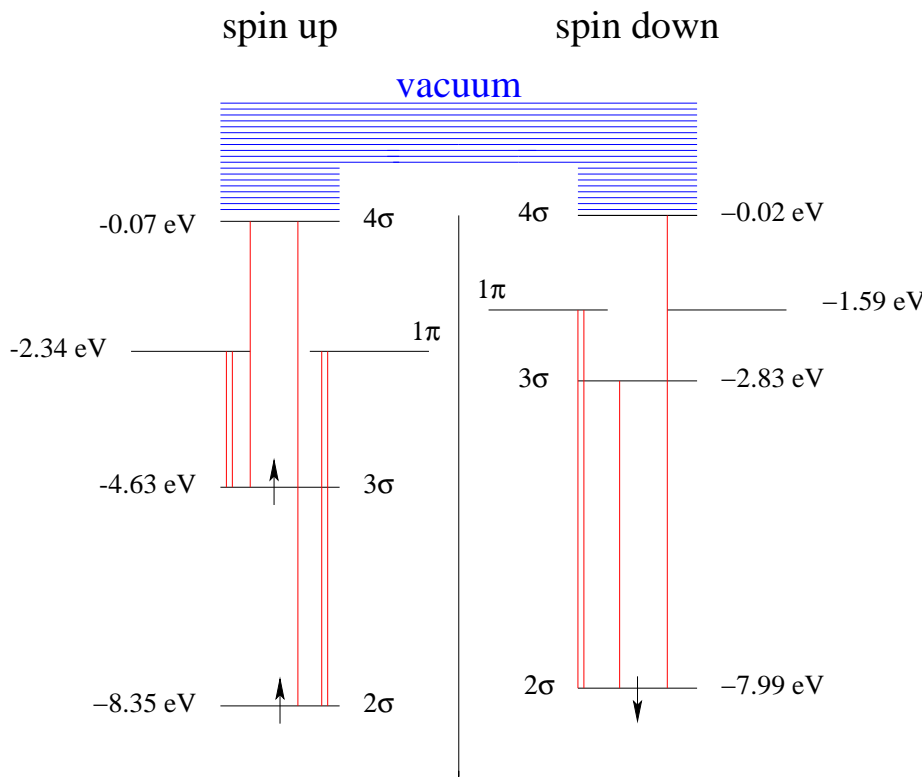} 
\caption{Schema degli autovalori di KS per il BeH calcolati con
         il programma ABINIT (DFT, approssimazione LSDA). Risultati
         ottenuti per acell=$50\times 50\times 50$ Bohr, ecut=$10$ Ha, nband=$50$.
         Per la classificazione degli stati si veda l'appendice A.}
\label{fig:BeH_orbitals}
\rule{\linewidth}{0.2 mm}
\end{center}
\end{figure}

La scelta di utilizzare questa molecola per testare il nostro programma
\`{e} motivata dal fatto che questa \`{e} la molecola pi\`{u} semplice che
abbiamo trovato a presentare un ground-state a spin polarizzato e dal fatto
che di questa molecola abbiamo a disposizione dati sperimentali \cite{Sper}
e dati teorici ottenuti nell'ambito dello stesso formalismo con cui
confrontarci \cite{Casida3,Casida4}.

Per lo studio dello spettro di eccitazione siamo partiti dal dato
sperimentale per la distanza d'equilibrio tra i due ioni della referenza
\cite{Sper} e abbiamo lanciato
il programma ABINIT in due step successivi: un primo step in cui vengono calcolati
in modo auto-consistente le funzioni d'onda degli stati occupati della molecola e
un secondo step in cui vengono calcolate anche le funzioni d'onda degli stati non
occupati, viene costruita la matrice $\Omega$ e viene diagonalizzata la matrice stessa.

\begin{figure}[t]
\begin{center}
\includegraphics[width=13 cm]{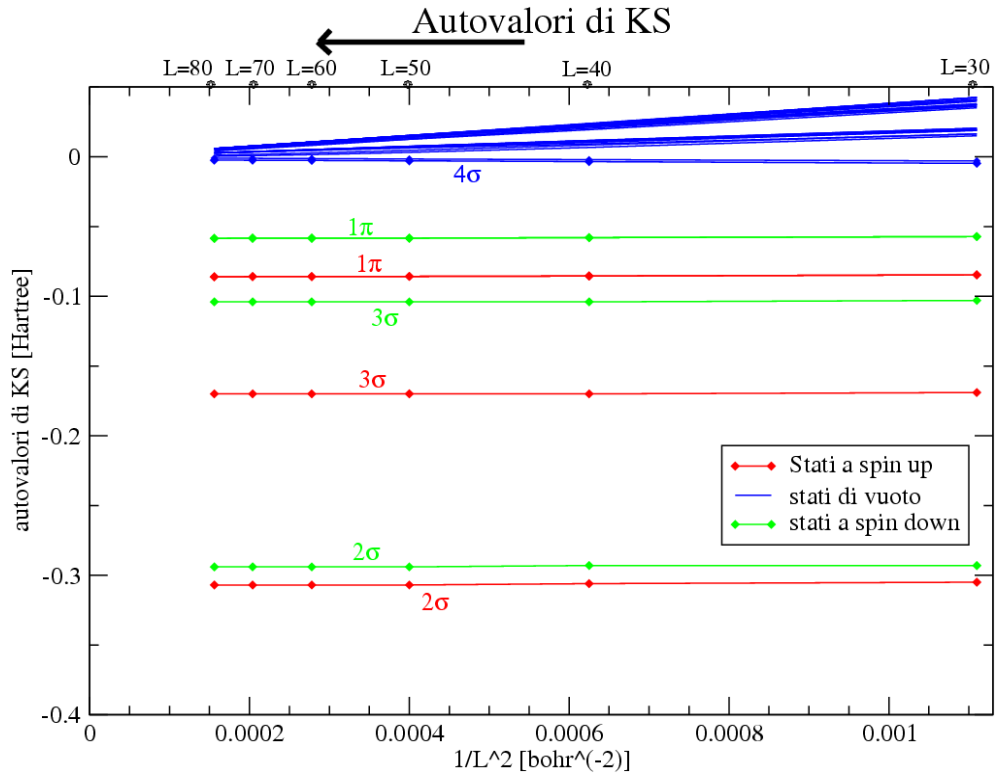}
\vspace{0.1 cm}
\caption{Effetti di taglia (L=lato dela supercella) sugli autovalori di KS.
         Energia di cut-off, ecut=10 Hartree}
\label{fig:KS_acell}
\rule{\linewidth}{0.2 mm} 
\end{center}
\end{figure}

Lo schema per lo stato fondamentale della molecola che abbiamo ottenuto \`{e}
$2\sigma^2 3\sigma^1$; non compare qui lo stato $1\sigma$ poich\'{e} il nostro programma
lavora solo con gli elettroni di valenza, inglobando gli elettroni di core in un
pseudopotenziale \cite{Abinit}.
Il diagramma completo degli autovalori di KS ottenuto dal secondo step \`{e} rappresentato
in figura \ref{fig:BeH_orbitals}.

Abbiamo due serie differenti di autovalori per gli elettroni a spin up e quelli
a spin down poich\'{e} il sistema \`{e} spin polarizzato. Dallo schema riportato
in figura possiamo iniziare ad analizzare innanzi tutto l'energia di ionizzazione
del sistema, $-\epsilon^{\uparrow}_{HOMO}=4.63$ eV, interpretando come tale 
l'energia dell'ultimo stato occupato riferita al livello di vuoto.
Vediamo che questo valore \`{e} di molto
inferiore al valore sperimentale, $-\epsilon^{\uparrow}_{HOMO}=8.21$ eV \cite{Sper},
ma in perfetto accordo con il valore riportato da Casida,
$-\epsilon^{\uparrow}_{HOMO}=4.60$ eV \cite{Casida3}. Vedremo come questo errore,
dovuto all'inadeguatezza del potenziale di scambio e correlazione LDA, si tradurr\`{a}
nell'impossibilit\`{a} di individuare alcune delle eccitazioni della molecola.

%******************************************************************************************************
\subsection{Convergenza per gli orbitali di KS}

Come prima cosa studiamo il comportamento degli autovalori di KS la variare di due parametri:
le dimensioni della supercella e l'energia di cut-off che definisce la base di onde piane.
Presentiamo innanzi tutto il grafico con la convergenza dei risultati rispetto alle dimensioni
della supercella (fig. \ref{fig:KS_acell}).
\begin{figure}[t] 
\begin{center}
\includegraphics[width=13 cm]{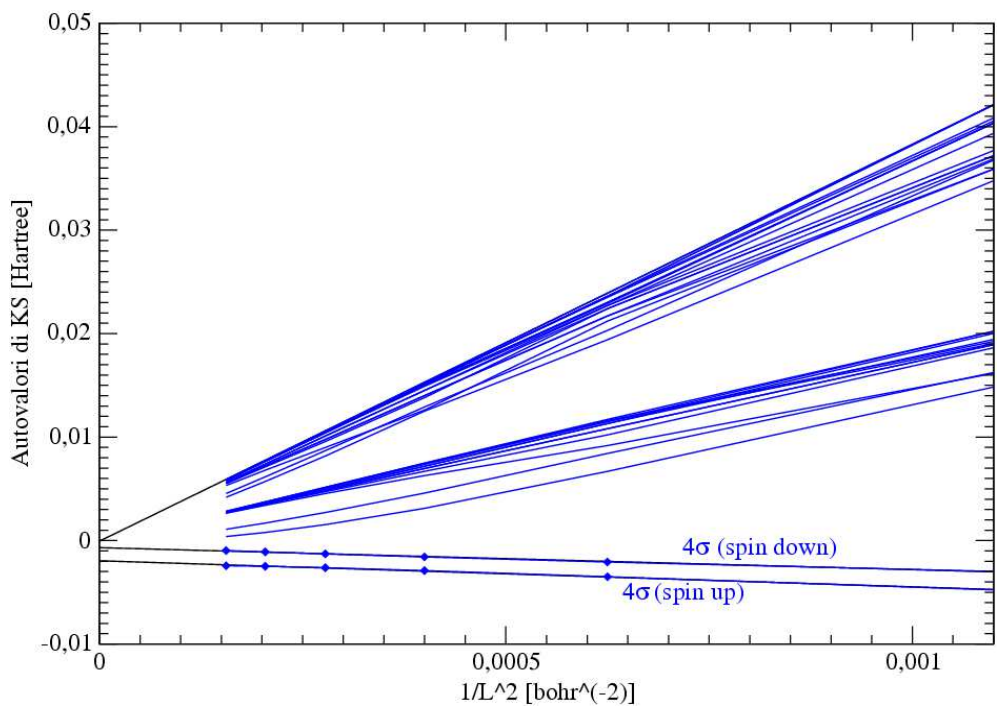}
\vspace{0.2 cm}
\caption{Ingrandimento della regione vicino al livello $0$ eV dell'energia della
         figura \ref{fig:KS_acell}}
\label{fig:KS_acell_vacuum}
\rule{\linewidth}{0.2 mm} 
\end{center}
\end{figure}
Abbiamo messo in grafico il variare dell'autoenergia di KS rispetto a $1/L^2$ poich\'{e}
in questo modo siamo in grado di individuare gli stati di vuoto. Uno stato di vuoto \`{e}
infatti caratterizzato da un'energia del tipo
\begin{equation}
\epsilon_j=\frac{\hbar^2 k_j^2}{2m}
\hspace{0.5 cm} \text{,}
\end{equation}
dove $m$ \`{e} la massa dell'elettrone, $j$ l'indice del livello eccitato e $k$ \`{e}
il vettore d'onda che, all'interno di una
cella di dimensioni finite, vale $\mathbf{k}_j=\frac{2\pi}{\lambda}\hat{\mathbf{k}}=
\frac{2\pi \mathbf{j}}{L}$, per $\mathbf{j}$ vettore a componenti intere,
e quindi l'energia di uno stato di vuoto risulta
\begin{equation}
\epsilon_j=\frac{4\pi^2\hbar^2 j^2}{2mL^2}
\end{equation}
da cui vediamo che gli stati di vuoto possono essere individuati come quegli stati che hanno una
dipendenza di tipo lineare dal quadrato dell'inverso del lato della supercella.
Nel grafico (fig. \ref{fig:KS_acell}) possiamo cos\`{i} individuare come stati di vuoto tutti
quegli stati che abbiamo rappresentato in blu.
\begin{figure}[t] 
\begin{center}
\includegraphics[width=12 cm]{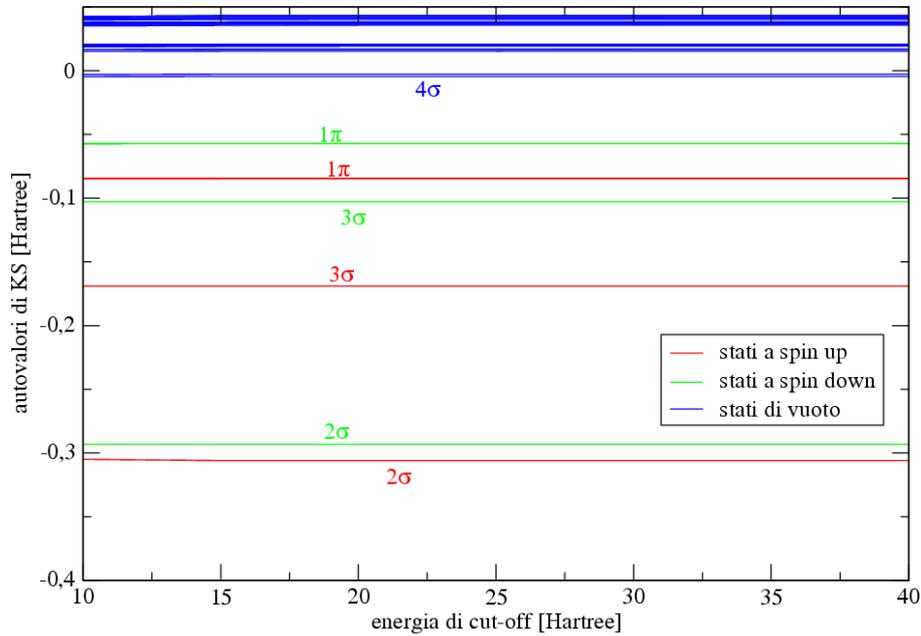}
\caption{Grafico di convergenza pr le auto-energie di KS rispetto al valore
         del cut-off sull'energia}
\label{fig:KS_ecut}
\rule{\linewidth}{0.2 mm}
\end{center}
\end{figure}
Questo diagramma giustifica la rappresentazione fatta 
dello schema dei livelli (fig. \ref{fig:BeH_orbitals}),
e in particolare permette di individuare il livello di vuoto
come il valore a cui tendono tali stati (fig. \ref{fig:KS_acell_vacuum}) e che
verr\`{a} preso come $0$ di riferimento della scala di energie.

Individuati gli stati di vuoto dal grafico (fig. \ref{fig:KS_acell}) possiamo invice
osservare come gli altri stati, che chiameremo stati di valenza per distinguerli dagli
stati di vuoto, convergono rapidamente rispetto alle dimensioni della supercella.
Allo stesso modo possiamo osservare nel grafico (fig. \ref{fig:KS_ecut}) come 
la convergenza sia stata raggiunta anche per l'energia di cut-off del sistema.

%*******************************************************************************************************
%*******************************************************************************************************
%*******************************************************************************************************
\section{Analisi dei dati}
\subsection{Confronto con i risultati ottenuti da Casida}

\begin{comment}
\begin{table}[t]
\begin{center}
%\rule{\linewidth}{0.2 mm}
\begin{tabular}{|l|l|l|l||l|}
\hline
Tipo di     &TDLDA      & TDLDA    &  TDLDA   &  Eccitazione   \\
eccitazione &Gaussian03 & DeMon2k  &  TDA     &  Principale    \\
\hline
$\Pi$    &  2.3657   &  2.2479  &  2.2860  & $3\sigma->1\pi\ \uparrow$            \\
$\Sigma$ &  4.6633   &  4.5103  &  4.5278  & $3\sigma->4\sigma\ \uparrow$         \\
$\Sigma$ &  4.7576   &  4.6300  &  4.6481  & $3\sigma->5\sigma\ \uparrow$         \\
$\Pi$    &  4.8451   &  4.7047  &  4.7104  & $3\sigma->2\pi\ \uparrow$            \\
$\Sigma$ &  5.1318   &  4.8049  &  4.8676  & $2\sigma->3\sigma\ \downarrow$       \\
$\Pi$    &  5.6338   &  5.1685  &  5.2953  & $2\sigma->1\pi\ \uparrow\downarrow$  \\
$\Sigma$ &  5.6414   &  5.4803  &  5.5030  & $3\sigma->6\sigma\ \uparrow$         \\
         &           &          &          & $2\sigma->3\sigma\ \downarrow$       \\
\hline
\end{tabular}
\caption{Spettro di eccitazione della molecola BeH. Dati presi dall'articolo di
         Casida \cite{Casida4}}
\label{eccitazioni-Casida}
\rule{\linewidth}{0.2 mm}
\end{center}
\end{table}
\end{comment}

Svolti i test di convergenza per lo studio delle autoenergie di KS possiamo passare allo studio
delle energie di eccitazione ottenute diagonalizzando la matrice di Casida (\ref{Omega})
e confrontare i risultati ottenuti con quelli a nostra disposizione.
Negli articoli \cite{Casida3,Casida4} sono riportate almeno le prime sette energie di eccitazione e in
particolare nel lavoro \cite{Casida4} possiamo trovare la transizione principale di KS associata
ad ogni eccitazione (tab. \ref{tab:eccitazioni-confronto-1})

\begin{table}[t]
\begin{center}
\begin{tabular}{|l|l|l|l|l||l|l|}
\hline
Tipo     & eccitaz.   &TDLDA          & TDLDA          & Presente & Calcoli       & Dati        \\
eccit.   & principale &Gaussian03     & DeMon2k        & Lavoro   & Quant.        & sper.       \\
     & \cite{Casida4} &\cite{Casida4} & \cite{Casida4} & TDDFT    & Chem.         & \cite{Sper} \\  
     &                &               &                &          & \cite{PTN_BeH}&             \\
\hline
$\Pi$     & $3\sigma->1\pi\ \uparrow$ &
2.3657  &  2.2479  &  2.40  & 2.56  & 2.48           \\
$\Sigma$  & $3\sigma->4\sigma\ \uparrow$ &
4.6633  &  4.5103  &  4.52  & 5.51  &                \\
$\Sigma$  & $3\sigma->5\sigma\ \uparrow$ &
4.7576  &  4.6300  &  4.69  & 5.61  &                \\
$\Pi$     & $3\sigma->2\pi\ \uparrow$ &
4.8451  &  4.7047  &  4.80  & 6.31  & 6.31           \\
$\Sigma$  & $2\sigma->3\sigma\ \downarrow$ &
5.1318  &  4.8049  &  ****  & 6.12  &                \\
$\Pi$     & $2\sigma->1\pi\ \uparrow\downarrow$ &
5.6338  &  5.1685  &  5.65  & 6.74  & $\sim$6.7      \\
$\Sigma$  & $3\sigma->6\sigma\ \uparrow$ &
5.6414  &  5.4803  &  ****  & 6.71  & $\sim$6.7      \\
$\Delta$  &  &       &          &        & 6.77  & $\sim$6.7      \\
$\Sigma$  &  &       &          &        &       & 6.71           \\
$\Sigma$  &  &       &          &        &       & 7.02           \\
$\Pi$     &  &       &          &        & 7.27  & 7.28           \\
\hline
\end{tabular}
\caption{Confronto dei dati riportati con alcuni valori teorici di riferimento
         e dati sperimentali}
\label{tab:eccitazioni-confronto-1}
\rule{\linewidth}{0.2 mm}
\end{center}
\end{table}

Facendo riferimento a questi dati possiamo scegliere, tra le eccitazioni che troviamo nel nostro
file di output
le stesse riportate da Casida, ovvero quelle a cui \`{e} associata la stessa transizione di KS
riportata nella referenza, ottenendo cos\`{i} i risultati riportati nella tabella
(\ref{tab:eccitazioni-confronto-1}).
Vediamo che l'accordo con i dati riportati da Casida \`{e} ottimo, unica eccezione \`{e} costituita
dalle eccitazioni che coinvolgono la transizione $2\sigma->3\sigma\ \downarrow$ che nei nostri
calcoli, per i parametri indicati in tabella (tab. \ref{tab:eccitazioni-confronto-1}) risulta
di difficile individuazione e che per il momento non riportiamo.
L'accordo non \`{e} invece altrettanto buono se confrontato con i valori teorici di riferimento,
tratti da un lavoro di I.D. Petsalakis \cite{PTN_BeH} (valori teorici presi come riferimento da
Casida nei lavori \cite{Casida3,Casida4})  e con alcuni dati sperimentali \cite{Sper}.

%****************************************************************************************************
%\subsection{Un'analisi pi\`{u} sistematica dei risultati}
Analizzando le eccitazioni ottenute confrontando i dati con quelli riportati da Casida abbiamo
per\`{o} trascurato di considerare due fatti:

\begin{figure}[t] 
\begin{center}
\includegraphics[width=13 cm]{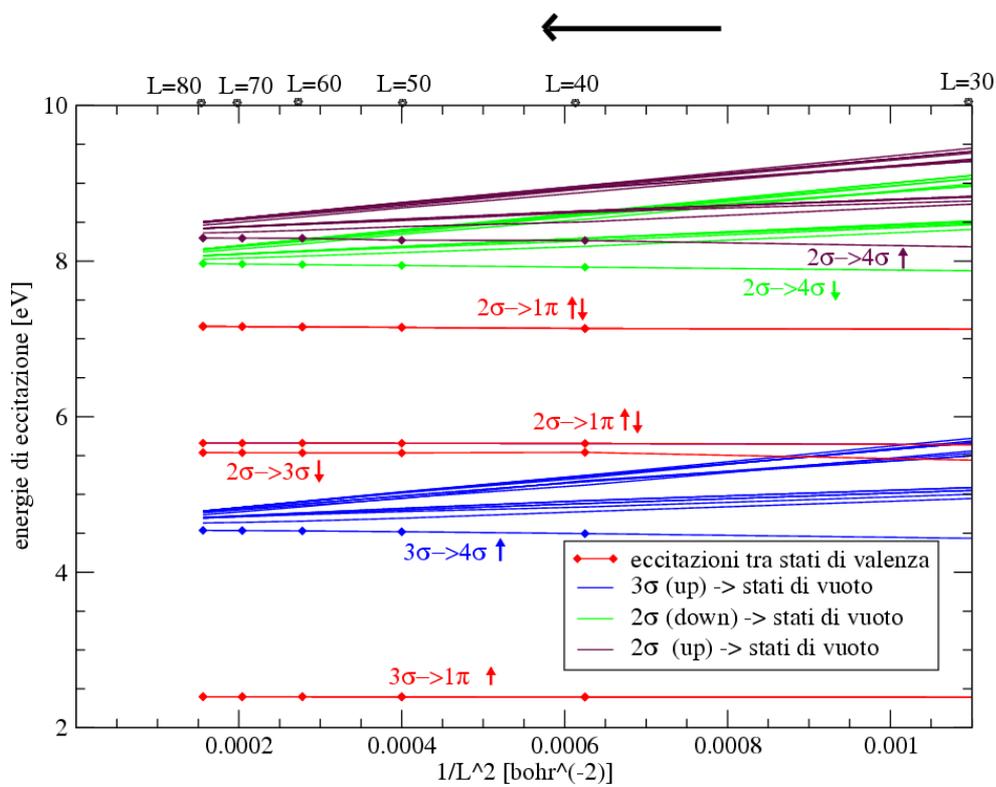}
\caption{Convergenza delle energie di eccitazione TDDFT rispetto alle dimensioni
         della supercella}
\label{fig:TDDFT_acell}
\rule{\linewidth}{0.2 mm} 
\end{center}
\end{figure}

\begin{enumerate}
\item L'approssimazione LDA \`{e} responsabile della sottostima dell'energia
di legame dei livelli molecolari occupati rispetto al livello di vuoto.
\item L'approssimazione adiabatica, come discusso nel capitolo in cui abbiamo presentato la teoria
di Casida, dovrebbe impedirci di ottenere, in linea di principio, alcune delle eccitazioni
del sistema e tra le eccitazioni trovate saremo costretti a scartare quelle per le quali lo spin
totale del sistema non viene conservato.
\end{enumerate}

La prima di queste due considerazioni ci porta a studiare nuovamente il problema degli stati di
vuoto gi\`{a} presentato per gli autovalori di KS e vedere come questo si riscontri nello studio
delle energie di eccitazione in TDDFT.
Guardando la figura \ref{fig:TDDFT_acell} possiamo osservare
che le eccitazioni che coinvolgono stati con energia superiore allo stato $4\sigma$ presentano
un andamento fortemente sensibile alle dimensioni della supercella e possono dunque
essere interpretate come eccitazioni verso stati di vuoto.
\begin{figure}[t] 
\begin{center}
\includegraphics[width=12 cm]{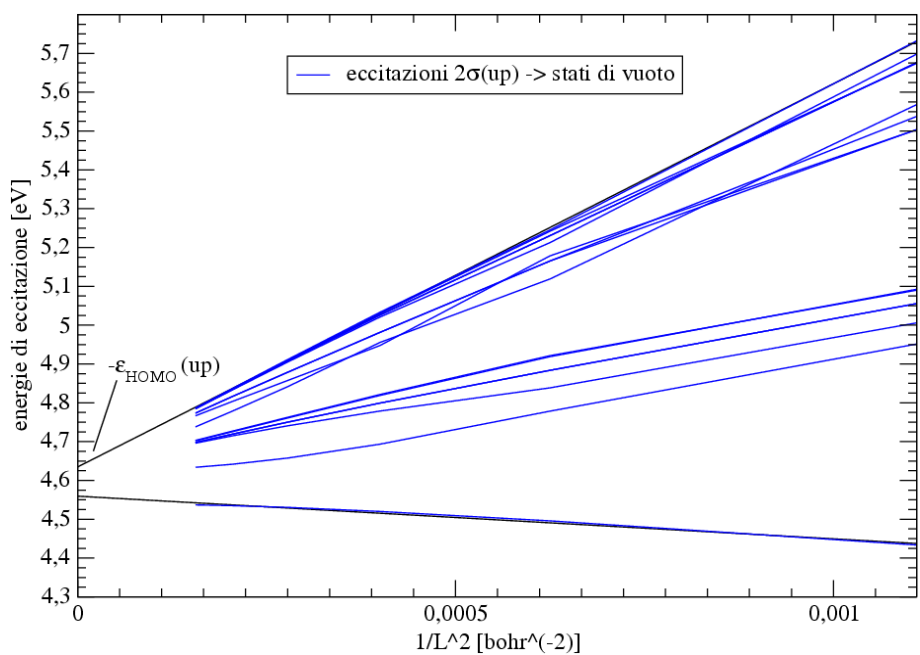}
\rule{\linewidth}{0.2 mm}
\caption{Grafici di convergenza delle energie di eccitazione verso stati
         di vuoto rispetto alle dimensioni della supercella.}
\label{fig:vacuum_study1}
\end{center}
\end{figure}
\begin{figure}[t] 
\begin{center}
\includegraphics[width=12 cm]{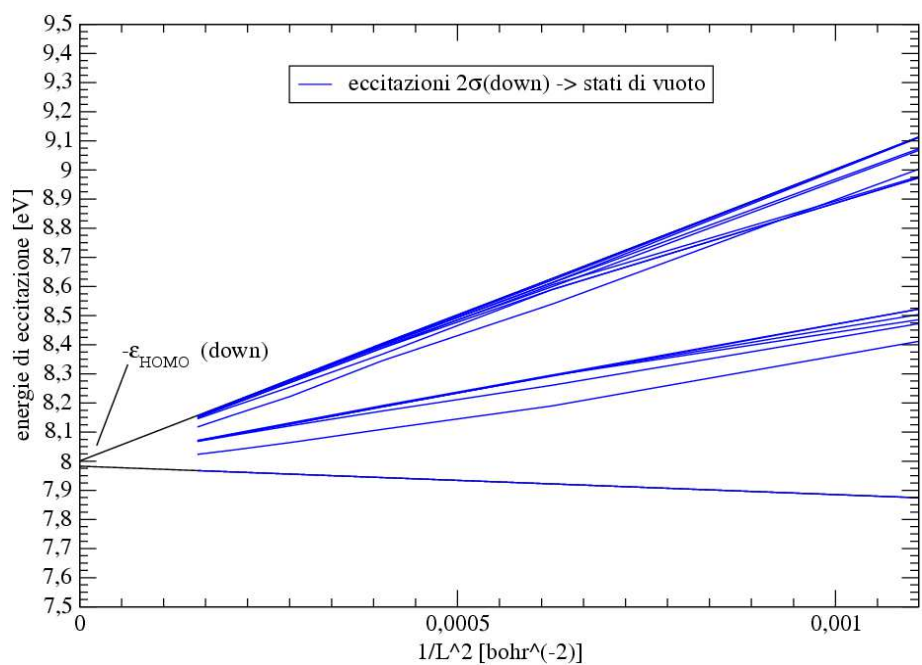}
\caption{Grafici di convergenza delle energie di eccitazione verso stati
         di vuoto rispetto alle dimensioni della supercella.}
\label{fig:vacuum_study2}
\rule{\linewidth}{0.2 mm}
\end{center}
\end{figure}
\begin{comment}
\begin{figure}[t] 
\begin{center}
\includegraphics[width=12 cm]{immagini/TDDFT_acell_inv_vacuum3_2.eps}
\caption{Grafici di convergenza delle energie di eccitazione verso stati
         di vuoto rispetto alle dimensioni della supercella.}
\label{fig:vacuum_study3}
\rule{\linewidth}{0.2 mm}
\end{center}
\end{figure}
\end{comment}

Come in precedenza possiamo studiare
l'andamento delle energie di eccitazione rispetto alla grandezza $1/L^2$ per mettere in evidenza
gli stati di vuoto. Nello spettro compaiono
delle serie differenti di energie di eccitazione a seconda di quale sia lo stato di
partenza da cui l'elettrone viene eccitato verso stati di valenza e
di vuoto (fig. \ref{fig:vacuum_study1},\ref{fig:vacuum_study2}).

Decidiamo quindi
di scartare tutte le eccitazioni fortemente sensibili alle condizioni al contorno e, per lo
stesso motivo,
di non considerare le eccitazioni che coinvolgono lo stato $4\sigma$, poich\'{e} anche quest'ultimo,
pur presentando un'autoenergia negativa, ha un comportamento molto pi\`{u} simile a uno stato di
vuoto (\ref{fig:KS_acell_vacuum}) che non ad uno stato di valenza%
\footnote{In proposito a questa considerazione possiamo anche osservare la serie di immagini,
che rappresentano gli orbitali di KS in spazio reale, raffigurate nella parte finale di questo
capitolo da cui \`{e} possibile osservare nuovamente come lo stato $4\sigma$ abbia un comportamento
differente dagli altri orbitali di valenza e molto pi\`{u} simile ad uno stato di vuoto.}.
La scelta di scartare le eccitazioni verso stati di vuoto viene in particolare fatta
in accordo con i dati presentati nel lavoro \cite{Tesi_dot_Myrta} per lo studio della molecola $N_2$.
In tale lavoro infatti viene mostrato come l'approssimazione LDA sia responsabile, a causa del
non corretto andamento asintotico del potenziale di scambio e correlazione, di uno spostamento verso
l'alto e dunque verso il vuoto di tutti i livelli di valenza. In questo modo si ha la
comparsa di una serie di eccitazioni verso stati di vuoto che
scompaiono studiando il sistema con un potenziale che presenti un andamento asintotico corretto
e che quindi l'autrice individua come non fisiche.

\begin{table}[t]
\begin{center}
\begin{tabular}{|l|l|l|l|l|}
\hline
Eccitazione  &  energia  & eccitazione & Configurazione & Contributo                        \\
             &  TDDFT    & principale  & elettronica    & dell'eccitazione                  \\
             &           &             & spaziale       & principale                         \\
\hline
$\Pi$    & 2.40 & $3\sigma\rightarrow 1\pi\ \uparrow$           & $2\sigma^2 1\pi^1$              & 99\% \\
%$\Sigma$ & 4.53 & $3\sigma\rightarrow 4\sigma\ \uparrow$        & $2\sigma^2 4\sigma^1$           & 99\% \\
$\Sigma$ & 5.54 & $2\sigma\rightarrow 3\sigma\ \downarrow$      & $2\sigma^1 3\sigma^2$           & 96\% \\
$\Pi$    & 5.66 & $2\sigma\rightarrow 1\pi\ \uparrow\downarrow$ & $2\sigma^1 3\sigma^1 1\pi^1$    & 99\% \\
$\Pi$    & 7.14 & $2\sigma\rightarrow 1\pi\ \uparrow\downarrow$ & $2\sigma^1 3\sigma^1 1\pi^1$    & 97\% \\
%$\Sigma$ & 7.96 & $2\sigma\rightarrow 4\sigma\ \downarrow$      & $2\sigma^1 3\sigma^1 4\sigma^1$ & 99\% \\
%$\Sigma$ & 8.28 & $2\sigma\rightarrow 4\sigma\ \uparrow$        & $2\sigma^1 3\sigma^1 4\sigma^1$ & 52\% \\
\hline
\end{tabular}
\caption{Schema completo delle eccitazioni a convergenza, le eccitazioni sono state
         scelte osservando i grafici (fig. \ref{fig:TDDFT_acell}, \ref{fig:vacuum_study1},
         \ref{fig:vacuum_study2})
         I parametri dei risultati riportati sono ecut 10 Hartree, 
         nband 50, acell $70\times70\times70$ Bohr}
\label{tab:eccitazioni_all}
\rule{\linewidth}{0.2 mm}
\end{center}
\end{table}

La seconda delle considerazioni ci porta a guardare in modo pi\`{u} dettagliato le eccitazioni
non sensibili alle condizioni al contorno e a cercare un metodo per individuare quali di queste
siano da scartare. A questo scopo scriviamo la tabella con le nostre eccitazioni
(\ref{tab:eccitazioni_all})%
\footnote{Abbiamo scelto di lavorare con una supercella di dimensioni maggiori
         poich\'{e} in questo modo siamo anche in gradi di ottenere l'eccitazione 
         $2\sigma^1 3\sigma^2$ che con questi parametri \`{e} pi\`{u} facilmente individuabile.}.
Come gi\`{a} detto scartiamo tutte le eccitazioni verso lo stato $4\sigma$.
Guardando alle altre vediamo che quelle che incontrerebbero il problema descritto nel capitolo
precedente sono le eccitazioni in cui abbiamo $3$ stati parzialmente occupati.
Osserviamo come per questo tipo di eccitazione ne compaiono sempre due che presentano lo
stesso tipo di occupazione degli orbitali spaziali. Di queste due dunque, immaginiamo che una
sar\`{a} l'eccitazione verso uno stato con spin totale maggiore e l'altra l'eccitazione verso uno
stato con spin totale $S=1/2$ come lo stato fondamentale del sistema utilizzando il fatto che
$\Delta S=0,1$ e rifacendoci a quanto detto nel capitolo precedente.

Ci affidiamo ora alla regola di Hund ed individuiamo l'eccitazione
a spin maggiore, cio\`{e} quella che vogliamo scartare, come quella ad energia minore e dunque 
tra le due eccitazioni $\Pi=5.657$ e $\Pi=7.145$ scegliamo di assegnare alla seconda
$S=1/2$.

\begin{comment}
Riscriviamo quindi ora la tabella con le sole eccitazioni che riteniamo da considerare
(tab. \ref{tab:eccitazioni_scelte}).
\begin{table}[t]
\begin{center}
\begin{tabular}{|l|l|l|l|l|}
\hline
Eccitazione  &  energia  & eccitazione & Configurazione & Contributo                        \\
             &           & principale  & elettronica    & dell'eccitazione                  \\
             &           &             & spaziale       & principale                         \\
\hline
$\Pi$    & 2.40 & $3\sigma\rightarrow 1\pi\ \uparrow$           & $2\sigma^2 1\pi^1$              & 99\% \\
$\Sigma$ & 5.54 & $2\sigma\rightarrow 3\sigma\ \downarrow$      & $2\sigma^1 3\sigma^2$           & 96\% \\
$\Pi$    & 7.14 & $2\sigma\rightarrow 1\pi\ \uparrow\downarrow$ & $2\sigma^1 3\sigma^1 1\pi^1$    & 97\% \\
\hline
\end{tabular}
\caption{Schema delle eccitazioni scelte}
\label{tab:eccitazioni_scelte}
\rule{\linewidth}{0.2 mm}
\end{center}
\end{table}
\end{comment}

Osserviamo che, riguardo ai due problemi segnalati, Casida sembra ignorare il
primo, probabilmente poich\'{e} lavorando con una base localizzata non \`{e} in gradi di distinguere
stati di vuoto da stati di valenza%
\footnote{Nella referenza \cite{Casida3} Casida individua in realt\`{a} il problema e cerca
di correggerlo studiando le energie di eccitazione ottenute grazie ad un potenziale con
andamento asintotico corretto. Tale studio non viene poi pi\`{u} menzionato nella referenza
sucessiva \cite{Casida4}.}.
Per il secondo invece elabora un modo molto pi\`{u} raffinato di
quello da me proposto in questo lavoro (facendo riferimento alla regola di Hund) e chiamato
``contaminazione di spin'' \cite{Casida4}.
Guardando proprio allo studio di tale parametro in effetti,
tra le eccitazioni da lui scelte, individua come maggiormante ``contaminata'' la stessa che noi
abbiamo scartato supponendo fosse a spin maggiore.

%*************************************************************************************************
\subsection{Altri test di convergenza}
Abbiamo riportato i test di convergenza fatti per il computo delle autoenergie di KS e quindi lo
studio delle energie di eccitazione rispetto alle dimensioni della supercella. Anche per
l'analisi delle eccitazioni del sistema sosno stati fatti dei test rispetto all'energia di
cut-off e rispetto ad un nuovo parametro, il numero di stati considerati per la costruzione
della matrice $\Omega$, che nel computo dello stato fondamentale del sistema non compariva.
\begin{figure}[t] 
%\rule{\linewidth}{0.2 mm} 
\begin{center}
\subfigure[TDDFT: convergenza rispetto a ecut]{\includegraphics[width=6.8 cm]{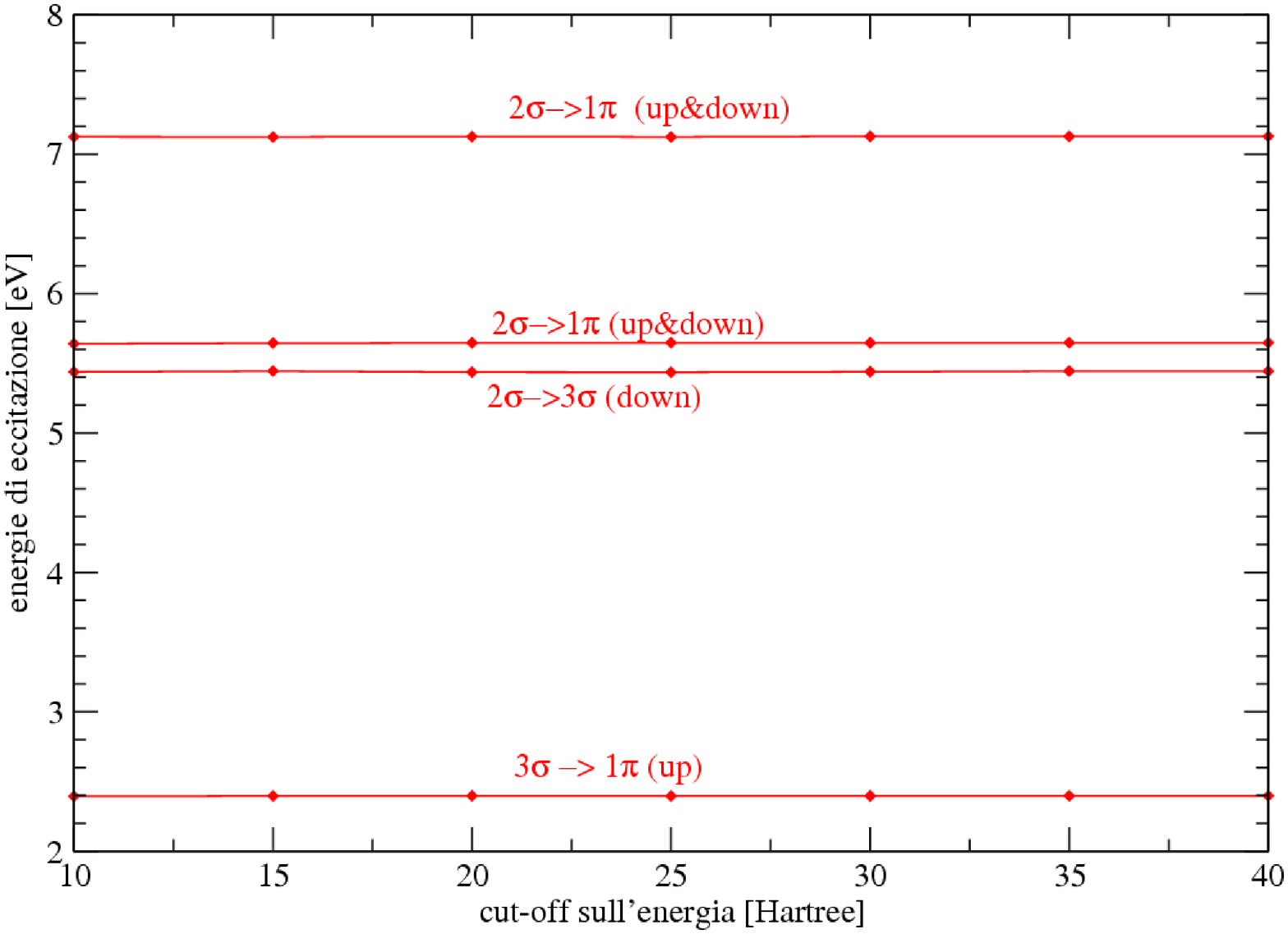}}
\subfigure[Particolare del grafico (a)]{\includegraphics[width=6.8 cm]{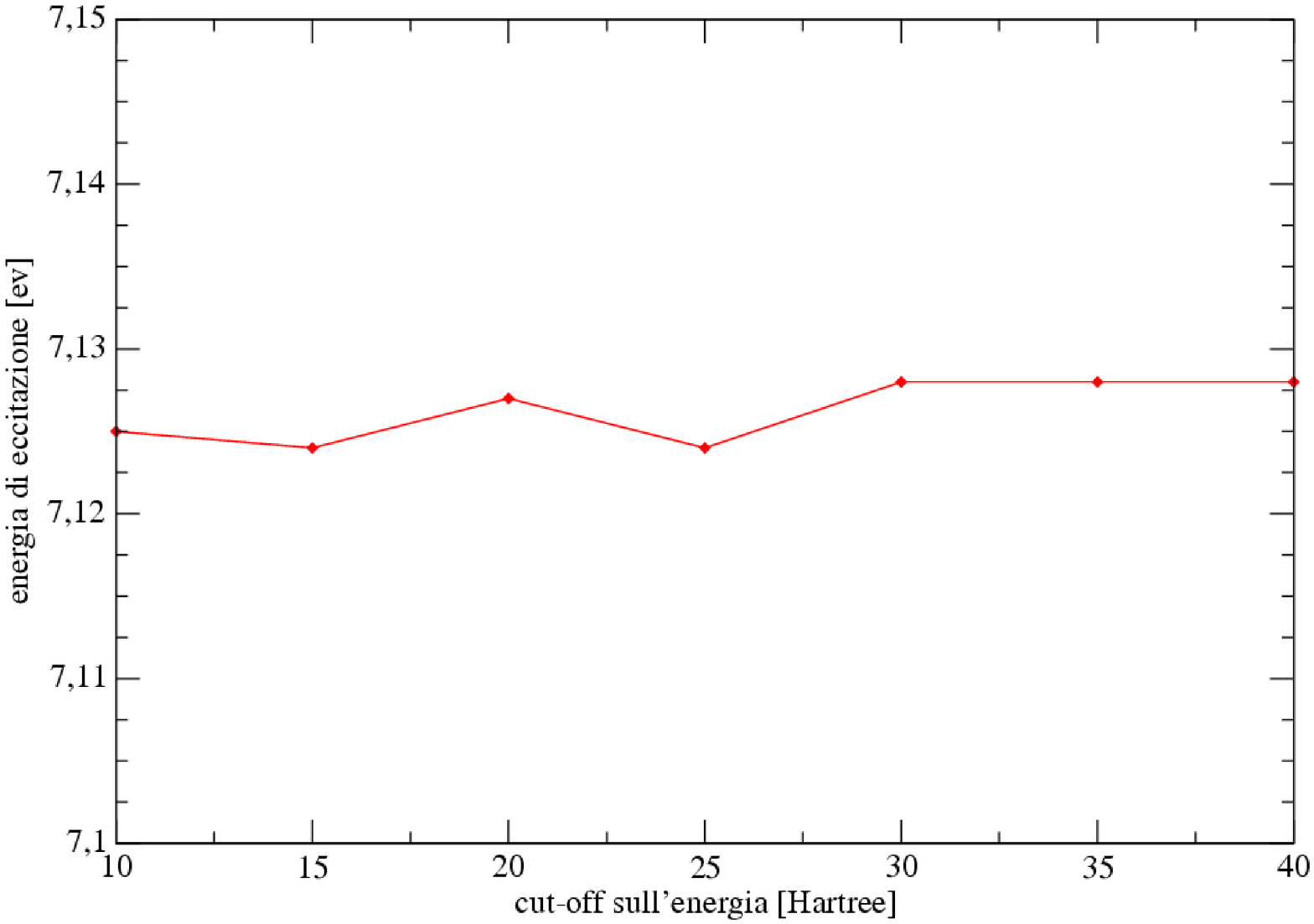}}
\subfigure[TDDFT: convergenza risetto al numero di stati]{\includegraphics[width=6.8 cm]{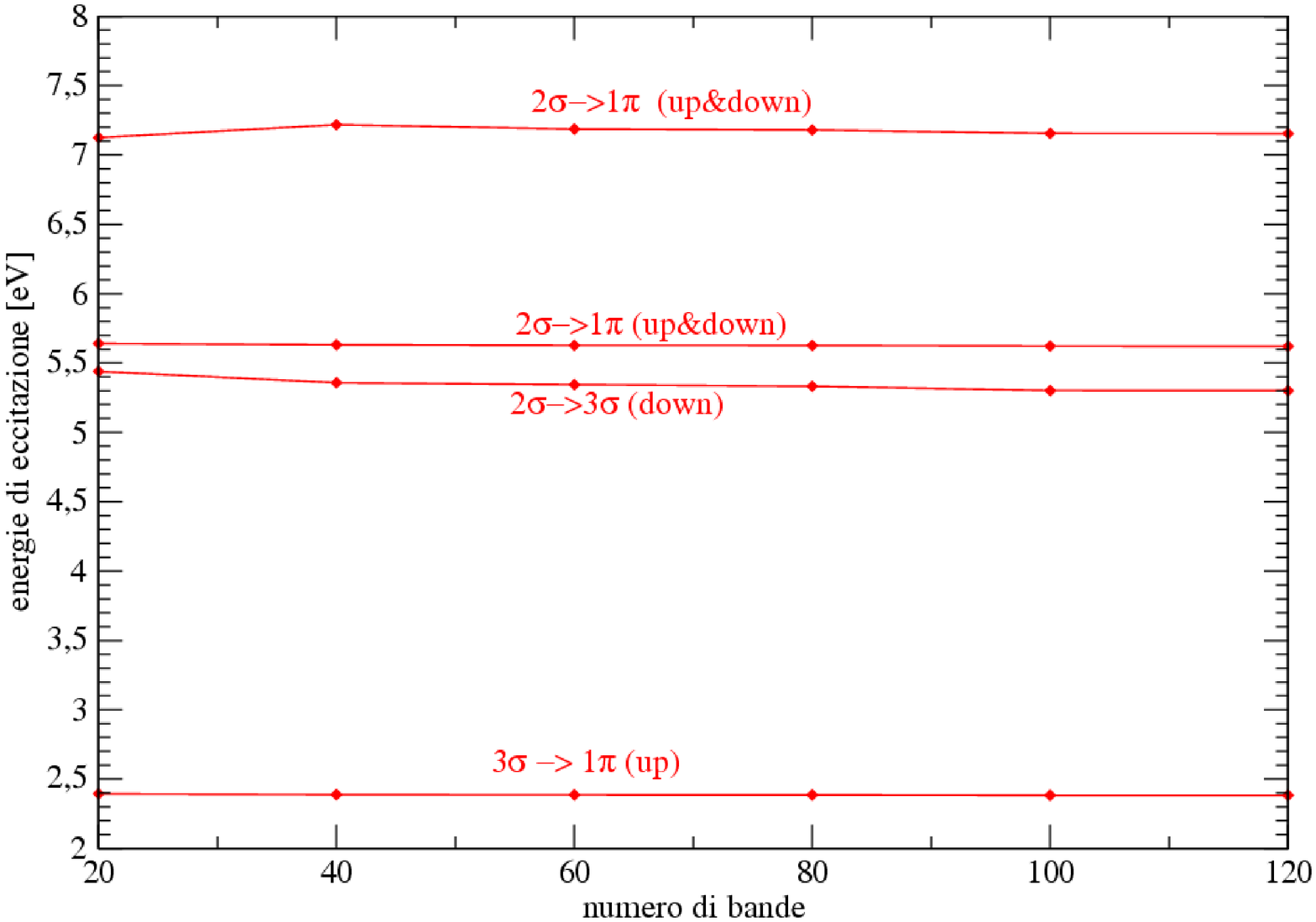}}
\subfigure[Particolare del grafico (c)]{\includegraphics[width=6.8 cm]{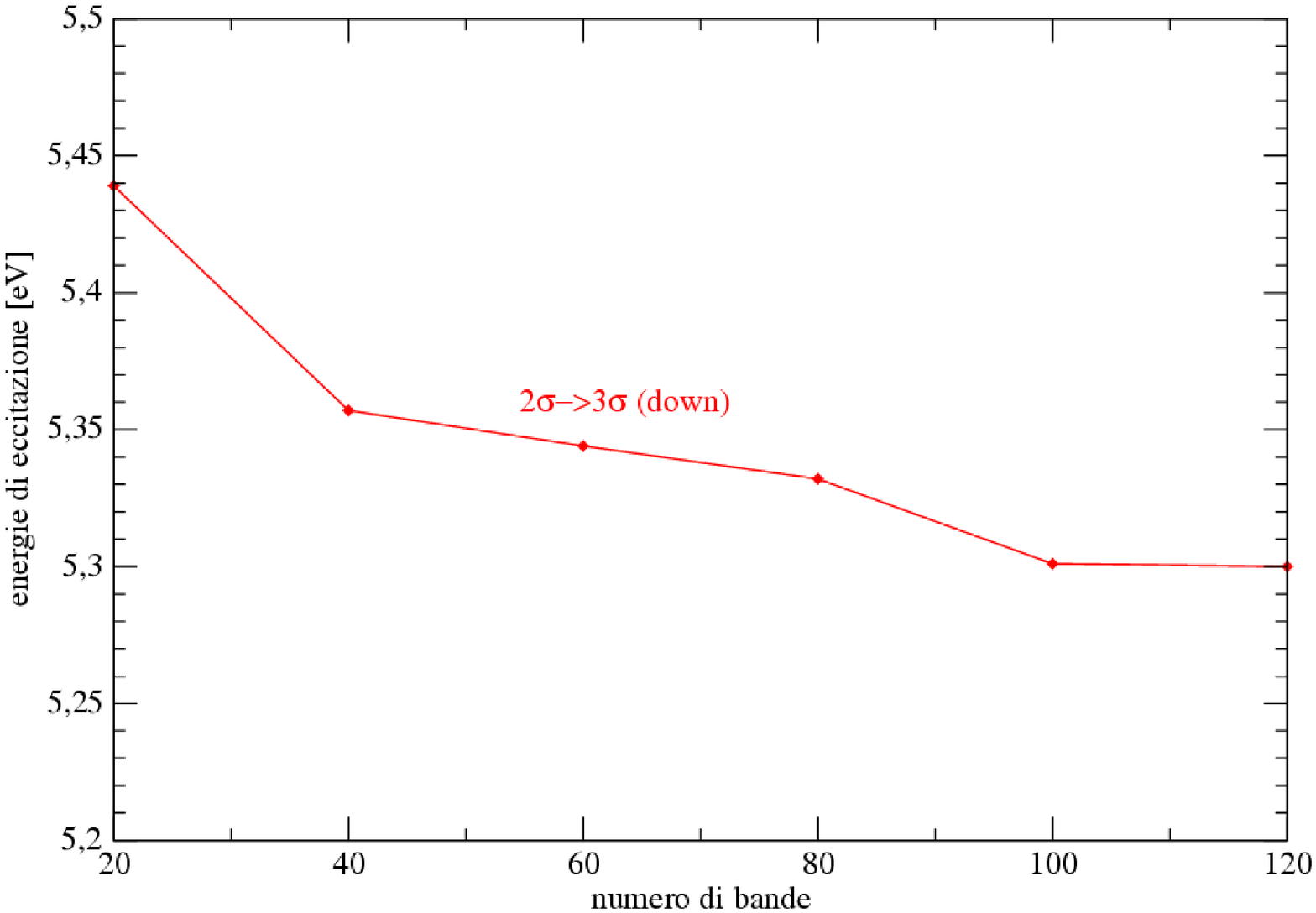}}
\caption{Test di convergenza per le energie di eccitazione rispetto all'energia
         di cut-off per la base di onde piane e rispetto al numero di stati considerati
         per la costruzione della matrice $\Omega$}
\label{fig:TDDFT_convergence}
\rule{\linewidth}{0.2 mm}
\vspace{-0.3 cm}
\end{center}
\end{figure}
Dai grafici (fig. \ref{fig:TDDFT_convergence})  
possiamo osservare come i risultati siano a convergenza per entrambi i parametri. La dimensione
della supercella resta, dal nostro punto di vista, il paramentro pi\`{u} importante per gli studi
di convergenza delle energie di eccitazione nell'ambito della TDDFT, soprattutto per l'individuazione
degli stati di vuoto.

%*************************************************************************************************
%*************************************************************************************************
%*************************************************************************************************
\section{Considerazioni finali}
Riportiamo ora una tabella in cui confrontiamo le eccitazioni considerate valide con i dati di
riferimento nella tabella (\ref{tab:eccitazioni_finale})

\begin{figure}[t] 
%\rule{\linewidth}{0.2 mm} 
\vspace{-0.5 cm}
\begin{center}
\includegraphics[width=13 cm]{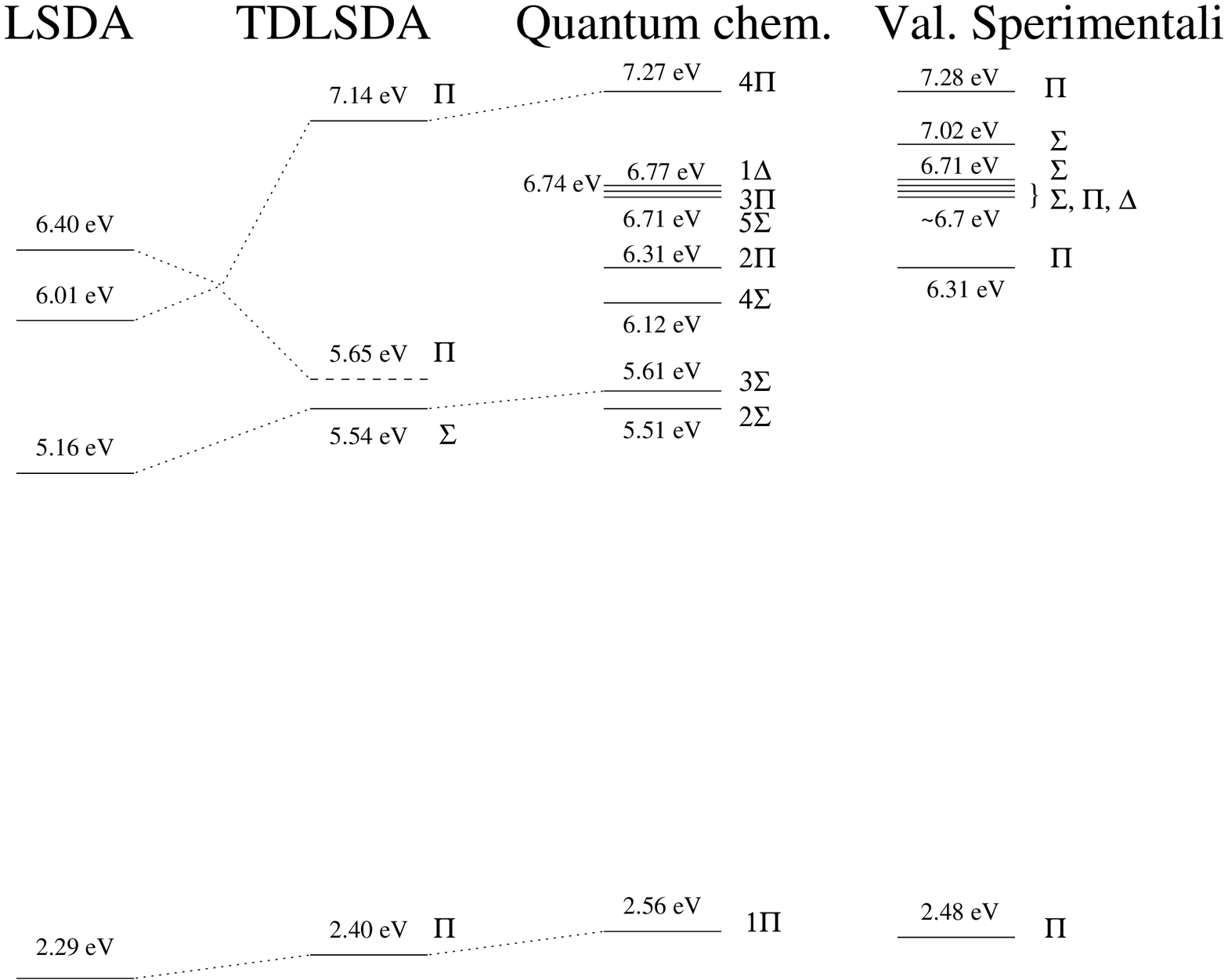}
\caption{Raffronto delle energie di eccitazione calcolate come differenze tra
         autoenergie di KS (ordine zero della nostra teoria), come energie di
         eccitazione in TDDFT (approssimazione LSDA/TDLSDA) e valori teorici e
         sperimentali presenti in letteratura \cite{Casida3,Sper}.
         L'eccitazione che rappresentiamo in 
         modo tratteggiato \`{e} quella che abbiamo individuato come da scartare.}
\label{fig:conclusioni}
\rule{\linewidth}{0.2 mm}
\vspace{-1 cm}
\end{center}
\end{figure}

\begin{table}[t]
\begin{center}
\begin{tabular}{|l||l|l||l|l|}
\hline
Eccitazione  & Dati     & Dati                & Presente & Calcoli              \\
& sperim. \cite{Casida3}& sperim. \cite{Sper} & lavoro   & quantum              \\
             &          &                     &          & chem. \cite{PTN_BeH} \\              
\hline
$\Pi$       &  2.484   &  2.48   &   2.40  & 2.56          \\
$\Sigma$    &          &         &   5.54  & 5.51          \\
$\Sigma$    &          &         &         & 5.61          \\
$\Sigma$    &          &         &         & 6.12          \\
$\Pi$       &  6.317   &  6.31   &         & 6.31          \\
$\Sigma$    &          &$\sim$6.7&         & 6.71          \\
$\Pi$       &          &$\sim$6.7&         & 6.74          \\
$\Delta$    &          &$\sim$6.7&         & 6.77          \\
$\Sigma$    &          &  6.71   &         &               \\
$\Sigma$    &          &  7.02   &         &               \\
$\Pi$       &  7.460   &  7.28   &   7.14  & 7.27          \\
\hline
\end{tabular}
\caption{Raffronto tra le energie di eccitazione (in eV) ottenute nel presente lavoro con
         dati sperimentali e teorici presenti in letteratura.}
\rule{\linewidth}{0.2 mm}
\end{center}
\label{tab:eccitazioni_finale}
\end{table}
Segnaliamo che lo scegliere per un valore ottenuto una determinata eccitazione
sia stato fatto in modo arbitrario, non avendo questa volta alcuna indicazione sul tipo di
eccitazione, salvo la distinzione $\Sigma$, $\Pi$. Sembra comunque promettente l'ottimo accordo
che in questo modo riusciamo ad ottenere con i valori di riferimento con una differenza media, per le 
3 eccitazioni calcolate, di circa 0.1-0.2 eV.
Uno schema riassuntivo
dei risultati ottenuti \`{e} infine riprodotto nella figura (fig \ref{fig:conclusioni})

Concludiamo l'analisi dei dati sottolineando come nel lavoro svolto le due approssimazioni eseguite,
l'approssimazione LDA
(e quindi l'errato andamento del potenziale di scambio e correlazione nella regione asintotica) per
il computo dello stato fondamentale e l'approssimazione adiabatica
per il computo del kernel hanno portato come conseguenza il
fatto che non siamo in grado di individuare la maggior parte delle eccitazioni del sistema. Sui tentativi
di migliorare l'approssimazione LDA abbiamo visto come una soluzione viene presentata in particolare
nella referenza \cite{Tesi_dot_Myrta} o come un approccio possibile sia quello di utilizzare l'equazione
di Sham-Schluter presentata all'interno di questa tesi (\ref{Sham-Schluter}).
Tale problema riteniamo sia stato gi\`{a} studiato in molte
altre occasioni nel computo dello stato fondamentale di differenti sistemi e ben delineato. Dal nostro
punto di vista crediamo possa essere un obiettivo di non sostanziale difficolt\`{a} implementare una delle
possibili soluzioni all'interno del software utilizzato \cite{Abinit}.

Il problema del miglioramento dell'approssimazione adiabatica invece \`{e} ad oggi uno degli argomenti di cui
si occupa attivamente la ricerca. Abbiamo dunque scelto di mostrare, nel prossimo capitolo,
un possibile approccio che permetterebbe di giungere alla soluzione di tale difficolt\`{a}
e che viene illustrato nelle referenze \cite{Tesi_dot_Fabien,Fabien_kernel}.
Dal nostro punto di vista l'implementazione di questa soluzione, o di altre possibili
forme dipendenti dalla frequenza del kernel della TDDFT, resta una sfida di maggiore
difficolt\`{a}.

\input{immagini.tex}
\clearpage

\newpage
\section{Il kernel della TDDFT}
\input{kernel.tex}

%% file: immagini.tex
%*************************************************
%Isosuperfici della densità del sistema
%*************************************************
\begin{figure}[t] 
\vspace{-0.5 cm}
\begin{center}
\subfigure[iso 0.041]{\includegraphics[width=5 cm, angle=-90]{immagini/BeH_ac16_0.041.epsf}}
\subfigure[iso 0.019]{\includegraphics[width=5 cm, angle=-90]{immagini/BeH_ac16_0.019.epsf}}
\subfigure[iso 0.013]{\includegraphics[width=5 cm, angle=-90]{immagini/BeH_ac16_0.013.epsf}}
\subfigure[iso 5e-6]{\includegraphics[width=5 cm, angle=-90]{immagini/BeH_ac16_5e-6.epsf}}
\subfigure[iso 5e-6]{\includegraphics[width=5 cm, angle=-90]{immagini/BeH_ac30_5e-6.epsf}}
\subfigure[iso 1e-8]{\includegraphics[width=5 cm, angle=-90]{immagini/BeH_ac30_1e-8.epsf}}
\caption{Densit\`{a} elettronica della molecola BeH.
         Isosuperfici a differenti densit\`{a} 0.041, 0.019, 0.013, 5.0e-6.
         \`E possibile notare come la nube elettronica sia principalmente localizzata
         attorno all'atomo di sinistra (atomo di Be). La figura (c) mostra
         la coda della densit\`{a} che va al di fuori delle dimensioni della supercella
         risentendo quindi degli effetti al contorno. Questa coda ha un valore molto piccolo
         ma risulta importante nel computo delle eccitazioni del sistema poich\'{e} viene
         moltiplicata per il kernel della TDDFT che \`{e} divergente nelle regioni di
         spazio a bassa densit\`{a}. In effetti nel lavoro svolto la convergenza delle
         energie di eccitazioni \`{e} risultata problematica soprattutto rispetto alle
         dimensioni della super-cella. La supercella visualizzata ha una dimensione di
         16 Bhor di lato. Nelle figure (d) ed (e) \`{e} invece possibile vedere come una
         supercella delle dimensioni di 30 Bohr sia sufficiente per eliminare gli effetti
         al contorno sulla distribuzione della densit\`{a}}
\label{fig:BeH_view} 
\end{center}
\vspace{-0.5 cm}
\end{figure}
\end{comment}

%*******************************************************
% Stato 2 sigma, banda 1
%*******************************************************
\begin{figure}[t]
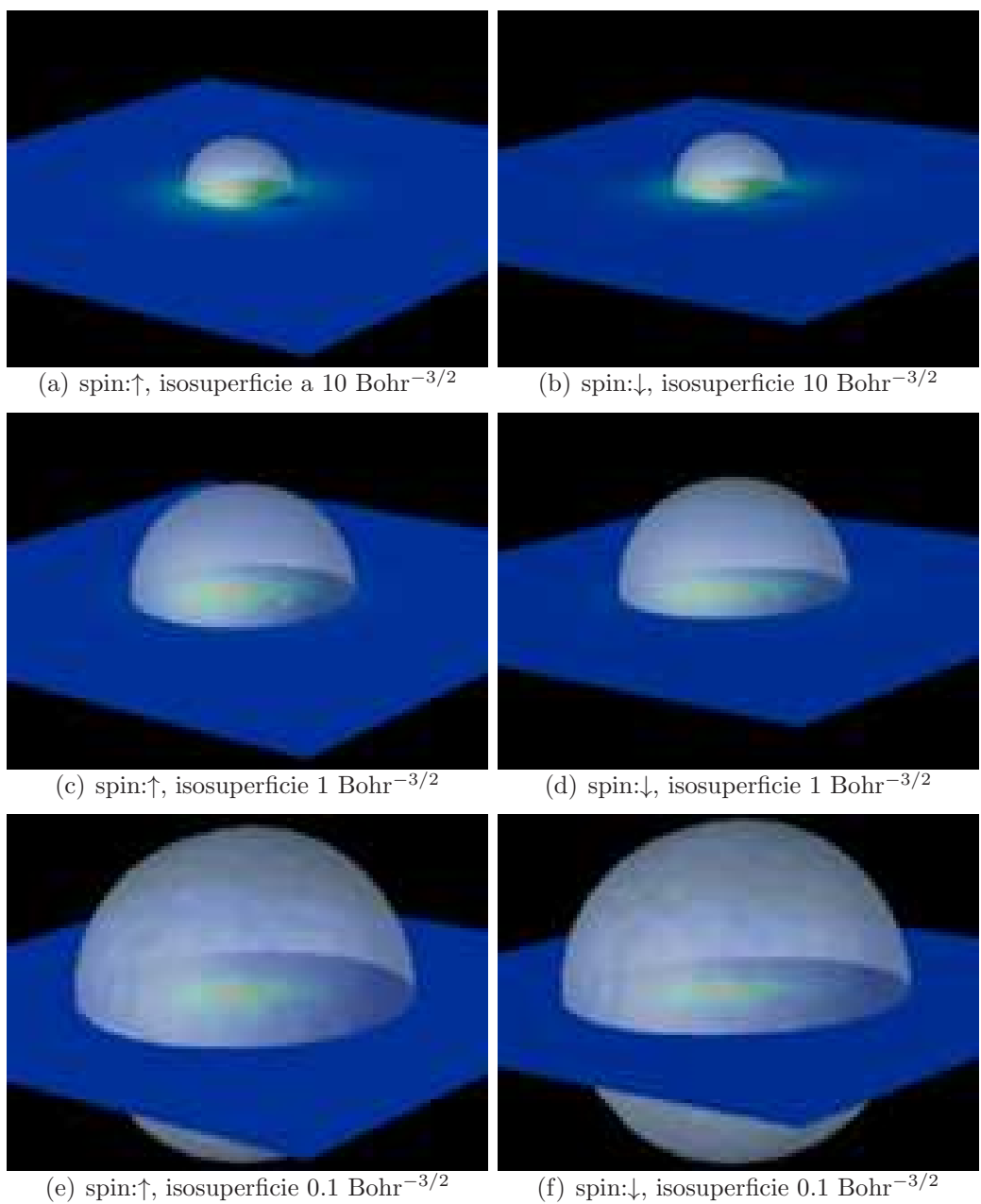
 
\begin{center}
\subfigure[spin:$\uparrow$, isosuperficie a $10$ Bohr$^{-3/2}$]{\includegraphics[width=5 cm, angle=-90]{immagini/BeH_b1s1_10.epsf}}
\subfigure[spin:$\downarrow$, isosuperficie $10$ Bohr$^{-3/2}$]{\includegraphics[width=5 cm, angle=-90]{immagini/BeH_b1s2_10.epsf}}
\subfigure[spin:$\uparrow$, isosuperficie $1$ Bohr$^{-3/2}$]{\includegraphics[width=5 cm, angle=-90]{immagini/BeH_b1s1_1.epsf}}
\subfigure[spin:$\downarrow$, isosuperficie $1$ Bohr$^{-3/2}$]{\includegraphics[width=5 cm, angle=-90]{immagini/BeH_b1s2_1.epsf}}
\subfigure[spin:$\uparrow$, isosuperficie $0.1$ Bohr$^{-3/2}$]{\includegraphics[width=5 cm, angle=-90]{immagini/BeH_b1s1_0.1.epsf}}
\subfigure[spin:$\downarrow$, isosuperficie $0.1$ Bohr$^{-3/2}$]{\includegraphics[width=5 cm, angle=-90]{immagini/BeH_b1s2_0.1.epsf}}
\caption{Stato $2\sigma$;
         Questa serie di figure, cos\`{i} come quelle delle pagiene successive, mostrano differenti
         isosuperfici della parte reale delle funzioni d'onda (la parte immaginaria \`{e} nulla).
         \`{E} possibile notare la differenza tra gli stati che abbiamo interpretato come di valenza ($2\sigma$,
         $3 \sigma$ e $4 \pi$) rispetto agli stati di vuoto. In particolare le funzioni d'onda di valenza sono
         localizzate nei pressi degli atomi, mentre gli stati di vuoto risentono delle condizioni al
         contorno. Tutte le immagini sono realizzate con una supercella cubica di 30 Bohr di lato} 
\label{fig:BeH_2sigma} 
\end{center}
\end{figure}

%******************************************************
%Stato 3 sigma, banda 2
%******************************************************
\begin{figure}[t] 
\begin{center}
\subfigure[spin:$\uparrow$, isosuperficie $-5$ Bohr$^{-3/2}$]{\includegraphics[width=5 cm, angle=-90]{immagini/BeH_b2s1_-5.epsf}}
\subfigure[spin:$\downarrow$, isosuperficie $-5$ Bohr$^{-3/2}$]{\includegraphics[width=5 cm, angle=-90]{immagini/BeH_b2s2_-5.epsf}}
\subfigure[spin:$\uparrow$, isosuperficie $-0.2$ Bohr$^{-3/2}$]{\includegraphics[width=5 cm, angle=-90]{immagini/BeH_b2s1_-0.2.epsf}}
\subfigure[spin:$\downarrow$, isosuperficie $-0.2$ Bohr$^{-3/2}$]{\includegraphics[width=5 cm, angle=-90]{immagini/BeH_b2s2_-0.2.epsf}}
\subfigure[spin:$\uparrow$, isosuperficie $0.2$ Bohr$^{-3/2}$]{\includegraphics[width=5 cm, angle=-90]{immagini/BeH_b2s1_0.2.epsf}}
\subfigure[spin:$\downarrow$, isosuperficie $0.2$ Bohr$^{-3/2}$]{\includegraphics[width=5 cm, angle=-90]{immagini/BeH_b2s2_0.2.epsf}}
\subfigure[spin:$\uparrow$, isosuperficie $5$ Bohr$^{-3/2}$]{\includegraphics[width=5 cm, angle=-90]{immagini/BeH_b2s1_5.epsf}}
\subfigure[spin:$\downarrow$, isosuperficie $5$ Bohr$^{-3/2}$]{\includegraphics[width=5 cm, angle=-90]{immagini/BeH_b2s2_5.epsf}}
\caption{Stato $3\sigma$}
\label{fig:BeH_3sigma}
\end{center}
\end{figure}

%***********************************************
% Stati 1 pi, bande 3 e 4
%***********************************************
% banda 3
\begin{figure}[t] 
\begin{center}
\subfigure[spin:$\uparrow$, isosuperficie $-5$ Bohr$^{-3/2}$]{\includegraphics[width=5 cm, angle=-90]{immagini/BeH_b3s1_-5.epsf}}
\subfigure[spin:$\downarrow$, isosuperficie $-5$ Bohr$^{-3/2}$]{\includegraphics[width=5 cm, angle=-90]{immagini/BeH_b3s2_-5.epsf}}
\subfigure[spin:$\uparrow$, isosuperficie $-0.5$ Bohr$^{-3/2}$]{\includegraphics[width=5 cm, angle=-90]{immagini/BeH_b3s1_-0.5.epsf}}
\subfigure[spin:$\downarrow$, isosuperficie $-0.5$ Bohr$^{-3/2}$]{\includegraphics[width=5 cm, angle=-90]{immagini/BeH_b3s2_-0.5.epsf}}
\subfigure[spin:$\uparrow$, isosuperficie $0.5$ Bohr$^{-3/2}$]{\includegraphics[width=5 cm, angle=-90]{immagini/BeH_b3s1_0.5.epsf}}
\subfigure[spin:$\downarrow$, isosuperficie $0.5$ Bohr$^{-3/2}$]{\includegraphics[width=5 cm, angle=-90]{immagini/BeH_b3s2_0.5.epsf}}
\subfigure[spin:$\uparrow$, isosuperficie $5$ Bohr$^{-3/2}$]{\includegraphics[width=5 cm, angle=-90]{immagini/BeH_b3s1_5.epsf}}
\subfigure[spin:$\downarrow$, isosuperficie $5$ Bohr$^{-3/2}$]{\includegraphics[width=5 cm, angle=-90]{immagini/BeH_b3s2_5.epsf}}
\caption{Stato $1\pi$}
\label{fig:BeH_1pix} 
\end{center}
\end{figure}

\begin{comment}
%**********************************************
% banda 4
\begin{figure}[t] 
\begin{center}
\subfigure[spin:$\uparrow$, iso -5]{\includegraphics[width=5 cm, angle=-90]{immagini/BeH_b4s1_-5.epsf}}
\subfigure[spin:$\downarrow$, iso -5]{\includegraphics[width=5 cm, angle=-90]{immagini/BeH_b4s2_-5.epsf}}
\subfigure[spin:$\uparrow$, iso -0.5]{\includegraphics[width=5 cm, angle=-90]{immagini/BeH_b4s1_-0.5.epsf}}
\subfigure[spin:$\downarrow$, iso -0.5]{\includegraphics[width=5 cm, angle=-90]{immagini/BeH_b4s2_-0.5.epsf}}
\subfigure[spin:$\uparrow$, iso 0.5]{\includegraphics[width=5 cm, angle=-90]{immagini/BeH_b4s1_0.5.epsf}}
\subfigure[spin:$\downarrow$, iso 0.5]{\includegraphics[width=5 cm, angle=-90]{immagini/BeH_b4s2_0.5.epsf}}
\subfigure[spin:$\uparrow$, iso 5]{\includegraphics[width=5 cm, angle=-90]{immagini/BeH_b4s1_5.epsf}}
\subfigure[spin:$\downarrow$, iso 5]{\includegraphics[width=5 cm, angle=-90]{immagini/BeH_b4s2_5.epsf}}
\caption{Stato $1\pi_y$}
\label{fig:BeH_1pi_y} 
\end{center}
\end{figure}
\end{comment}

%*******************************************************
% Stati di vuoto
%*******************************************************
% banda 5, spin up
\begin{figure}[t]
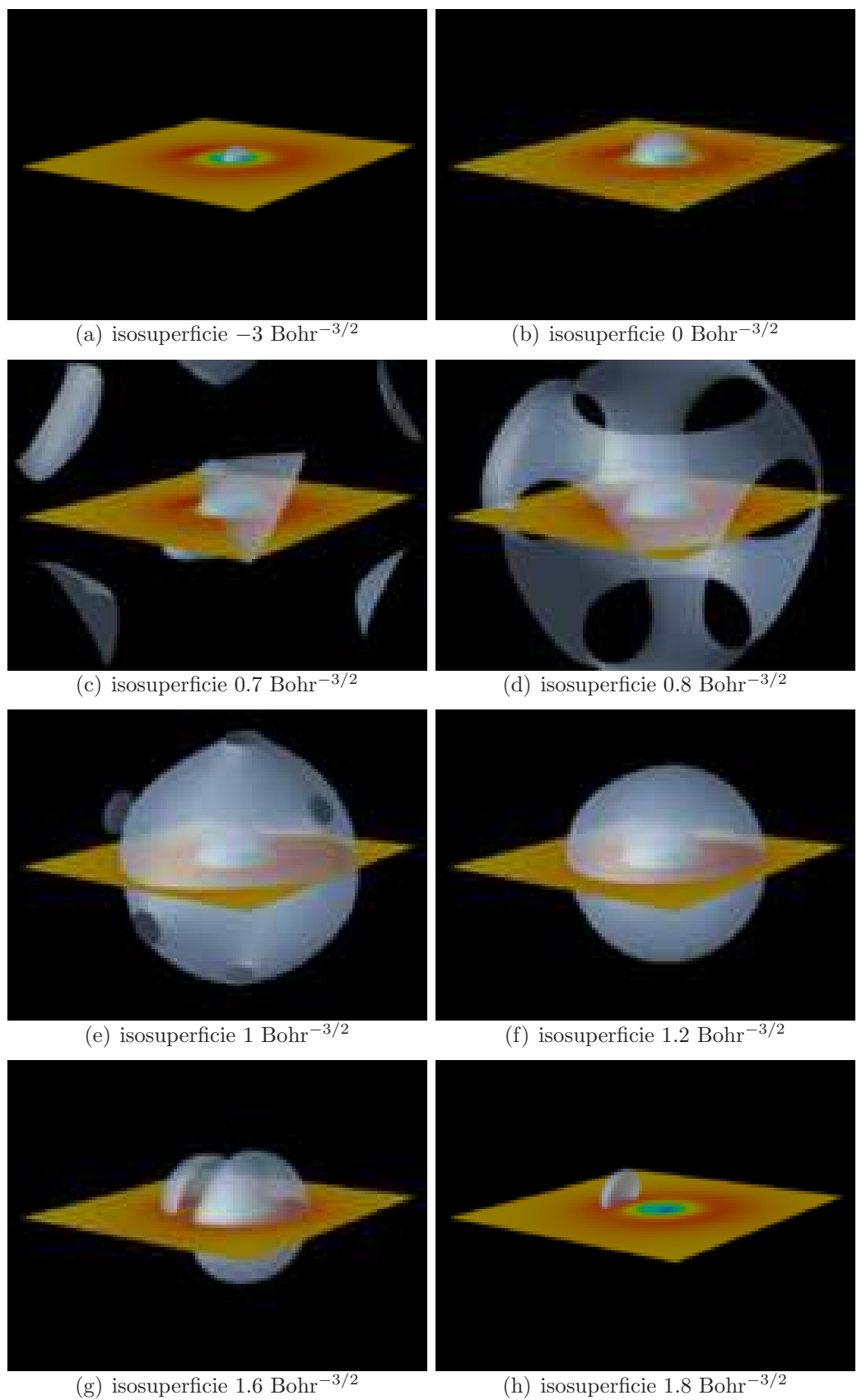
 
\begin{center}
\subfigure[isosuperficie $-3$ Bohr$^{-3/2}$]{\includegraphics[width=5 cm, angle=-90]{immagini/BeH_b5s1_-3.epsf}}
\subfigure[isosuperficie $0$ Bohr$^{-3/2}$]{\includegraphics[width=5 cm, angle=-90]{immagini/BeH_b5s1_0.epsf}}
\subfigure[isosuperficie $0.7$ Bohr$^{-3/2}$]{\includegraphics[width=5 cm, angle=-90]{immagini/BeH_b5s1_0.7.epsf}}
\subfigure[isosuperficie $0.8$ Bohr$^{-3/2}$]{\includegraphics[width=5 cm, angle=-90]{immagini/BeH_b5s1_0.8.epsf}}
\subfigure[isosuperficie $1$ Bohr$^{-3/2}$]{\includegraphics[width=5 cm, angle=-90]{immagini/BeH_b5s1_1.epsf}}
\subfigure[isosuperficie $1.2$ Bohr$^{-3/2}$]{\includegraphics[width=5 cm, angle=-90]{immagini/BeH_b5s1_1.2.epsf}}
\subfigure[isosuperficie $1.6$ Bohr$^{-3/2}$]{\includegraphics[width=5 cm, angle=-90]{immagini/BeH_b5s1_1.6.epsf}}
\subfigure[isosuperficie $1.8$ Bohr$^{-3/2}$]{\includegraphics[width=5 cm, angle=-90]{immagini/BeH_b5s1_1.8.epsf}}
\caption{Stato $4\sigma \uparrow$, questo stato pur presentando autoenergia negativa
         appare sensibile alle condizioni al contorno con un comportamento simile ad uno stato
         di vuoto.}
\label{fig:BeH_vacuum_5up} 
\end{center}
\end{figure}

%*******************************************************
% banda 5, spin down
\begin{figure}[t]
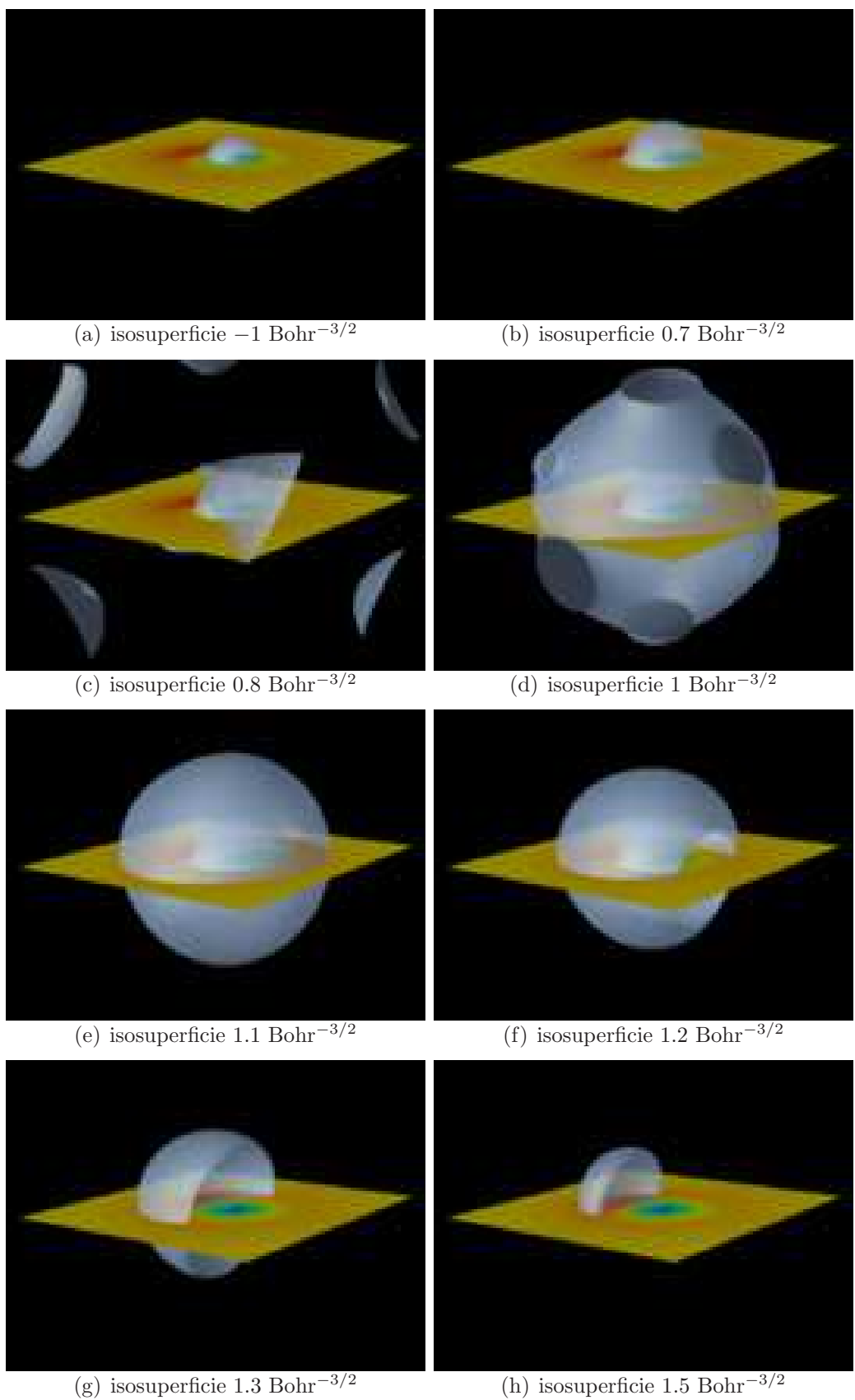
 
\begin{center}
\subfigure[isosuperficie $-1$ Bohr$^{-3/2}$]{\includegraphics[width=5 cm, angle=-90]{immagini/BeH_b5s2_-1.epsf}}
\subfigure[isosuperficie $0.7$ Bohr$^{-3/2}$]{\includegraphics[width=5 cm, angle=-90]{immagini/BeH_b5s2_0.7.epsf}}
\subfigure[isosuperficie $0.8$ Bohr$^{-3/2}$]{\includegraphics[width=5 cm, angle=-90]{immagini/BeH_b5s2_0.8.epsf}}
\subfigure[isosuperficie $1$ Bohr$^{-3/2}$]{\includegraphics[width=5 cm, angle=-90]{immagini/BeH_b5s2_1.epsf}}
\subfigure[isosuperficie $1.1$ Bohr$^{-3/2}$]{\includegraphics[width=5 cm, angle=-90]{immagini/BeH_b5s2_1.1.epsf}}
\subfigure[isosuperficie $1.2$ Bohr$^{-3/2}$]{\includegraphics[width=5 cm, angle=-90]{immagini/BeH_b5s2_1.2.epsf}}
\subfigure[isosuperficie $1.3$ Bohr$^{-3/2}$]{\includegraphics[width=5 cm, angle=-90]{immagini/BeH_b5s2_1.3.epsf}}
\subfigure[isosuperficie $1.5$ Bohr$^{-3/2}$]{\includegraphics[width=5 cm, angle=-90]{immagini/BeH_b5s2_1.5.epsf}}
\caption{Stato $4\sigma \downarrow$, vedi il commento per la serie di immagini dello
         stato $4\sigma \uparrow$.}
\label{fig:BeH_vacuum_5down} 
\end{center}
\end{figure}

%*******************************************************
% banda 6, spin up
\begin{figure}[t]
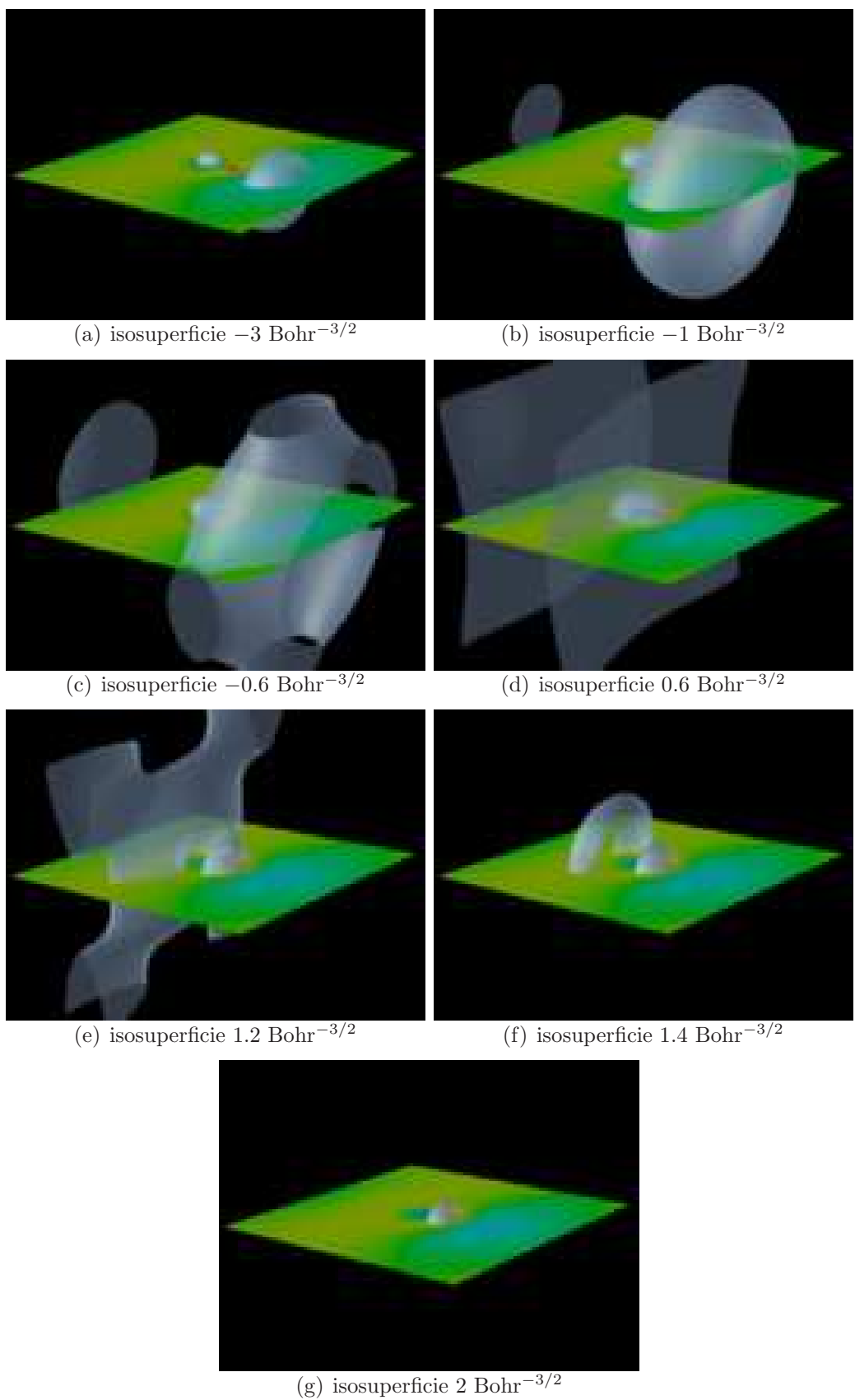
 
\begin{center}
\subfigure[isosuperficie $-3$ Bohr$^{-3/2}$]{\includegraphics[width=5 cm, angle=-90]{immagini/BeH_b6s1_-3.epsf}}
\subfigure[isosuperficie $-1$ Bohr$^{-3/2}$]{\includegraphics[width=5 cm, angle=-90]{immagini/BeH_b6s1_-1.epsf}}
\subfigure[isosuperficie $-0.6$ Bohr$^{-3/2}$]{\includegraphics[width=5 cm, angle=-90]{immagini/BeH_b6s1_-0.6.epsf}}
\subfigure[isosuperficie $0.6$ Bohr$^{-3/2}$]{\includegraphics[width=5 cm, angle=-90]{immagini/BeH_b6s1_0.6.epsf}}
\subfigure[isosuperficie $1.2$ Bohr$^{-3/2}$]{\includegraphics[width=5 cm, angle=-90]{immagini/BeH_b6s1_1.2.epsf}}
\subfigure[isosuperficie $1.4$ Bohr$^{-3/2}$]{\includegraphics[width=5 cm, angle=-90]{immagini/BeH_b6s1_1.4.epsf}}
\subfigure[isosuperficie $2$ Bohr$^{-3/2}$]{\includegraphics[width=5 cm, angle=-90]{immagini/BeH_b6s1_2.epsf}}
%\subfigure[iso 5]{\includegraphics[width=5 cm, angle=-90]{immagini/BeH_b6s1_5.epsf}}
\caption{Stato $5\sigma \uparrow$, esempio del comportamento di uno sato di vuoto.}
\label{fig:BeH_vacuum_6up} 
\end{center}
\end{figure}

%******************************************************************
%\end{document}

%% file: kernel.tex
\subsection{Introduzione}
Nella presente sezione presenteremo quello che \`{e} un possibile approccio per superare
le difficolt\`{a} da noi incontrate nell'ambito dell'approssimazione adiabatica per il
kernel della TDDFT.
Tale approccio \`{e} esposto nelle referenze \cite{Fabien_kernel,Tesi_dot_Fabien}
e nasce dal tentativo di scrivere un kernel a due punti per la teoria MBPT.
Esporremo qui solo la formulazione esatta poich\'{e} siamo semplicemente interessati
a mostrare come sia possibile superare le difficolt\`{a} da noi incontrate e non a
svolgere calcoli espliciti.

%******************************************************************************************
%******************************************************************************************
%******************************************************************************************
%\section{L'equazione di S.S. per il kernel della TDDFT}
\subsection{Un kernel a due punti per la teoria molti corpi}
Vogliamo cercare di formulare un'equazione che leghi il kernel della TDDFT con
il kernel della MBPT in modo analogo a quanto fatto per il potenziale
di scambio e correlazione e la self energia, eq. (\ref{Sham-Schluter}), per
la teoria dello stato fondamentale. A differenza di quanto
fatto in quest'ultimo caso, in cui si considera la self-energia esatta del sistema per
ottenere il potenziale, la grandezza many-body che verr\`{a} presa in considerazione
non sar\`{a} il kernel dell'equazione di Bethe-Salpeter ma una sua versione a due punti.

Per ottenere tale kernel ripartiamo da quanto mostrato nel capitolo $5$ nella
costruzione dell'equazione di Bethe-Salpeter. In particolare ripartiamo dell'equazione
(\ref{Polarizzazione-Dyson}), questa volta per\`{o} lavoriamo con la polarizzazione
irriducibile
\begin{multline}
\Pi^{\star}(1,2)= G(1,2)G(2,1) + \\
          G(1,3)G(1,4)\frac{\delta \Sigma^{\star}(3,4)}{\delta G(5,6)}
          G(5,7)G(6,8) \tilde{\Gamma}(7,8;2)
\hspace{0.5 cm} \text{.}
\end{multline}

Come abbiamo gi\`{a} visto tentare di costruire un'equazione di Dyson a partire da questa si
incontrano difficolt\`{a} legate con il fatto che il kernel \`{e} una grandezza a quattro punti.
D'altra parte nel capitolo sulle eccitazioni abbiamo rinunciato al tentativo di continuare
a lavorare con grandezze a due punti poich\'{e} abbiamo individuato l'importanza fisica dell'interazione
particella-buca in tale ambito.

Cercando invece di lavorare comunque con grandezze a due punti possiamo definire l'ordine zero
della polarizzabilit\`{a} come
$\Pi_0(1,2)=G(1,2)G(2,1)$ e riscrivere l'equazione moltiplicando l'ultimo termine per 
$\Pi_0(1,1') \Pi_0^{-1}(3,1')=\delta(1,3)$ ottenendo:
\begin{multline}
\Pi^{\star}(1,2)= \Pi_0(1,2) + \Pi_0(1,1')\Pi_0^{-1}(3,1')G(3,4)G(3',5) \\
          \frac{\delta \Sigma^{\star}(4,5)}{\delta G(6,7)}
          G(6,8)G(7,9) \tilde{\Gamma}(8,9;2)
\hspace{0.5 cm} \text{.}
\end{multline}
Come in precedenza consideriamo l'equazione di Hedin per il vertice (\ref{vertice}) che per\`{o}
riscriviamo utilizzando le regole di derivazione a catena
\begin{equation}
\begin{split}
\tilde{\Gamma}(1;2,3)&=\ \delta(1,2)\delta(1,3)+\frac{\Sigma^{\star}(2,3)}{\delta V(1)}     \\
                     &=\ \delta(1,2)\delta(1,3)+\frac{\Sigma^{\star}(2,3)}{\delta \rho(4)}
                         \frac{\delta \rho(4)}{\delta V(1)}
\hspace{0.5 cm} \text{.}
\end{split}
\end{equation}
Utilizzando infine la definizione di funzione risposta possiamo scrivere un'equazione di
tipo Dyson per la polarizzazione ridotta a due punti
\begin{equation}
\Pi^{\star}(1,2)=\Pi_0(1,2)+\Pi_0(1,3) f_{xc}^{eff}(3,4) \Pi^{\star}(4,2)
\hspace{0.5 cm} \text{,}
\end{equation}
dove abbiamo definito il kernel dell'equazione usando la notazione della referenza
\cite{Fabien_kernel}
\begin{equation}
f_{xc}^{eff}(1,2)= -i\Pi_0^{-1}(1,3)G(3,4)G(5,3)\frac{\delta \Sigma(4,5)}{\delta \rho(2)}
\hspace{0.5 cm} \text{.}
\end{equation}

%********************************************************************************************
\subsection{Un kernel per la TDDFT}
Il kernel $f_{xc}^{eff}$ viene costruito nella MBPT per cercare di ottenere un'approssimazione
successiva a quella GW che possa essere inclusa come grandezza a due punti
\cite{Tesi_dot_Fabien,Fabien_kernel}. Un'altra importante
applicazione risulta per\`{o} dal fatto che il kernel ottenuto pu\`{o} essere utilizzato per
ottenere il kernel $f_{xc}$ della TDDFT in modo esatto, almeno in linea di principio
\cite{Tesi_dot_Fabien}.

Per vedere questo partiamo nuovamente dall'equazione di Dyson per la funzione di Green
(\ref{Green2}) scrivendo la self energia completa come $\Sigma_H^{\star}=V_H+\Sigma^{\star}$
e quindi il potenziale di Hartree come $V_H=v_{KS}-v_{xc}$:
\begin{equation}
G(1,2)=G_0(1,2)+G_0(1,3)\left(\delta(3,4)(v_{KS}(4)-v_{xc}(4))+\Sigma^{\star}(3,4)\right)G(4,2)
\hspace{0.5 cm} \text{,}
\end{equation}
scriviamo quindi la corrispondente equazione per la funzione $G^{-1}$
\begin{equation}
G^{-1}(1,2)= G_0^{-1}(1,2)-\delta(1,2)(v_{KS}(2)-v_{xc}(2))-\Sigma^{\star}(1,2)   \\
\hspace{0.5 cm} \text{,}
\end{equation}
e deriviamo rispetto alla, usando il fatto che la variazione della funzione $G_0^{-1}$
\`{e} nulla:
\begin{equation}
\frac{\delta G^{-1}(1,2)}{\delta \rho(3)}=
-\delta(1,2)(\frac{\delta v_{KS}(2)}{\delta \rho(3)}-\frac{\delta v_{xc}(2)}{\delta \rho(3)})
-\frac{\delta \Sigma^{\star}(1,2)}{\delta \rho(3)}   \\
\hspace{0.5 cm} \text{.}
\end{equation}
Moltiplichiamo quindi l'espressione risultante per $iG(4,1)G(2,4^+)$ e
usiamo la relazione (\ref{rel-Green}) e le
definizioni di $\Pi_0$, $\chi_{KS}$ e $f_{xc}^{eff}$ ottenendo:
\begin{equation}
\delta(3,4)= \Pi_0(4,2)(\chi_{KS}^{-1}(2,3)-f_{xc}(2,3))-iG(4,1)G(2,4^+)\frac{\delta \Sigma^{\star}(1,2)}{\delta \rho(3)}
\hspace{0.5 cm} \text{.}
\end{equation}
Infine raccogliamo la funzione $\Pi_0$ e usiamo la definizione introdotta di $f_{xc}^{eff}$
ottenendo cos\`{i} una relazione esatta tra il kernel della TDDFT e $f_{xc}^{eff}$:
\begin{equation}
f_{xc}(1,2)=\left(\chi_{KS}^{-1}(1,2)-\Pi_0^{-1}(1,2) \right)+f_{xc}^{eff}(1,2)
\hspace{0.5 cm} \text{.}
\end{equation}

Poich\'{e} siamo interessati alla correzione all'approssimazione adiabatica dobbiamo ora
applicare la trasformata di Fourier all'equazione scritta. Omettendo per semplicit\`{a}
di notazione gli indice delle variabili spaziali otteniamo
\begin{equation} \label{kernel-Fabien}
f_{xc}(\omega)= \left(\chi_{KS}^{-1}(\omega)-\Pi_0^{-1}(\omega) \right)+f_{xc}^{eff}(\omega)
\hspace{0.5 cm} \text{.}
\end{equation}
Quella appena scritta \`{e} un'equazione esatta per il kernel della TDDFT che pu\`{o}
essere utilizzata per superare l'approssimazione adiabatica utilizzata nell'ambito
di questa tesi.

%% file: conclusioni.tex
%%%%%%%%%%%%%%%%%%%%%%%%%%%%
%Stile nell'impaginazione
%%%%%%%%%%%%%%%%%%%%%%%%%%%%
\pagestyle{fancy}
\fancyhf{}
\fancyhead[LE,RO]{\bfseries\thepage}
%\fancyhead[LO]{\bfseries\rightmark}
%\fancyhead[RE]{\bfseries\leftmark}
%\renewcommand{\headrulewidth}{0.5pt}
%\renewcommand{\footrulewidth}{0pt}
%\addtolength{\headheight}{0.5pt}
%\fancypagestyle{plain}{%
%              \fancyhead{}
%              \renewcommand{\headrulewidth}{0pt}}

Il computo dello spettro di eccitazione e degli stati eccitati di un sistema fisico
tramite l'utilizzo di software numerico richiede una quantit\`{a} di risorse molto
maggiori di quelle necessarie per calcoli di stato fondamentale \cite{Onida}.
Nella presente tesi \`{e} stato utilizzato un approccio che da questo punto di vista
si \`{e} rivelato piuttosto agile, perlomeno per i test eseguiti, che sono stati lanciati
su di un singolo calcolatore e che nel peggiore dei casi hanno richiesto un tempo di calcolo
di qualche ora. I risultati crediamo siano incoraggianti, ma la teoria necessita ancora
ulteriori sviluppi.

L'analisi del formalismo di Casida mette in luce come il computo
dell'energia di stato fondamentale e degli autovalori di Kohn e Sham possa essere
legato alle energie di eccitazione del sistema preso in considerazione. L'utilizzo
di tale formalismo evidenzia come lo spettro di eccitazione del sistema sia ottenuto
a partire dalla differenza tra le autoenergie di Kohn e Sham tramite due correzioni:
la prima dovuta al kernel della TDDFT e quindi agli effetti di campo medio e
di scambio e correlazione,
la seconda dovuta al mix tra le transizioni di Kohn e Sham necessario per rispettare
le simmetrie di spin del sistema. Questo secondo punto, in particolare, delinea la
differenza principale tra sistemi a spin equilibrato e sistemi con stato fondamentale
spin-polarizzato.

Abbiamo messo in evidenza quali siano allo stato attuale dell'arte i limiti di questo tipo
di approccio legati alle approssimazioni eseguite.
Innanzi tutto l'approssimazione LDA per il computo dello stato fondamentale del sistema
porta ad un errore nella stima dell'energia degli stati occupato rispetto al livello
di vuoto e come conseguenza all'impossibilit\`{a} di individuare alcune delle eccitazioni
del sistema. Il problema era gi\`{a} stato riscontrato in letteratura \cite{Tesi_dot_Myrta,
Casida3} e in particolare \`{e} dovuto al non corretto andamento del potenziale di scambio
e correlazione nella regione asintotica. Una possibile soluzione \`{e} quella di utilizzare 
un potenziale di scambio e correlazione ricavato a partire dalla MBPT, soluzione presentata
nel Capitolo 4, equazione (\ref{Sham-Schluter});
sempre in letteratura possiamo trovare calcoli TDDFT eseguiti con un potenziale
corretto \cite{Tesi_dot_Myrta}. Rispetto ai calcoli trovati in letteratura, in cui il computo
degli stati viene eseguito con software che sviluppano le funzioni d'onda su una base di orbitali
localizzate, il lavoro della presente tesi \`{e} per\`{o} stato realizzato lavorando con un
codice a onde piane, il che ha permesso di individuare in modo pi\`{u} chiaro, dal punto di
vista dell'autore, la presenza di stati di vuoto.

Il secondo problema \`{e} invece legato all'approssimazione adiabatica per il kernel di scambio
e correlazione per il computo degli stati eccitati. Questo tipo di approssimazione infatti
distrugge la simmetria di spin nel caso di un sistema con stato fondamentale spin-polarizzato
e porta come conseguenza alla mancanza di alcune eccitazioni e
alla comparsa di eccitazioni che non presentano la corretta simmetria di spin.
Nell'ambito di questa tesi abbiamo proposto, per lo spettro calcolato, un criterio che ci 
permetta di individuare quali siano le energie di eccitazione soggette a tale problema e
nella parte finale, equazione (\ref{kernel-Fabien}), abbiamo illustrato una possibile
soluzione per superare il problema incontrato
partendo dal kernel dell'equazione di Bethe-Salpeter.
Rispetto all'approssimazione LDA, la correzione dell'approssimazione adiabatica
\`{e} uno degli argomenti di cui si occupa attivamente la ricerca e, a nostra conoscenza, non
sono ancora stati eseguiti calcoli teorici su sistemi isolati in cui i problemi connessi
all'applicazione di tale approssimazione ai sistemi spin-polarizzati
siano stati superati con successo. In letteratura viene comunque proposto un metodo per
individuare le eccitazioni
che non presentano la corretta simmetria di spin \cite{Casida4}.
Sempre in letteratura \`{e} possibile
trovare punti di partenza
alternativi a quello da noi proposto per superare tale approssimazione \cite{Casida2}.

All'interno di questa tesi \`{e} stato infine mostrato come il formalismo di Casida
dovrebbe essere modificato per lo studio di sistemi con stato fondamentale spin polarizzato,
polarizzazione non collineare, e per lo studio di una perturbazione di carattere pi\`{u}
generale di quella considerata da Casida (parte conclusiva del Capitolo 6).

%% file: appendix.tex
%%%%%%%%%%%%%%%%%%%%%%%%%%%%
%Stile nell'impaginazione
%%%%%%%%%%%%%%%%%%%%%%%%%%%%
\pagestyle{fancy}
\renewcommand{\chaptermark}[1]
             {\markboth{#1}{}}
\renewcommand{\sectionmark}[1]
             {\markright{\thesection\ #1}}
\fancyhf{}
\fancyhead[RE,RO]{\bfseries\thepage}
\fancyhead[LO]{\bfseries\leftmark}
\fancyhead[LE]{\bfseries\leftmark}
\renewcommand{\headrulewidth}{0.5pt}
\renewcommand{\footrulewidth}{0pt}
\addtolength{\headheight}{0.5pt}
\fancypagestyle{plain}{%
              \fancyhead{}
              \renewcommand{\headrulewidth}{0pt}}

\chapter{Descrizione di molecole biatomiche}
%**************************************************************************************
\section{Classificazione degli orbitali molecolari}
Gli autostati elettronici di una qualsiasi molecola non possono essere
classificati sulla base del valore del momento angolare totale, in modo analogo
a quanto viene fatto per gli atomi isolati,
poich\'{e} il campo a cui sono soggetti gli elettoni non gode di simmetria centrale e
dunque il momento angolare non \`{e} una grandezza conservata del sistema.
Per le molecole lineari in generale e
biatomiche in particolare  \`{e} per\`{o} conservata la proiezione del momento angolare
lungo l'asse che unisce i due atomi; gli stati elettronici possono dunque essere
classificati in base a questa proiezione: $m_z$.
Nelle molecole come negli atomi vengono utilizzati come riferimento
gli orbitali a particella singola, anche quando la molecola presenta un numero superiore ad
$1$ di elettroni. Tale situazione riflette anche l'impiego di autofunzioni a particella
singola nello schema teorico della DFT, Kohn e Sham (sistema ausiliario non interagente), 
e nello schema delle teorie di campo medio (Hartree e Hartee-Fock).

Vengono dunque classificati come orbitali di tipo $\sigma$ quegli orbitali che hanno
proiezione del momento angolare nulla, orbitali di tipo $\pi$ gli orbitali con proiezione
$m_z=\pm 1$, orbitali $\delta$ quegli orbitali con proiezione $m_z=\pm 2$; raramente
vengono considerati valori maggiori della proiezione del momento angolare \cite{Landau3}.
Osserviamo che tutti gli stati con momento angolare $0$ avranno degenerazione pari ad $1$,
mentre tutti gli altri stati sono doppiamente degeneri in assenza di campi magnetici.

\begin{comment}
Le funzioni d'onda degli orbitali di tipo $\sigma$ pu\`{o} inoltre essere classificata in
base all'autovalore dell'operatore di riflessione rispetto a qualsiasi piano passante
per l'asse della molecola: vengono indicati con $\sigma^+$ 
\end{comment}
%**************************************************************************************
\section{Classificazione delle eccitazioni molecolari}
Per quanto concerne la classificazione delle eccitazioni a molti elettroni
per molecole biatomiche
esse verrano di conseguenza indicate in base al valore della proiezione del
momento angolare totale dello stato finale. Poich\'{e} quasi per la
maggior parte delle molecole lo stato fondamentale presenta $m_z=0$%,
il valore $m_z$ dello stato finale coincider\`{a} con la variazione
dello stesso nell'eccitazione.
Un'eccitazione che coinvolge due stati di tipo $\sigma$ potr\`{a}
essere solo un'eccitazione con variazione del momento angolare pari a $0$ e sar\`{a}
dunque indicata come eccitazione $\Sigma$, dove viene utilizzata la lettera greca
maiuscola corrispondente allo stato con lo stesso momento angolare.
Un'eccitazione che coinvolge uno stato $\sigma$ ed uno stato $\pi$ pu\`{o} avere una
variazione del momento angolare pari al solo valore $1$ e dunque avremo un'eccitazione
di tipo $\Pi$. Possiamo cos\'{i} costruire la seguente tabella per alcune eccitazione
molecolari.
\begin{table}[!ht]
\begin{center}
\begin{tabular}{|c|c|c|c|c|}
\hline
Stato      &  Stato   & $\delta m_z$&  tipo di      & degenerazione      \\
iniziale   &  finale  & possibile        &  eccitazione  & dell'eccitazione   \\
\hline
$\sigma$   &  $\sigma$  &  0     & $\Sigma$             &  1      \\
$\sigma$   &  $\pi$     &  1     & $\Pi$                &  2      \\
$\sigma$   &  $\delta$  &  2     & $\Delta$             &  2      \\
$\pi$      &  $\sigma$  &  1     & $\Pi$                &  2      \\
$\pi$      &  $\pi$     &  0,2   & $2\ \Sigma$,$\Delta$ &  1,1,2  \\
%.....     &  .....     & .....  & .....                & .....   \\ 
\hline
\end{tabular}
\end{center}
\end{table}

%***************************************************************************************
%***************************************************************************************
%***************************************************************************************
\chapter{Differenti rappresentazioni di un operatore}
In meccanica quantistica una qualsiasi grandezza fisica pu\`{o} essere rappresentata da
un operatore auto-aggiunto che agisce sullo spazio di Hilbert delle funzioni d'onda.
Tale operatore pu\`{o} essere rappresentato in modi differenti a seconda che si scelga di
utilizzare il formalismo della prima o della seconda quantizzazione e che si scelga di
rappresentare tale operatore ``nel punto'' o come nella sua forma ``completa''. Scopo di
quest'appendice \`{e} chiarire la notazione che abbiamo utilizzato nel presente lavoro; a
questo scopo consideriamo l'operatore potenziale di Hartree per la TDDFT.

In prima quantizzazione l'operatore viene rappresentato come
\begin{equation}
\hat{v}^I_H[\rho](\mathbf{x},t)=\int d^3\mathbf{x'}
\frac{\rho(\mathbf{x'},t)}{|\mathbf{x'}-\hat{\mathbf{x}}|}
\hspace{0.5 cm} \text{,}
\end{equation}
mentre in seconda quantizzazione la sua rappresentazione nel puntuale \`{e}
\begin{equation}
\begin{split}
\hat{v}_H[\rho](\mathbf{x},\mathbf{y},t)&=\ 
\langle\mathbf{x}|\hat{v}^I_H[\rho](\mathbf{z},t)|\mathbf{y}\rangle
\hat{\psi}^{\dag}(\mathbf{x})\hat{\psi}(\mathbf{y}) \\
   &=\ v_H[\rho](\mathbf{x},t)\langle \mathbf{x}|\mathbf{y}\rangle 
       \hat{\psi}^{\dag}(\mathbf{x})\hat{\psi}(\mathbf{y})  \\
   &=\ v_H[\rho](\mathbf{x},t)\delta^3(\mathbf{x}-\mathbf{y})
       \hat{\psi}^{\dag}(\mathbf{x})\hat{\psi}(\mathbf{x})
\hspace{0.5 cm} \text{,}
\end{split}
\end{equation}
il che mette in modo naturale in evidenza il fatto che stiamo lavorando con
un operatore locale, ovvero diagonale in spazio $\mathbf{x}$. Le formule appena
scritte chiariscono inoltre quale sia la relazione tra $\hat{v}_H(\mathbf{x},t)$
e la grandezza $v_H(\mathbf{x},t)$ senza cappuccio. 

L'apice all'operatore scritto in prima quantizzazione ``I'' serve a distinguerlo
dall'operatore scritto in seconda quantizzazione. In effetti $\hat{v}^I_H(\mathbf{x})$
va inteso
in modo differete dall'operatore $\hat{v}_H(\mathbf{x})$, poich\'{e} quest'ultimo agisce soltanto
nel punto $\mathbf{x}$ mentre $\hat{v}^I_H(\mathbf{x})$ agisce su tutto lo spazio.
Altra differenza fondamentale \`{e} data dal fatto che l'operatore in prima quantizzazione
\`{e} un operatore a particella singola, mentre l'operatore in seconda quantizzazione
\`{e} un operatore a molti corpi.
Ha dunque senso parlare di operatore puntuale e operatore completo solo in seconda quantizzazione,
mentre in prima quantizzazione esiste una forma unica dello stesso, che per\`{o} pu\`{o}
essere utilizzata solo per lo studio di un singolo elettrone%
\footnote{L'estensione ad un operatore a un corpo per $n$ elettroni sarebbe
dato dalla somma di $n$ operatori differenti}.

La forma completa dell'operatore in seconda quantizzazione \`{e} invece rappresentata
dalla somma sulla base degli operatori di campo della forma puntuale%
\footnote{Notiamo come la variabile temporale venga trattata in
modo differente da quelle spaziali. La differenza nasce dal fatto che il formalismo scritto
abbia come punto di partenza il formalismo hamiltoniano che tratta in modo differente lo
spazio ed il tempo.}%
:
\begin{equation}
\begin{split}
\hat{v}_H[\rho](t)=\int \int d^3\mathbf{x} d^3\mathbf{y}
             v_H[\rho](\mathbf{x},t)\delta^3(\mathbf{x}-\mathbf{y})
             \hat{\psi}^{\dag}(\mathbf{x})\hat{\psi}(\mathbf{x})
\hspace{0.5 cm} \text{.}
\end{split}
\end{equation}
Lo scrivere l'operatore completo anzich\`{e} semplicemente la sua versione puntuale nasce
in modo del tutto naturale in seconda quantizzazione se si cerca di cambiare base.
Per scrivere infatti un operatore su di una base qualsiasi non dobbiamo far altro che 
proiettare l'operatore su tale base. Il risultato \`{e} il seguente:
\begin{equation}
\begin{split}
\hat{v}_H[\rho](t)&=\sum_{ij}
\langle i|\hat{v}_H(t)[\rho]|j\rangle 
\hat{\psi}_i^{\dag}\hat{\psi}_j  \\
&=\sum_{i,j} v^H_{ij}[\rho](t)\hat{\psi}_i^{\dag}\hat{\psi}_j
\hspace{0.5 cm} \text{.}
\end{split}
\end{equation}
dove evidentemente siamo dovuti partire dall'operatore completo e non dalla sua versione
puntuale per ottenere le componenti in spazio $(i,j)$; se fossimo partiti dalla sua forma
puntuale avremmo ottenuto un potenziale $v^H_{ij}[\rho](\mathbf{x},t)$, rappresentazione
che non \`{e} n\'{e} in spazio $(\mathbf{x})$ n\'{e} in spazio delle
configurazioni.

\begin{comment}
Scrivendo infine il valor medio dell'operatore rispetto allo stato fondamentale in
prima e seconda quantizzazione,
\begin{equation}
\begin{split}
\langle\hat{v}^I_{H}\rangle&=\ \int d^3\mathbf{x}_1\dots d^3\mathbf{x}_n
\Psi^*_0(\mathbf{x}_1,\dots,\mathbf{x}_n,t)
\hat{v}^I_H(\mathbf{x},t)\Psi_0(\mathbf{x}_1,\dots,\mathbf{x}_n,t)  \\
 &=\ int d^3\mathbf{x} v_H(\mathbf{x},t)\rho(\mathbf{x},t)
\end{split}
\end{equation}
\end{comment}

%************************************************************************************************
%************************************************************************************************
%************************************************************************************************
\begin{comment}
\chapter{Nuovo set di equazioni di Hedin}
\begin{flalign}
\label{Hedin's equations new}
&G(1,2) = g_{Hxc}(1,2) + g_{Hxc}(1,3) \tilde{\Sigma}^{\star}(3,4)(3,4)G(4,2)   \\
&W(1,2) = w(1,2) + \big(w(1,3)+f_{xc}(1,3)\big) \tilde{\Pi}(3,4) W(4,2)   \\
&\hbar \tilde{\Pi}(1,2) = -i \tilde{\Gamma}(1;3,4) G(2,3) G(4,2)     \\
&\hbar \tilde{\Sigma}^{\star}(1,2) = i \tilde{\Gamma}(4;1,3) G(3,2) W(2,4)-\delta(1,2)v_{xc}(2)  \\
&\tilde{\Gamma}(1;2,3)=\delta(1,2) \delta(1,3)+   \nonumber \\
&\phantom{\Gamma(1;2,3)=\delta(1,2)}\hbar^{-1}\tilde{\Gamma}(1;4,5) G(6,4) G(5,7)
              \frac{\delta \tilde{\Sigma}^{\star}(2,3)}{\delta G(6,7)} 
\hspace{0.5 cm} \text{.}
\end{flalign}

\end{comment}

%% file: ringraziamenti.tex
\chapter*{Note finali}
\addcontentsline{toc}{chapter}{Note Finali}
Il lavoro della presente tesi \`{e} stato realizzato grazie al supporto che mi \`{e} stato dato
da molte persone che fanno parte del \emph{Nanoquanta Network of excellence} \cite{Nanoquanta}.
L'esperienza di quest'anno di lavoro mi ha permesso infatti di conoscere, almeno in parte, la realt\`{a}
della ricerca all'interno di differenti universit\`{a} europee e grazie al mio relatore, il prof. Giovanni
Onida, sono stato messo in contatto con differenti progetti di livello europeo e anche mondiale
che sono alla base del mondo stesso della ricerca.

Innanzi tutto il progetto Nanoquanta appunto, progetto fondato nell'ambito del ``Sixth Framework Programme''
della Comunit\`{a} Europea e che coinvolge dieci centri di ricerca dislocati in sette stati del'Unione
Europea. Sono venuto in contatto con questo progetto soprattutto partecipando, in qualit\`{a} di spettatore,
al ``Nanoquanta Workshop'' tenutosi a Houffalize (Belgium) nel settembre 2006, dove ho potuto assistere
alla presentazione di numerosi articoli e lavori di ricerca tra i quali quello dello stesso prof. Mark Casida,
autore del formalismo che \`{e} stato studiato all'interno di questa tesi%
\footnote{Grazie ai contatti con lo stesso Casida \`{e} stato possibile ottenere un preprint del lavoro che
egli sta attualmente svolgendo e che abbiamo quindi utilizzato per un confronto con i risultati
della presente tesi.}. Quindi il progetto ``ETSF'' (European Theoretical Spectroscopy Facility
\cite{ETSF}), che potremmo definire il futuro del progetto
Nanoquanta e che si pone come obiettivo quello di sviluppare una serie di strumenti teorici e sperimentali,
all'interno dei centri di ricerca che ne fanno parte, che permettano di descrivere le propriet\`{a}
spettroscopiche e non solo di un qualsiasi materiale.

Grande importanza hanno rivestito inoltre per questa tesi, cos\`{i} come la rivestono
per la comunit\`{a} scientifica, i
software open sources, che vengono sviluppati e condivisi gratuitamente%
\footnote{E pensare che Bill Gates \`{e} diventato uno degli uomini pi\`{u} ricchi del modo 
vendendo un prodotto di cui esiste una \emph{valida} alternativa distribuita in modo gratuito!},
tra i quali Linux \cite{Linux}, che costituisce il sistema operativo sul quale ho costantemente lavorato
e sul quale funziona in particolare Abinit \cite{Abinit}. Quest'ultimo \`{e} un software che permette
di realizzare calcoli di struttura della materia di differente tipo e costituisce appunto il codice
a cui ho in parte lavorato e che mi ha permesso di ottenere i risultati che sono riportati all'interno
di questa tesi. In particolare ho partecipato, come sviluppatore, al Workshop Abinit tenutosi a
Liège (Belgium) nel Gennaio 2007 mostrando durante una breve presentazione orale le modifiche da me
implementate nel programma e soprattutto venendo a contatto con una parte
della comunit\`{a} mondiale degli sviluppatori e degli utenti del programma.

Un punto di riferimento importante \`{e} stato infine costituito dai professori Xavier Gonze,
leader del progetto Abinit, e
Gian-Marco Rignanese con i quali ho lavorato per circa cinque mesi a Louvain la Neuve (Belgium) grazie
al sostegno del progetto Erasmus che ha permesso il mio soggiorno all'estero mettendo in contatto 
l'Universit\`{a} degli Studi di Milano con L'UCL (Université Catholique de Louvain), al contributo 
economico del progetto CNISM, che ha sostenuto il presente progetto di ricerca e infine, nuovamente,
al prof. Giovanni Onida che mi ha messo in contatto con l'universit\`{a} del Belgio.

%\section*{Fisica, Matematica e altre scienze}
%********************************************************************************************
%********************************************************************************************
%********************************************************************************************
\chapter*{Ringraziamenti}
\addcontentsline{toc}{chapter}{Ringraziamenti}
Ci\`{o} che \`{e} stato esposto all'interno di questa tesi \`{e} il risultato di circa un anno di
lavoro e la conclusione di un percorso di studio di cinque anni. Tale lavoro \`{e} stato reso
possibile grazie al sostegno di molti che mi sono stati compagni lungo il cammino svolto. Il primo
grazie spetta ai miei genitori e a mio fratello, ma anche ai cugini agli zii e alle nonne,
che sono stati la mia famiglia non solo in questi
cinque anni di universit\`{a}; a loro
devo non solo il supporto economico datomi ma soprattutto il sostegno e la sicurezza offertami.
Per tutto il mio periodo di studi la mia famiglia ha infatti rappresentato un posto in cui rifugiarmi
tutte le volte che ne avevo bisogno, un posto sereno lontano dalle preoccupazioni e dalle difficolt\`{a}
che ognuno di noi \`{e} chiamato ad affrontare nel proprio percorso. Grazie.

Dopo la mia famiglia un punto di riferimento importante sono stati gli amici di Gessate, coloro con i
quali sono cresciuto e ho diviso molte delle esperienze che ho vissuto: le uscite la sera, le
gite in montagna e a sciare, le prime vacanze al mare con gli amici e le esperienze con l'Oratorio 
e le lunghe discussioni sui massimi sistemi. Sono trascorsi molti anni e con alcuni di coloro che
sono stati i miei migliori amici ormai non mi sento che ogni tanto, \`{e} strano come
tutto sia destinato a cambiare. A Marco, Daniele, Fabio, Nino, Marco, Valeria, Marzia,
Irene, Anna, Don Stefano e tanti altri. Grazie.

Ancora un grazie a tutti coloro che sono stati miei compagni e amici anche se solo per un breve periodo
di tempo: gli amici della montagna che per qualche ragione sono capitati a Narro a trascorrere con me
numerose estati; i compagni di calcio dell' A.C.Gessate che per tante partite hanno indossato con me la
stessa maglia; tutti gli Erasmus del CLL che con me hanno condiviso l'esperienza di un semestre a Louvain
la Neuve; gli esponenti dello "zoccolo duro" compagni di un'esperienza che dura fin dalle scuole medie e
infine gli amici del liceo.
Grazie.

Un ringraziamento speciale ai miei compagni di universit\`{a} che forse pi\`{u} di tutti hanno condiviso con
me le difficolt\`{a}, le gioie e le esperienze di questi cinque anni. Quando penso al mio periodo di studi in
universit\`{a} ricordo soprattutto le due settimane bianche, le estati in barca, l'inter-rayl, la vacanza in
Scozia la laurea triennale e molti altri momenti insieme. Sono forse stati questi gli avvenimenti che pi\`{u}
mi hanno segnato e mi hanno cambiato negli ultimi cinque anni. A Marco (Bingo), Giorgio (Jo), Luca (Ter1),
Ricccardo (Ricky), Giovanni (Gio), Guglielmo (Piccoletto) e Luca (Trippy) e ad altri compagni di
universit\`{a}. Grazie.

Un grosso grazie a colei con la quale per la prima volta ho provato l'emozione di dividere molto pi\`{u}
che qualche esperienza, dividere la propria vita in ogni suo aspetto ed in ogni sua forma e che pi\`{u} di tutti
in questo ultimo anno \`{e} stata con me nei momenti belli e in quelli pi\`{u} difficili. Grazie.

Un grazie ai professori e alle persone che mi hanno seguito in questo lavoro di tesi
per circa un anno. Un grazie infine a tutti coloro che non ho nominato ma che per un
motivo o per l'altro mi sono stati di sostegno. Grazie.